\documentclass[fleqn,usenatbib]{mnras}
\DeclareUnicodeCharacter{03A6}{$\Phi$}

\usepackage{newtxtext,newtxmath}

\usepackage[T1]{fontenc}

\DeclareRobustCommand{\VAN}[3]{#2}
\let\VANthebibliography\thebibliography
\def\thebibliography{\DeclareRobustCommand{\VAN}[3]{##3}\VANthebibliography}
\DeclareUnicodeCharacter{2212}{-}
\DeclareUnicodeCharacter{0301}{-}

\usepackage{graphicx}	
\usepackage{amsmath}	

\usepackage{longtable}
\usepackage{booktabs}   
\usepackage{caption}

\newcommand{\blue}{\textcolor{blue}}

\title[MIGHTEE: Evolving Radio Luminosity Functions]{MIGHTEE: The evolving radio luminosity functions of star-forming galaxies to $z\sim 4.5$ and the cosmic history of star formation}

\author[N. J. Thykkathu et al.]{
Nijin J. Thykkathu$^{1,2}$\thanks{E-mail: nijin.thykkathu@physics.ox.ac.uk},
Matt J. Jarvis$^{1,3}$,
Imogen H. Whittam$^{1,3}$,
C. L. Hale$^{1,4}$,
A. M. Matthews$^{5}$,
\and
I. Heywood$^{1,6,7,8}$,
Eliab Malefahlo$^{3,9}$,
R. G. Varadaraj$^{1}$,
N. Stylianou$^{1}$,
Chris Pearson$^{1,2,6}$,
\and
Nick Seymour$^{11}$,
Mattia Vaccari$^{12,13,14}$
\\
$^{1}$Astrophysics, Department of Physics, University of Oxford, Keble Road, Oxford, OX1 3RH, UK \\
$^{2}$RAL Space, UKRI STFC Rutherford Appleton Laboratory, Chilton, Didcot, Oxfordshire OX11 0QX, UK\\
$^{3}$Department of Physics and Astronomy, University of the Western Cape, Robert Sobukwe Road, 7535 Bellville, Cape Town, South Africa\\
$^{4}$Institute for Astronomy, University of Edinburgh, Royal Observatory, Blackford Hill, Edinburgh, EH9 3HJ, UK\\
$^{5}$Carnegie Observatories, 813 Santa Barbara Street, Pasadena, CA 91101, USA\\
$^{6}$Department of Physics and Electronics, Rhodes University, PO Box 94, Makhanda, 6140, South Africa\\
$^{7}$SKA Observatory, Jodrell Bank, Lower Withington, Macclesfield, SK11 9FT, UK \\
$^{8}$South African Radio Astronomy Observatory, Liesbeek House, River Park, Gloucester Road, Mowbray, 7700, South Africa \\
$^{9}$UNISA Centre for Astrophysics and Space Sciences, College of Science, Engineering and Technology, University of South Africa, Cnr Christian de Wet Rd \\ \quad and Pioneer Avenue, Florida Park, 1709 Roodepoort, South Africa\\
$^{10}$Department of Physical Sciences, The Open University, Milton Keynes MK7 6AA, UK\\
$^{11}$International Centre for Radio Astronomy Research, Curtin University, GPO Box U1987, Bentley, WA 6845, Australia\\
$^{12}$ Inter-University Institute for Data Intensive Astronomy, Department of Astronomy, University of Cape Town, 7701 Rondebosch, Cape Town, South Africa \\
$^{13}$ Department of Physics and Astronomy, University of the Western Cape, 7535 Bellville, Cape Town, South Africa \\
$^{14}$ INAF - Istituto di Radioastronomia, via Gobetti 101, 40129 Bologna, Italy \\
}

\date{Accepted XXX. Received YYY; in original form ZZZ}

\pubyear{2026}

\begin{document}
\label{firstpage}
\pagerange{\pageref{firstpage}--\pageref{lastpage}}
\maketitle

\begin{abstract}
A key question in extragalactic astronomy is how the star-formation rate density (SFRD) evolves over cosmic time. A powerful way of addressing this question is using radio-continuum observations, where the radio waves are unaffected by dust and are able to reach sufficient resolution to resolve individual galaxies.
We present an investigation of the 1.4 GHz radio luminosity functions (RLFs) of star-forming galaxies (SFGs) and Active Galactic Nuclei (AGN) using deep radio continuum observations in the COSMOS and XMM–LSS fields, covering a combined area of $\sim 4$\,deg$^2$. These data enable the most accurate measurement of the evolution in the SFRD from mid-frequency radio continuum observations.
We model the total RLF as the sum of evolving SFG and AGN components, negating the need for individual source classification.
We find that the SFGs have systematically higher space densities at fixed luminosity than found in previous radio studies, but consistent with more recent studies with MeerKAT. We attribute this to the excellent low-surface brightness sensitivity of MeerKAT.

We then determine the evolution of the SFRD. Adopting the far-infrared -- radio correlation results in a significantly higher the SFRD at $z > 1$, compared to combined UV and far-infrared measurements. However, using more recent relations for the correlation between star-formation rate and radio luminosity, based on full spectral energy distribution modelling, can resolve this apparent discrepancy. Thus radio observations provide a powerful method of determining the total SFRD, in the absence of dust-sensitive far-infrared data.
\end{abstract}

\begin{keywords}
galaxies: evolution -- galaxies: active -- galaxies: star formation -- radio continuum: galaxies
\end{keywords}



\section{Introduction} \label{introduction}

The cosmological evolution of star-forming galaxies (SFGs) and active galactic nuclei (AGN) is critically important for our general understanding of the evolution of galaxies and the build-up of structure in the Universe.
Galaxy evolution is a combination of long-term interactions between internal and external environmental processes, involving the stars, gas clouds, morphological structure, and the small and large-scale environment. All of these can influence the level of star formation (SF) and AGN activity. 

 In the last decade, multi-wavelength observations of galaxy populations have increased dramatically but there is still no clear understanding of the core mechanisms that control the star formation rate histories of individual galaxies. One of the reasons is that it is not possible to follow individual galaxies throughout their evolution and statistical methods that investigate many galaxies at different epochs have to be used. These studies are further complicated by the vast array of selection biases depending on the wavelength that is used for the observations, coupled with how measurements at these wavelengths translate to estimates of physical properties, such as star-formation rate (SFR) \citep[e.g.][]{Kennicutt_1998,Kennicutt_2012}.

The radio continuum emission observed in SFGs primarily arises from synchrotron radiation, a phenomenon originating from the acceleration of high-energy electrons within supernova remnants. These electrons gain energy through interactions within the remnants, particularly during the cataclysmic explosions that occur when massive stars undergo supernova events. This synchrotron emission mechanism tends to dominate the radio emission spectrum at rest-frame frequencies of $<$ 30 GHz \citep{1992ARA&A..30..575C}.
The radio emission emanating from SFGs exhibits a strong correlation with the far-infrared emission, a relationship commonly referred to as the FIR-radio correlation \citep{Yun2001,Jarvis2010, Read2018,Algera2020} and, given the far-infrared emission is a good tracer of the star-formation rate in galaxies,  the SFR -- radio correlation \citep[e.g.][]{Bell2003,Wang2019,Smith2021,Gurkan2021,Cook2024}.

Furthermore such measurements are not affected by dust obscuration, thus potentially leading to an unbiased view of the star-formation history of the Universe, providing semi-independent measurements of the the cosmic star-formation rate density up to $z\sim 2$ that can also be traced by conbining UV, optical and far-infrared wavelengths \citep[e.g.][]{Karim2011,Novak_2017,Malefahlo_2021,Gentile2025}. Recently, using extremely deep radio continuum observations, \cite{Cochrane2023,matthews2024confirmationsubstantialdiscrepancyradio}  have shown that the SF history traced by radio emission may actually exceed that from UV and far-infrared tracers at $z<1.5$. 

As we move to higher redshifts  ($z>3$), inverse Compton scattering of the Cosmic Microwave Background photons may also become important \citep[e.g.][]{Murphy2009}. Indeed, evidence for this has recently been found from stacking the radio emission from high-redshift Lyman-break galaxies, leading to a relative decrease in the observed radio continuum emission at a given SFR \citep{Whittam2025}. As such, determining the evolving star-formation rate density (SFRD) at even earlier epochs may present different challenges.

As efforts are made to utilise radio emission as a reliable indicator of star formation activity, a significant challenge arises when attempting to discern the contribution of AGN to the observed radio emission. This radio emission can often be mis-associated with the emission originating from star formation processes (and vice versa) \citep[e.g.][]{Kimball2011, Condon2013, White2015, White2017, Malefahlo2020,Macfarlane2021, Yue2025}, all of which highlight the problem of measuring the star-formation contribution in optically-selected radio-quiet quasars, which are known to have an AGN, are not selected via their radio emission, but still often have detectable radio emission. The situation can become even more complicated for radio-selected samples, where it is not known {\em a priori} whether the radio sources have an AGN at all.
This potential confusion necessitates careful consideration and analysis to disentangle the distinct sources of radio emission and to accurately interpret the underlying astrophysical phenomena. The most robust method to account for this is with sensitive high-resolution studies with very-long baseline interferometry \citep[e.g.][]{Muxlow2020,Radcliffe2021, Morabito2025}. However, such data is often limited to small areas and relatively poor surface brightness sensitivity.

It has been known for several decades that high radio luminosity AGN ($L_{\rm rad} \gtrsim 10^{25}$\,W\,Hz$^{-1}$) evolve positively, meaning they become more common at higher redshift \citep[e.g.][]{DP90,Jarvis2001RLF,Willott2001,Rigby2011}, in a similar way to the cosmic SFRD. These AGN can also have a significant impact on galaxy evolution, where AGN outflows are often posited as being responsible for controlling or terminating star formation \citep[e.g.][]{2009Natur.460..213C,Fabian_2012,Best_2012,2014ARA&A..52..589H, Kondapally2023, Heckman2024, Hardcastle_2020}. On the other hand, the evolution of low luminosity ($L_{\rm rad} \lesssim 10^{25}$) AGN is not as well understood and some studies suggest that there is no evolution of the radio luminosity function \citep{Clewley_2004}, whereas some suggested they do evolve but slowly compared to the high luminosity AGN \citep{2009ApJ...696...24S,2f494631963a431cb58dfea47f6ad8e5, Kondapally2022}.

At low-redshift ($z \lesssim 0.2$), volume-limited samples of radio sources with complete redshift information and classification of sources into star-forming or AGN-dominated can be achieved using optical emission-lines. \cite{Mauch_2007} used 7824 radio sources divided into SFGs and AGNs using optical spectroscopy,  revealing that SFGs dominate the local radio luminosity function below $L_{1.4\text{GHz}} < 10^{23}$ WHz$^{-1}$, whereas AGN dominate above this threshold.
However, at higher redshifts, complete spectroscopy becomes more difficult to obtain and the emission lines used for distinguishing SFGs and AGN can be redshifted beyond the optical window, particularly at $z>1$. However, recent work using the overlap of the Dark Energy Spectroscopic Instrument \citep[DESI;][]{DESI} and the LOFAR two-metre sky survey deep fields \citep{Sabater2021,Tasse2021}, shows that this is now possible to $z\sim 1$ \citep[e.g.][]{Arnaudova2025}.

One method of avoiding the pitfalls in classifying radio sources into AGN and/or star-forming galaxies, in order to measure their evolution, is by measuring the complete radio luminosity function with a flexible enough parametrisation that allows for the contributions from both populations to be represented. This has been done at both radio wavelength \citep[e.g][]{McAlpine2013} and for the rest-frame UV luminosity function \citep[e.g.][]{Adams2023}.

Modelling the total RLF provides a statistical framework for separating SFGs and AGN based on their radio emission. By constructing separate RLF models for SFGs and AGN, but by fitting for them jointly, we can characterise the luminosity distribution for each population across different redshifts. Importantly, this method does not rely on classifications for every sources, but aims to understand the evolution of the populations as a whole, through statistical separation. In this paper, we adopt this strategy to measure the evolving radio-luminosity function using data from the MeerKAT International \,GHz Tiered International Exploration \citep[MIGHTEE;][]{Jarvis2016} survey.

This paper is organised as follows. In Section~\ref{sec:data}, we describe the MeerKAT data and the multi-wavelength ancillary data used throughout this work. Section~\ref{sec:completeness} outlines the completeness corrections applied. In Section~\ref{sec:radio_luminosity}, we describe the methods to measure and model RLFs and their evolution over cosmic time. In Section~\ref{sec:results_discussion}, we present the results of our fits to the RLFs of both SFGs and AGN and discuss how this translates to measurements of the evolution in the cosmic SFRD. Finally, in Section~\ref{sec:methods} we summarise our results and present our conclusions.

Throughout this paper we adopt the following cosmological parameters: $H_{0} = 70~\mathrm{km\,s^{-1}\,Mpc^{-1}}$, 
$\Omega_{\mathrm{M}} = 0.3$, and $\Omega = 0.7$. All magnitudes are AB magnitudes unless stated otherwise \citep{1983ApJ...266..713O}. We assume a spectral index defined as $S_{\nu} \propto \nu^{\alpha}$, with $\alpha = -0.7$, when converting flux density ($S_{\nu}$) to luminosity at a rest-frame frequency of 1.4\,GHz. Logarithms are base 10 unless stated otherwise.

\section{Data and sample selection}\label{sec:data}

In this paper we use the Early Science data in the COSMOS and the XMM-LSS fields from the MIGHTEE survey. For more details about the data and the observations, please refer to \cite{Heywood_2021}. The COSMOS observations consist of a single pointing with the MeerKAT telescope, covering $\sim$1.6 deg$^2$ down to a limiting thermal noise of $\sim$2\,$\mu$Jy\,beam$^{-1}$ at L-band. The larger XMM-LSS field consists of three overlapping pointings, covering $\sim$3.5\,deg$^{2}$. Two versions of the Stokes I image (and associated data products) are produced, one with higher sensitivity and slightly poorer resolution, and one for higher resolution at the expense of sensitivity. 
The maximum sensitivity image was produced with Briggs robust $= 0.0$, has synthesised beam size = 8.6 $\times$ 8.6 arcsec$^2$ and thermal noise $\sim$2\,$\mu$Jy\,beam$^{-1}$.
Although due to classical confusion, these observations are limited to a total noise of approximately 4$\mu$Jy beam$^{-1}$ at the centre of the primary beam.  The higher resolution image was produced with Briggs robust $= -1.2$, has a synthesised beam size $\sim $5 $\times$ 5 arcsec$^2$ and thermal noise $\sim$6\,$\mu$Jy\,beam$^{-1}$. Due to the wide bandwidth of the MeerKAT $L-$band receiver, the effective frequency varies across the primary beam and \cite{Heywood_2021} provide an effective frequency map for each image, which gives the effective frequency at each pixel.
Source extraction was conducted using the Python Blob Detection and Source Finder \citep[PyBDSF;][] {Mohan2015PyBDSFPB} with the default parameters in \cite{Heywood_2021}. The extracted catalogue contains 30{,}170 radio components across the two fields: 9{,}896 in COSMOS and 20{,}274 in XMM-LSS.

We use the cross-matched catalogues provided by \cite{whittam2023mighteemultiwavelengthcounterpartscosmos} for the COSMOS field and \cite{Zhu_2023} for the XMM-LSS field, to associate the sources with an optical counterpart. We associate these sources with the photometric redshifts determined using the Le Phare \citep{Ilbert2009} spectral energy distribution fitting code combined with machine learning photometric redshifts using GPz \citep{Almosallam2016a, Almosallam_2016b} as detailed in \citet{Hatfield_2022}, following similar work \citep{Duncan2018,Duncan2021}. 

 Specifically, we use optical data from the Canada–France–Hawaii-Telescope Legacy Survey \citep[CFHTLS;][]{Cuillandre2012}, and the HyperSuprimeCam Strategic Survey Programme \citep[HSC DR2;][]{Aihara2018a,Aihara2018b,Aihara2019}. The corresponding near-infrared data in the XMM-LSS field are from the VISTA Deep Extragalactic Observations (VIDEO) survey \citep{Jarvis2013}, while UltraVISTA DR6 \citep[][]{McCracken2012} provides the near-infrared coverage in COSMOS. For further details we refer the reader to \citet{Bowler2021,Adams2023,Varadaraj2023,Varadaraj2026}.
We use the full probability distribution function (PDF) for the photometric redshifts to measure the RLF accounting for uncertainties in redshift.

Spectroscopic redshifts are available over both fields from a range of surveys and we use the updated compilation from \cite{Khostovan2025} in COSMOS and also from \cite{Vaccari2022}, which include more recent data.  We also include additional spectroscopic redshifts from the Dark Energy Spectroscopic Instrument \citep[DESI; ][]{DESI}. For radio sources where spectroscopic redshifts are available, we use these, otherwise photometric redshifts are adopted.

We are unable to obtain reliable redshifts in regions where the optical/near-infrared imaging is compromised by bright stars and associated artefacts. We therefore apply a bright–star and coverage mask to the radio footprints in the MIGHTEE Early-Science COSMOS and XMM–LSS fields, excluding radio sources that fall inside masked pixels. The resulting effective area for each field is $\simeq0.83\ \mathrm{deg}^2$ (COSMOS) and $\simeq3.23\ \mathrm{deg}^2$ (XMM–LSS), giving a combined area of $\simeq4.06\ \mathrm{deg}^2$.

\begin{figure*}
    \centering
    \includegraphics[width=\textwidth]{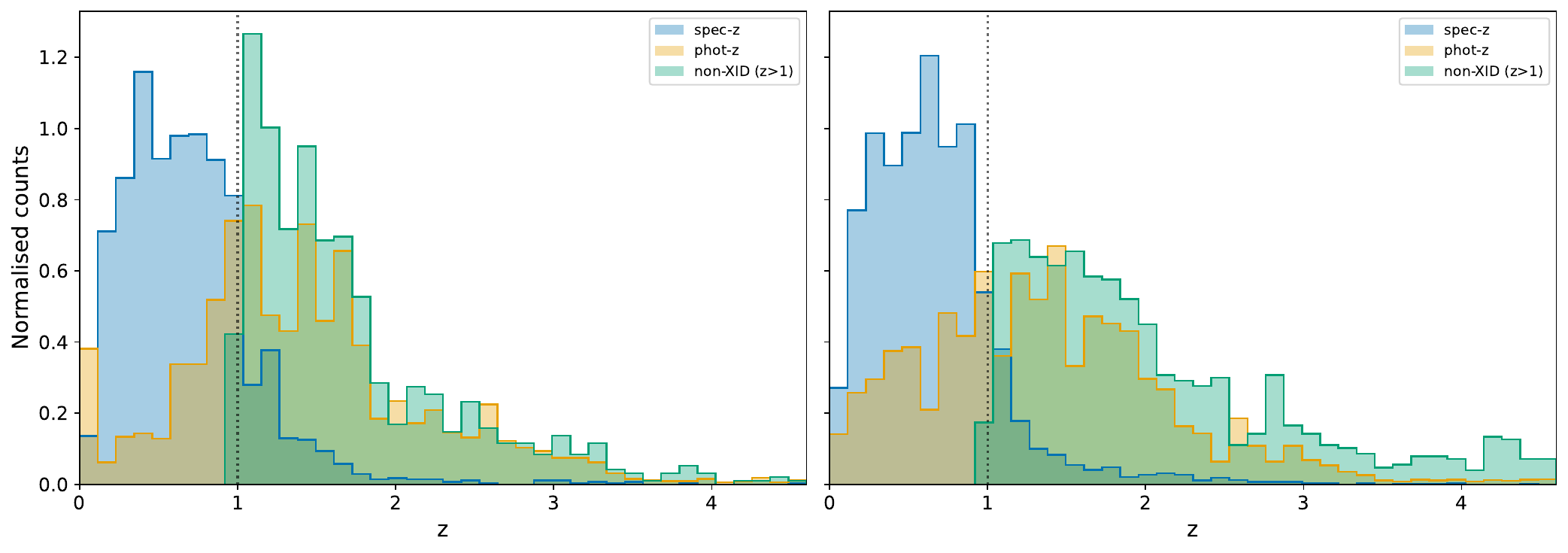}
    \caption{
    Normalised Redshift distributions for radio sources in the COSMOS (left) and XMM-LSS (right) fields.
    Each panel shows the normalised redshift histograms for sources with spectroscopic redshifts (spec-z; blue), photometric redshifts where no spec-z is available (phot-z; orange), and radio sources without optical counterparts plotted at $z$>1 (non‑XID; green). Non-XID sources are assigned redshifts using the flux-dependent method described in Section~\ref{sec:opticalcompleteness} and are shown here restricted to $z$>1.
    }
    \label{fig:histogram}
\end{figure*}

We adopt a conservative flux-density cut at 40$\mu$Jy (around 8-10$\sigma$ across both fields) to mitigate against confusion and significant completeness corrections (see Section~\ref{sec:completeness}). With this flux-density limit  and after masking, 6{,}039 radio sources remain in COSMOS. Of these, 830 lack an optical/NIR counterpart from the visually inspected catalogue of \cite{whittam2023mighteemultiwavelengthcounterpartscosmos}, yielding $C_{\rm opt}= 86.3$ percent. In the XMM–LSS field, 17{,}666 of 18{,}796 sources at $S>40\mu$Jy have an optical/NIR counterpart from the catalogue of \cite{Zhu_2023}. We note that the higher completeness in the XMM-LSS field arises from the different approaches to the cross-matching, with the approach taken in \cite{whittam2023mighteemultiwavelengthcounterpartscosmos} being more conservative than that used by \cite{Zhu_2023}. However, we note that we correct for the incompleteness due to missing counterparts in both fields in a similar way (see Section~\ref{sec:opticalcompleteness}), and the difference between the source cross-matching has a negligible effect on our results. The redshift distributions for both the COSMOS and XMM-LSS fields, denoting the spectroscopic versus photometric redshifts, are shown in Fig.~\ref{fig:histogram}

\section{Completeness corrections}\label{sec:completeness}

Completeness is a key concern in constructing the luminosity function from a flux-density limited survey. Completeness corrections become imperative for properly including faint radio sources in the analysis, where their detection is strongly correlated with the noise properties of the image. Neglecting this aspect can lead to a substantial underestimation of the space density. Furthermore, flux-density limited surveys are susceptible to the Malmquist bias, where more luminous objects at higher redshifts and fainter objects at lower redshifts are preferentially detected. This bias introduces a potential distortion in the observed sample, leading to skewed representations of source distributions and space densities. Addressing completeness and mitigating Malmquist bias is therefore paramount for ensuring the accuracy and reliability of the derived luminosity function. 

\subsection{Radio completeness correction}
The radio completeness quantifies the fraction of sources that are detected as a function of flux density. The noise properties of the images used in this analysis are not constant, with the noise increasing towards the edges due to the primary beam correction, so detectability varies across the field. We therefore use the results from \cite{Hale_2022}, who measured the completeness by injecting simulated radio sources into the residual maps and re-ran the source detection.

\begin{figure}
	\includegraphics[width=0.5\textwidth]{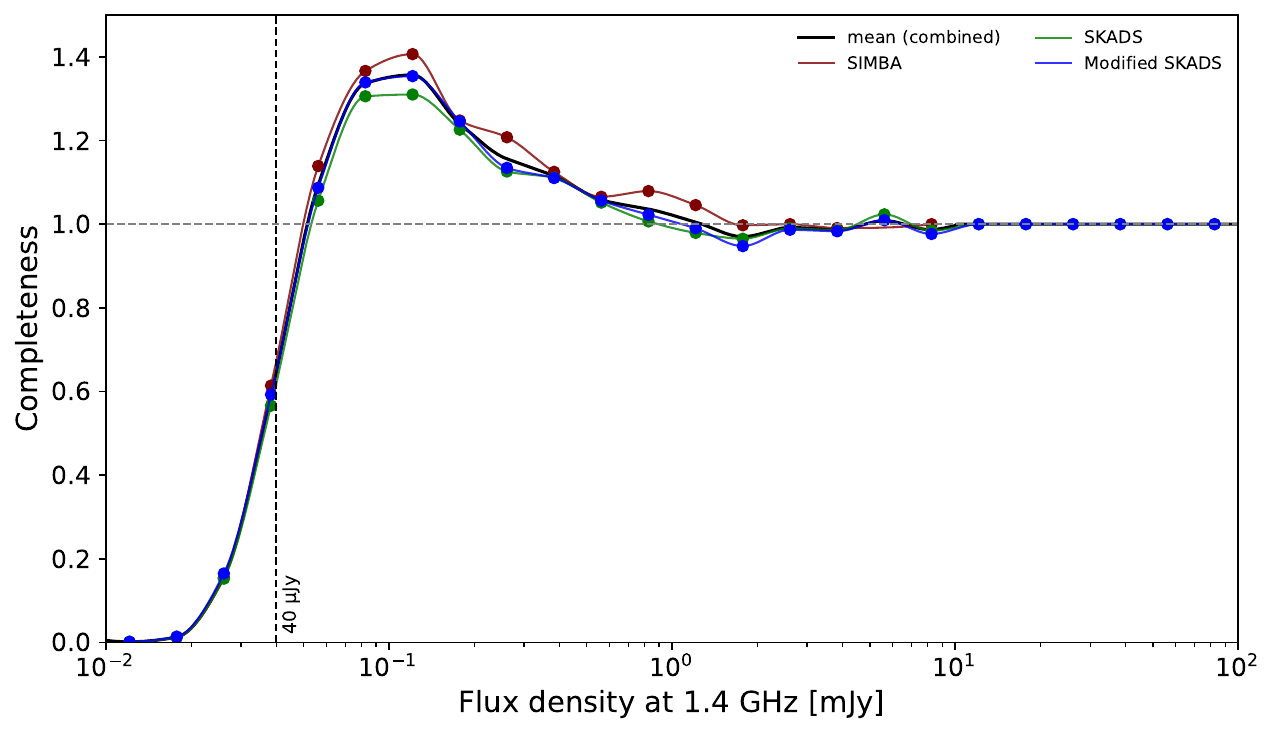}
	\centering
    \caption{Completeness of the 1.4\,GHz COSMOS catalogue as a function of input flux density. SKADS (green), modified SKADS (blue), SIMBA (maroon), the mean completeness is the back line. The vertical dashed line marks the uniform flux–density cut at 40\,$\mu$Jy; the horizontal dashed line marks 100\% completeness. Completeness values exceeding one intermediate flux densities result from noise shifting some sources from adjacent, lower-flux bins into brighter bins \citep[see][for details]{Hale_2022}.}
    \label{fig:CompletenessCurve}
\end{figure}

To summarise, three different input catalogues were used, based on SIMBA \citep{Dave_2019}; SKADS \citep[][]{Wilman_2008}; and modified SKADS. These sources were  injected and recovered in flux-density bins and the completeness was calculated as the ratio of the number of recovered to injected of sources in each bin. We then fit a smooth function to the binned completeness measurements using a monotonic cubic interpolation in $\log_{10} (S)$. At bright flux densities ($S> 10$~mJy), where the data are fully complete, we explicitly set the completeness to 1. The three separate simulation-based curves, shown in Fig.~\ref{fig:CompletenessCurve}, were then combined into a single mean completeness function as a function of 1.4-GHz flux density, which we apply as the radio correction in the following analysis.

In this work, we choose to use flux-density limit of $S \geq 40\mu$Jy, at which the completeness is 0.65. At this threshold, uncertainties associated with rms noise variations are minimized, which are more significant at lower flux-density limits and the measured flux densities becomes less affected by confusion noise. 

\subsection{Optical completeness correction}\label{sec:opticalcompleteness}

Within the unmasked area, some radio sources lack optical/NIR counterparts and thus have no redshift available, either photometric or spectroscopic. 
Moreover, the ancillary imaging depths differ between COSMOS and XMM–LSS, so simply discarding unmatched radio sources would bias the space densities (especially against dust-obscured and high-redshift systems) and could do so unevenly between fields. In our baseline analysis we include these sources (hereafter referred to as non-XID) by assigning statistical redshifts. Assigning statistical redshifts to radio-only detections in each field preserves total number counts, avoids counterpart-driven selection biases, and puts both fields on a more uniform footing before combining them to determine the RLF.

To achieve this, we started with the complete radio component catalogue, which  contains all detected components. 
Within each field we take the matched catalogue, divide it into log-spaced radio flux-density bins, and in each bin construct an empirical redshift PDF from the matched sources, smoothed with a kernel density estimator (KDE). For every unmatched radio source we locate its flux-density bin and draw a redshift from the corresponding KDE, if that bin is empty we use the KDE from the nearest non-empty flux bin. Redshifts assigned to non-XID sources are drawn once from these flux-density dependent KDEs and are held fixed across all Monte-Carlo realisations, although we note that resampling the non-XID sources does not make any difference to our results.

In our baseline analysis, we force all sources lacking optical identifications to be at $z>1$. Whilst approximately 50 per cent of radio sources with optical/NIR counterparts and brighter than 40$\mu$Jy typically reside at $z<1$, optically unidentified systems in this flux range are more likely to be faint galaxies that fall below the optical detection limits due to them residing at higher redshift \citep[see e.g.][]{Gentile2024,Gentile2025}. Adopting $z>1$ for these non-XID sources therefore provides a conservative estimate of their radio luminosities and avoids biasing our analysis toward low-redshift, optically bright populations. 
The assigned redshifts (shown in Fig.~\ref{fig:histogram}) are carried forward in the analysis alongside those objects with spectroscopic or photometric redshifts.

As a check on the impact of our baseline strategy for assigning redshifts to non-XID sources, we also repeat the analysis using a second redshift–assignment approach. We repeat the method of assigning redshifts to the non-XID sources, but without forcing them to be at $z> 1$. In this case the overall redshift distribution is similar to that of the redshift distribution of sources of similar flux-density distribution which have an optical/NIR counterparts. Adopting this strategy does not change our results significantly, as once the shape of the RLF is measured well at low redshift with a large fraction of spectroscopic redshifts, then how we spread the $\sim 10-15$ per cent of non-XID sources does not strongly affect evolution parameters.

Finally, to check how our method of sampling from the photometric redshift PDF may effect our results, we also adopt a `single-$z$' approach, in which every source is assigned a single "best" redshift taken directly from the catalogue (spectroscopic or the best-fit photometric redshift), rather than being sampled from its full redshift PDF. In this approach, all non-XID sources are placed at $z>1$, following a similar method to our baseline strategy.

The number of sources with spectroscopic and photometric redshifts in each radio luminosity and redshift bin is given in Table.~\ref{tab:combined_vmax_counts}, alongside the number of non-XID sources from our baseline model where they are forced to have $z>1$. 

\section{The Radio luminosity function}\label{sec:radio_luminosity}

In this section, we describe the form of the radio luminosity function (RLF) that we adopt in this analysis, how we calculate the RLF in redshift bins, and how we model its evolution.

\subsection{The $\mathbf{1/V_{\rm max}}$ Method}
\label{sec:vmax}

To measure the cosmic evolution of radio sources, we determine the RLFs for different redshift bins using the $1/V_{\rm max}$ method \citep{Schmidt_1968}.

\begin{figure*}
	\includegraphics[width=0.48\textwidth]{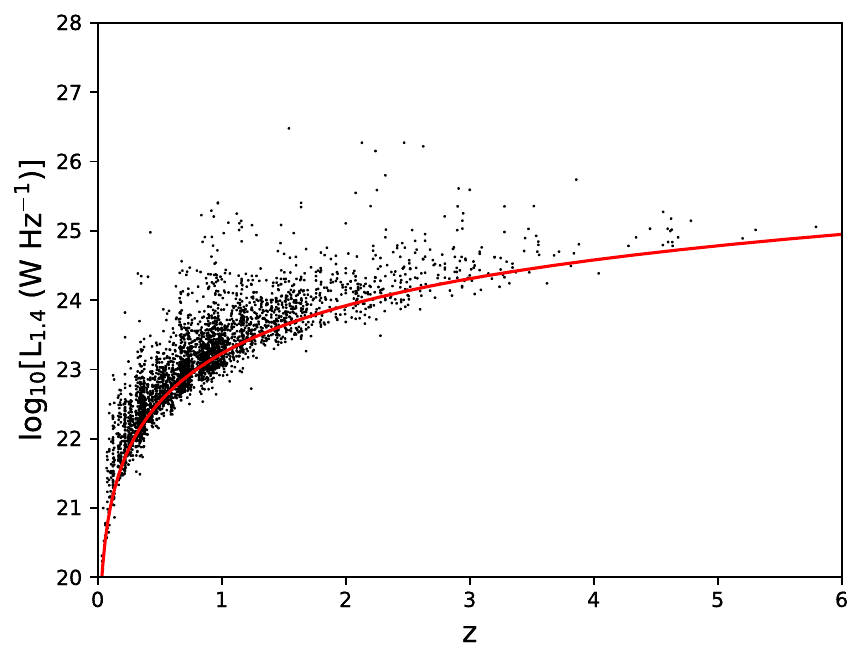}	
    \includegraphics[width=0.48\textwidth]{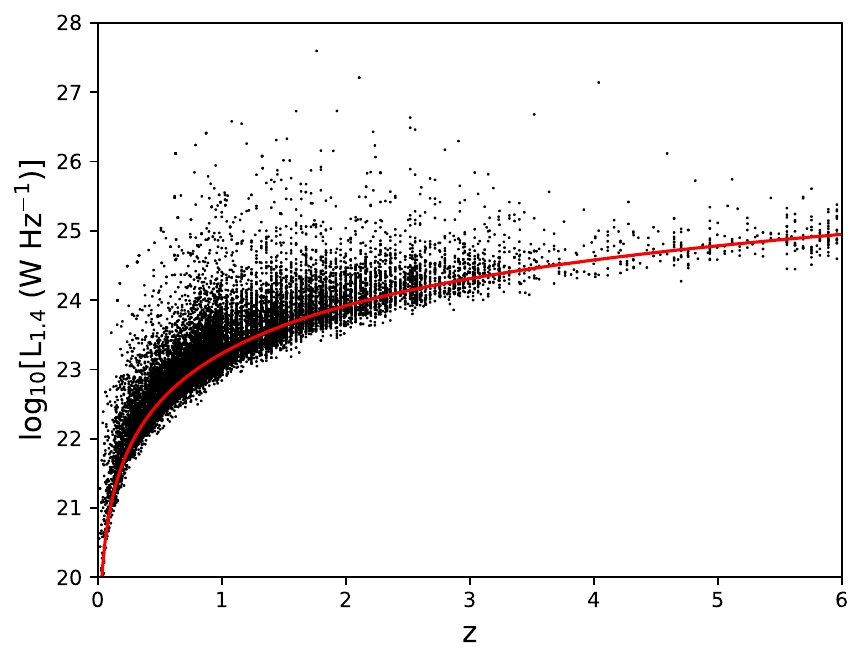}
	\centering
	\centering
    \caption{Rest-frame 1.4\,GHz luminosity versus redshift for COSMOS (left) and XMM–LSS (right). Black points are individual sources. The solid red curve corresponds to the adopted 40$\mu$Jy flux-density cut at the frequency.}
    \label{fig:cosomos_lookup}
\end{figure*}

We first calculate the rest-frame radio luminosity at 1.4~GHz, corresponding to the observed-frame flux density. The MIGHTEE data has a varying effective frequency across the image, with a mean effective frequency of $\nu_{\rm eff} \sim 1.28$~GHz. To determine the rest-frame luminosity at 1.4~GHz ($L_{1.4_{\rm rf}}$) we therefore use,
\begin{equation}
L_{1.4_{\rm rf}} = \frac{4\pi D_L^2}{(1+z)^{1+\alpha}} S_{\nu_{\rm eff}} \left(\frac{1.4}{\nu_{\rm eff}}\right)^{\alpha}
\end{equation}
where $D_{L}$ is the luminosity distance, $\alpha$ is the spectral index of the source\footnote{We assume $\alpha = −0.7$, which is typical for SFGs \citep{Condon_1992, Bell2003}} and $z$ is the redshift of the source. In practice we use the effective frequency associated with the position of the source in the continuum images as described in \cite{Heywood_2021}.
The rest-frame 1.4 GHz luminosity as a function of redshift for our sample is shown in  Fig.~\ref{fig:cosomos_lookup} for both the COSMOS and XMM-LSS fields.

The radio luminosity function (number density per unit \(\log L\)) is given by:
\begin{equation}\label{eq:rlf}
\phi (L,z) = \frac{1}{\Delta \log L} \sum_{i=1}^N  \frac{1}{V_{\rm max_{i}}} ,
\end{equation}
with the corresponding uncertainty given by \cite{1985ApJ...299..109M}:
\begin{equation}
\sigma_{\phi(L,z)}= \frac{1}{\Delta \log L} \sqrt{\sum_{i=1}^N  \frac{1}{V_{\rm max_i}^2}},
\end{equation}
where \(V_{\max,i}\) is the maximum comoving volume accessible to source \(i\) in the given \((L,z)\) bin.

We calculate $V_{\rm max}$ by dividing the sample into ten redshift bins spanning $0.2<z<4.6$. We limit the analysis to $z>0.2$ as we do not have the necessary cosmic volume to accurately measure the luminosity function at $z<0.2$. We adopt a luminosity–bin width of $\Delta\log L = 0.25$\,dex to strike a balance of preserving useful resolution across a reasonable range in luminosity while limiting the Poisson noise, and limiting leakage between adjacent bins due to photometric redshift uncertainties and noise fluctuations. At the bright end, where sources are sparse, we widen to $\Delta\log L = 0.4$ for $\log L \ge 26$ to compensate for the low source density.

For each source \(i\), \(V_{\max}\) is computed by summing the accessible comoving volume element within each redshift bin from \(z_{\min}\) to \(z_{\max}\), where \(z_{\max}\) is set by the redshift at which the observed flux density at 1.4~GHz drops to the survey selection limit ($S_{1.4} \geq 40\mu$Jy), and is capped at the upper edge of the relevant redshift bin. We also apply a completeness correction \(C(S_{1.4})\) using the empirical completeness curves shown in Fig.~\ref{fig:CompletenessCurve}, to obtain a completeness corrected measurement of the maximum accessible comoving volume for each source:

\begin{equation}\label{eq:Vmax}
V_{\max,i}\;=\;\sum_{z=z_{\min}}^{z_{\max}}\big[\,V(z+\Delta z)-V(z)\,\big]\;C(S_{1.4})\,.
\end{equation}

The incompleteness due to some radio sources lacking an optical/NIR counterpart is, as described in Section~\ref{sec:opticalcompleteness}, accounted for by including unmatched sources with statistical redshift values.

\subsection{Parameterisation of the Radio Luminosity Function}\label{sec:local_rlf}

 To quantify the combined RLF for both AGN and SFGs we consider the parametric forms used in previous work. Specifically, we adopt a double power law for the AGN, following \cite{Mauch_2007}:

\begin{equation}\label{agnlocal}
\phi_{0}^{\mathrm{AGN}}(L) = \frac{\phi_{*}^{\rm AGN}}{\left(\frac{L}{L_{*}^{\rm AGN}}\right)^{a} + \left(\frac{L}{L_{*}^{\rm AGN}}\right)^{b}}
\end{equation}
where $\phi_*^{\rm AGN}$ is the normalisation, $L_*^{\rm AGN}$ is the knee or break in the RLF, and $a$ and $b$ are the bright and faint end slopes.

For the SFG population we use the four parameter analytical form used by \cite{Novak_2017}, which is comprised of a power-law plus lognormal distribution from \cite{Saunders1990The6A}:

\begin{equation}\label{sflocal}
\phi_{0}^{\mathrm{SF}}(L) = \phi_{*}^{\rm SF}\left(\frac{L}{L_{*}^{\rm SF}}\right)^{1-\delta}
\exp{\left[-\frac{1}{2 \sigma^2}\log_{10}^2\left(1+\frac{L}{L_{*}^{\rm SF}}\right)\right]}
\end{equation}
where $\phi_*^{\rm SF}$ is the normalisation, $L_*^{\rm SF}$ is the knee of the distribution, $\delta$ is the faint-end slope and $\sigma$ describes the exponential fall off beyond the knee. 

Both the AGN and SFG populations are known to evolve strongly with redshift. This evolution is traditionally modelled as either; (i) pure luminosity evolution, whereby the specific population changes in luminosity with redshift, and the comoving space density of the population remains constant; (ii) pure density evolution, where the comoving space density evolves with redshift; or (iii) a combination of both of these evolutionary forms. In practice the data are often not sufficient to fully disentangle pure luminosity from pure density evolution or combinations thereof, as the data need to  sample below and above the knee or break in the luminosity function. This means that both deep and wide areas are necessary. This is particularly true for radio surveys, where we have a combination of both SFGs and AGN, which have different shaped luminosity functions and may evolve in different ways.

\begin{table}
\caption{Priors adopted for the free parameters in the model fitting.
Log-normal priors are assumed for the local SFG RLF parameters $\phi_{*}^{\rm SF}$ and $L_{*}^{\rm SF}$, and uniform priors are adopted for the remaining RLF parameters.
The local AGN RLF parameters are held fixed in all fits (see Table~\ref{tab:best_fit_params}).}

\label{tab:priors}
\renewcommand{\arraystretch}{1.5}
\begin{tabular}{|l@{\hspace{25pt}}|c|}
\hline
\textbf{Parameters} & \textbf{Priors} \\
\hline\hline
$\log_{10}(\phi_{*}^{\rm SF} / \mathrm{Mpc}^{-3}\,\mathrm{dex}^{-1})$
    & Gaussian $\sim (\mu = -2.45,\ \sigma = 0.30)$ \\
$\log_{10}(L_{*} ^{\rm SF} / \mathrm{W\,Hz}^{-1})$
    & Gaussian $\sim (\mu = 21.25,\ \sigma = 0.25)$ \\
$\delta$                      & Uniform $\in [-2, 5]$ \\
$\sigma$                      & Uniform $\in [0, 1]$ \\
$\alpha_L^{\mathrm{SF}}$               & Uniform $\in [0, 5]$ \\
$\beta_L^{\mathrm{SF}}$                & Uniform $\in [-1, 0]$ \\
$\alpha_D^{\mathrm{AGN}}$              & Uniform $\in [0, 5]$ \\
$\beta_D^{\mathrm{AGN}}$               & Uniform $\in [-2, 0]$ \\
\hline
\end{tabular}
\end{table}

For this study we adopt an approach driven by our core science aims, to measure the evolution of the SFGs using the radio continuum data, whilst accounting for the contribution to the total radio luminosity function from the AGN. We therefore fix the shape of the AGN luminosity function using the local AGN luminosity function from \cite{Novak_2018}, which is consistent with \cite{Mauch_2007}. We allow the AGN luminosity function to evolve using a pure density only evolution model, due to the fact that we do not have enough cosmological volume to accurately determine the break of the AGN part of the RLF, which is important for decoupling pure density from pure luminosity evolution. We note that our results for the SFG luminosity function do not change significantly if we adopt a pure luminosity evolution model for the AGN. We parameterise the evolution with,

\begin{equation}\label{sf_ple}
\phi(\mathrm{AGN}) = \phi_{0}^{\mathrm{AGN}}{(1+z)^{\alpha_{D}^{\mathrm{AGN}} + z\beta_{D}^{\mathrm{AGN}}}},
\end{equation}
where $\alpha_{D}^{\mathrm{AGN}}$ and $\beta_D^{\mathrm{AGN}}$ provide the flexibility to model the expected increase in comoving space density from low to high redshift, whilst also permitting a turnover beyond a given redshift.

For the SFGs, which are the focus of this work, we allow all the parameters of the local luminosity function (Eq.~\ref{sflocal}) to be free parameters, and adopt a pure luminosity evolution model, due to the fact that we predominantly sample beyond the knee in the luminosity function, given by:

\begin{equation}\label{agn_pde}
\phi(\mathrm{SF}) = \phi_{0}^{\mathrm{SF}}\left[\frac{L}{(1+z)^{\alpha_{L}^{\mathrm{SF}} + z\beta_{L}^{\mathrm{SF}}}}\right]
\end{equation}
where $\alpha_{L}^{\mathrm{SF}}$ and $\beta_L^{\mathrm{SF}}$ again provide flexibility for the luminosity evolution to increase and then decrease beyond a given redshift. 

We then fit for both the AGN evolution parameters ($\alpha_{D}^{\mathrm{AGN}}$ and $\beta_D^{\mathrm{AGN}}$) and the SFG evolution parameters ($\alpha_{L}^{\mathrm{SF}}$ and $\beta_L^{\mathrm{SF}}$) along with the overall shape of the SFG RLF, similar to \cite{McAlpine2013}.
One advantage of fitting the total RLF is that the inference does not depend on the specific galaxy classification scheme, provided the RLF shapes for the SFG and AGN populations are adequately specified.

\begin{table*}
\centering
\caption{Best-fit parameters for RLF of SFGs and AGN. We show the results using the two different approaches of assigning redshift statistically for those sources without an optical/NIR counterpart (non-XID), with those restricted to be at $z> 1$ shown in the first row, with row 2 showing the results when the non-XID sources are allowed to populate all redshift ranges. The third colum shows the results when the best-fit photometric redshift is used for those sources without a spectroscopic redshift. We fit a pure luminosity evolution (PLE) model for SFGs ($\alpha_L^{\mathrm{SF}},\beta_L^{\mathrm{SF}}$) and a pure density evolution (PDE) model for AGN ($\alpha_D^{\mathrm{AGN}},\beta_D^{\mathrm{AGN}}$). The local AGN LF parameters were fixed:
$\log_{10}(\phi_*^{\rm AGN} /\mathrm{Mpc^{-3}\,dex^{-1}}) = -5.1$, $\log_{10}(L_*^{\rm AGN} / \mathrm{W\,Hz^{-1}}) = 24.59$, $a = -1.27$ and 
$b = -0.49$.}
\label{tab:best_fit_params}
\setlength{\tabcolsep}{5.5pt}
\begin{tabular}{|l|c|c|c|c|c|c|c|c|}
\hline
Method &
$\log_{10}(\phi_*^{\rm SF}/\mathrm{Mpc^{-3}\,dex^{-1})}$ &
$\log_{10}(L_*^{\rm SF}/\mathrm{W\,Hz^{-1})}$ &
$\delta$ &
$\sigma$ &
$\alpha_L^{\mathrm{SF}}$ &
$\beta_L^{\mathrm{SF}}$ &
$\alpha_D^{\mathrm{AGN}}$ &
$\beta_D^{\mathrm{AGN}}$ \\
\hline\hline

$z>1$ &
$-2.00_{-0.11}^{+0.10}$ &
$21.32_{-0.20}^{+0.18}$ &
$1.48_{-0.26}^{+0.20}$ &
$0.514_{-0.029}^{+0.027}$ &
$3.87_{-0.11}^{+0.10}$ &
$-0.416_{-0.034}^{+0.034}$ &
$2.61_{-0.10}^{+0.09}$ &
$-0.640_{-0.044}^{+0.043}$ \\
\hline

Uniform-$z$ &
$-1.89_{-0.10}^{+0.08}$ &
$21.17_{-0.13}^{+0.17}$ &
$1.50_{-0.21}^{+0.19}$ &
$0.563_{-0.025}^{+0.028}$ &
$3.96_{-0.078}^{+0.079}$ &
$-0.423_{-0.025}^{+0.024}$ &
$2.26_{-0.12}^{+0.13}$ &
$-0.690_{-0.060}^{+0.058}$ \\
\hline

Single-$z$ &
$-1.89_{-0.07}^{+0.06}$ &
$21.09_{-0.11}^{+0.16}$ &
$1.32_{-0.21}^{+0.21}$ &
$0.527_{-0.026}^{+0.027}$ &
$4.19_{-0.13}^{+0.13}$ &
$-0.540_{-0.046}^{+0.049}$ &
$2.55_{-0.08}^{+0.08}$ &
$-0.546_{-0.031}^{+0.029}$ \\
\hline
\end{tabular}

\vspace{6pt}
\begin{minipage}{0.95\textwidth}
\small

\end{minipage}
\end{table*}

\subsection{Multinest Sampling}

To explore the multi-parameter likelihood space, in order to find the best-fit RLF parameters and associated posteriors, we use \textsc{MultiNest} \citep{2008MNRAS.384..449F}.
\textsc{MultiNest} is a Bayesian inference algorithm designed to efficiently calculate the Bayesian evidence and explore complex parameter spaces. \textsc{MultiNest} returns the full posterior distribution, enabling robust parameter estimation and accurate uncertainty quantification. We use the Python interface \textsc{PyMultiNest} \citep{Buchner_2014}.

In Bayesian inference we seek the posterior
\begin{equation}\label{eq:posterior}
P(\boldsymbol{\theta}\!\mid\!D,H)=\frac{\mathcal{L}(D\!\mid\!\boldsymbol{\theta},H)\,\pi(\boldsymbol{\theta}\!\mid\!H)}{Z(D\!\mid\!H)}\,,
\end{equation}
where \(\boldsymbol{\theta}\) are the model parameters, \(D\) the data, \(H\) the model, \(\mathcal{L}\) the likelihood, and \(\pi\) the prior. The Bayesian evidence ($z$) given by:

\begin{equation}\label{eq:evidence}
Z(D\!\mid\!H) = \int \mathcal{L}(D\!\mid\!\boldsymbol{\theta},H)\,\pi(\boldsymbol{\theta}\!\mid\!H)\,{\rm d}\boldsymbol{\theta}
\end{equation}
normalises the posterior and enables quantitative model comparison.

We adopt uniform priors with bounds set by physical plausibility on all local–SFG luminosity function and all evolution parameters (Table~\ref{tab:priors}), apart from $L_{*}^{\rm SF}$ and $\phi_{*}^{\rm SF}$, where we adopt log-normal priors centred around the local values given by \cite{Novak_2017}. This is because our survey has limited volume at low redshift and both the normalisaton and knee in the luminosity function are relatively well-defined at low redshift. As we fit for the faint-end slope it also removes some of the degeneracy whilst retaining the flexibility for a flatter or steeper faint-end slope for the SFG luminosity function. We marginalise over the photometric–redshift uncertainty via a per–resample Monte–Carlo strategy. For each of \(N_{\rm rep}=100\) realisations (where $N_{\rm rep}$ is the number of photometric–redshift resampling iterations), the redshift of every sources is sampled from the PDF of its photometric redshift, while spectroscopic redshifts and non-XID values are kept fixed. The full binned RLF across all redshifts is recomputed, and a single \textsc{MultiNest} fit is run to obtain posterior samples for each realisation. The final photometric redshift posterior is formed by equal weight concatenation of the posterior samples from all \(N_{\rm rep}\) runs. This numerically integrates over the photometric redshift PDFs without reweighting and preserves the correct overall likelihood weight.

\section{Results and Discussion}

\begin{figure*}
    
    \includegraphics[width=\textwidth]{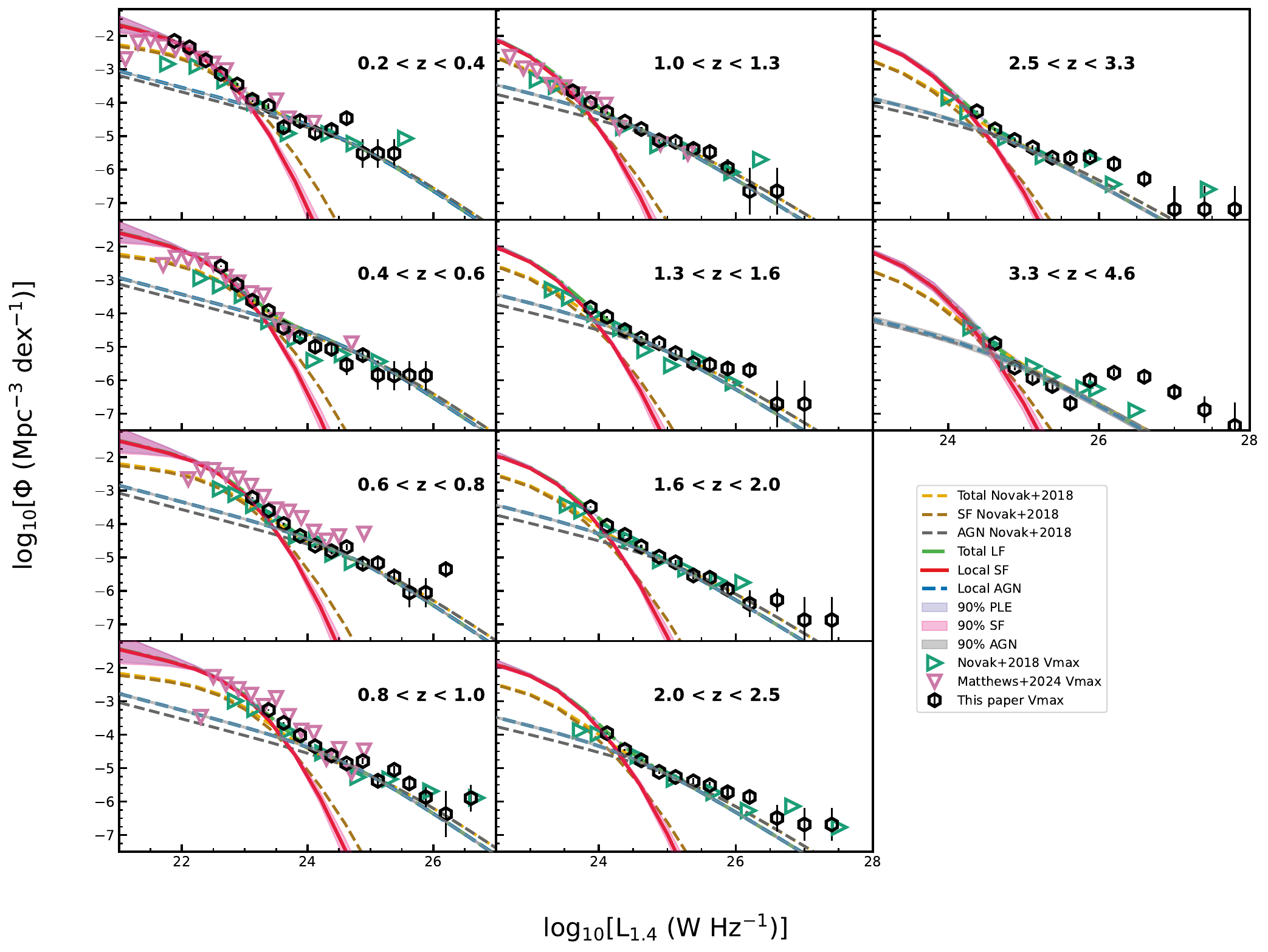}
    \centering
    \caption{
    Total 1.4\,GHz RLF in ten redshift bins from our 
    photometric redshift PDF-based analysis, including statistically assigned redshifts for the non-XID sources restricted to be at $z>1$.
    Black hexagons show the $1/V_{\rm max}$ measurements from this work.
    Red solid and blue dashed curves show the best-fit SF (PLE) and AGN (PDE) components, respectively, with magenta and light-blue bands show their 90 per cent confidence intervals.
    The yellow, brown and grey dashed curves denote the total, SFG, and AGN RLFs, respectively from \citet{Novak_2018}.
    For reference, the $1/V_{\rm max}$ points from \citet{Novak_2018} and \citet{matthews2024confirmationsubstantialdiscrepancyradio} are also shown as green right-pointing and magenta down-pointing triangles, respectively.
    }
    \label{fig:pdf_rlf_evolution}
\end{figure*}

\begin{figure*}
	\includegraphics[width=\textwidth]{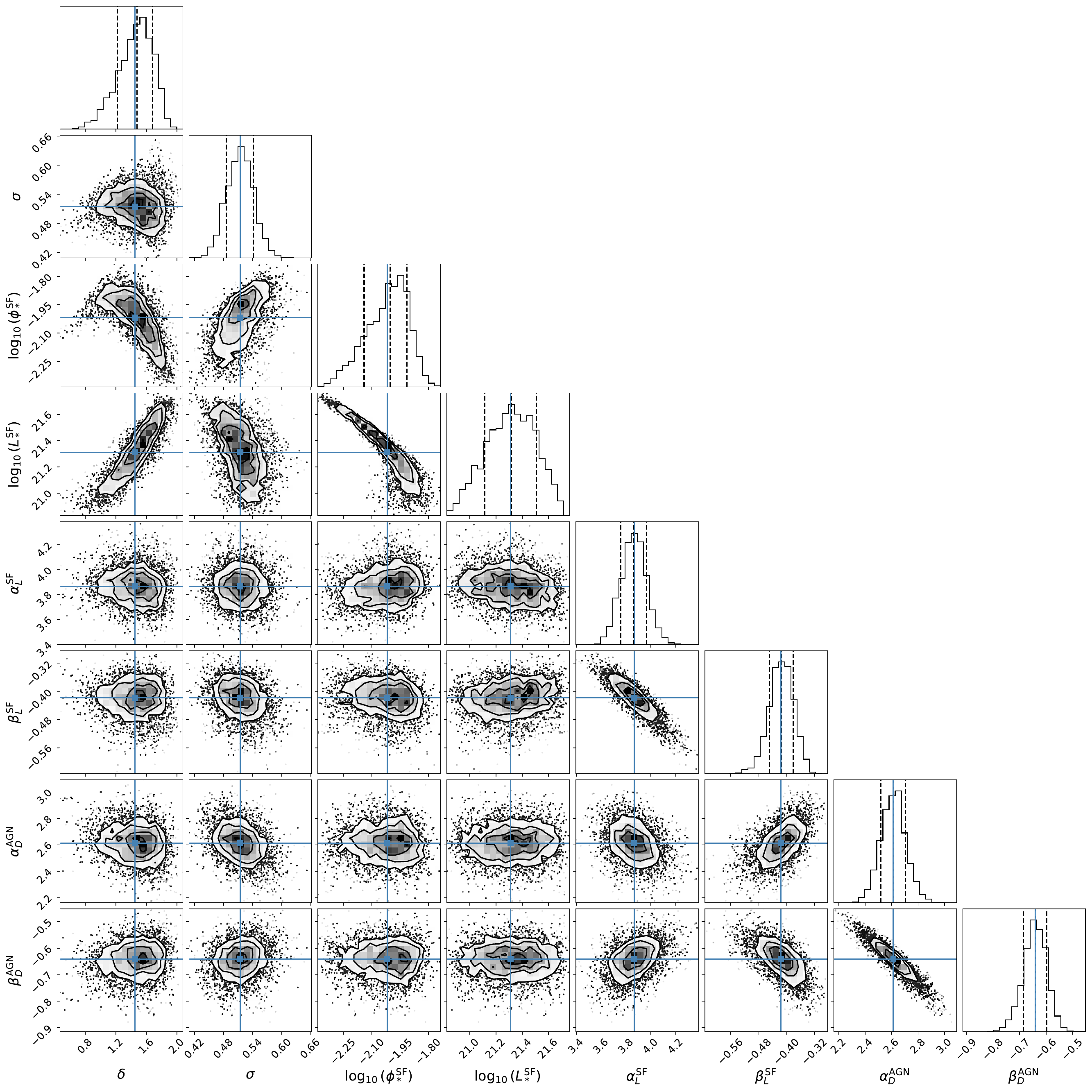}
	\centering
    \caption{
    Posterior distributions for the eight parameters of the total RLF model:
    SFG Schechter function parameters $(\delta, \sigma, \phi_{*}^{\rm SF}, L_{*}^{\rm SF})$,
    SFG luminosity evolution ($\alpha_L^{\mathrm{SF}}$, $\beta_L^{\mathrm{SF}}$), and AGN density evolution ($\alpha_D^{\mathrm{AGN}}$, $\beta_D^{\mathrm{AGN}}$).
    Contours show the 68th and 95th percentile confidence regions; vertical lines in the 1D marginals mark the median and 16th/84th percentiles.
    The SFG (PLE) and AGN (PDE) evolution pairs exhibit the expected anti–correlations, while cross–component couplings remain weak, indicating that the two channels are separately constrained.
    }
    \label{fig:pdf_corner_plot}
\end{figure*}

In Fig.~\ref{fig:pdf_rlf_evolution} we present the total RLF in redshift bins, showing our $1/V_{\max}$ measurements and the modelled redshift evolution of the SFG and AGN components alongside data points from previous studies. The corresponding posterior distributions of the model parameters for the case where we distribute non-XID radio sources at $z>1$ are shown in Fig.~\ref{fig:pdf_corner_plot}. For the other approaches, the total RLFs and the corresponding corner plots  
are shown in Figs. \ref{fig:rlf_uniform} and  ~\ref{fig:corner_plot_1} (uniform-$z$ for non-XID sources) and Figs.~\ref{fig:rlf_single_appendix} and \ref{fig:single_corner_plot_appendix} for the case of a single best redshift.

\label{sec:results_discussion}

\subsection{The local radio luminosity function of SFGs}
The first thing that is noticeable with our measured binned luminosity function values (Fig.~\ref{fig:pdf_rlf_evolution}) is that at low redshift and at the lowest luminosities they are significantly higher than the luminosity function point from \cite{Novak_2017}, who use data from the 3~GHz survey over the COSMOS field. However, our data are consistent with the other very deep MeerKAT survey over the DEEP2 field by \cite{matthews2024confirmationsubstantialdiscrepancyradio}. The key differences between the VLA-3GHz survey and both ours and the DEEP2 MeerKAT data are the observed frequency ($\sim 1.3$~GHz vs 3~GHz), but probably more importantly is the sensitivity of MeerKAT to low-surface brightness emission. The latter is likely the more important consideration, as many studies have shown that the spectral index distribution between 1.4 and 3~GHz is well behaved with a mean spectral index of $\alpha \approx 0.7$. Indeed, \cite{Hale_2022} showed that the MIGHTEE data detected lower surface brightness, but relatively high flux density ($\gtrsim$ 100\,$\mu$Jy), sources that were completely invisible in the VLA-3GHz survey data, even though the flux-density sensitivity would suggest that such objects should have been detected at $>50\sigma$. This is due to the core-dominated baseline distribution of the MeerKAT array, which results in both high sensitivity to diffuse emission, coupled with long enough baselines to image at $\sim 5$\,arcsec resolution. On the other hand, the VLA-3GHz survey was designed to maximise resolution using the VLA A-Array configuration to allow more robust cross-identification to optical counterparts and enabling studies of the radio morphology to help separate AGN from SFGs.
However, this survey design comes at the cost of losing low-surface brightness sensitivity due to the absence of high enough sensitivity on short baselines.  Although the VLA-3GHz survey complemented the A-array data with shallower C-array data, this clearly was insufficient to detect the faint low-surface brightness emission that is seen with MeerKAT.
All of this essentially leads to both more low-surface brightness galaxies being detected for a given total flux density and more emission being detected for galaxies already in the sample, thus moving them to higher radio luminosities, in the MeerKAT data.

Another difference between the MIGHTEE data used here, particular in the COSMOS field, and the VLA-3GHz work of \cite{Novak_2017}, is the availability of many more spectroscopic redshifts since 2017 (see Table.~\ref{tab:combined_vmax_counts}), particularly in these low-redshift bins, from DESI \citep[][]{DESI}. This means that our measured luminosity function is more robust at these redshifts and luminosities. Thus, the significantly higher normalisation at the faint end is indeed real, providing confirmation of similar results using MeerKAT data \citep{matthews2024confirmationsubstantialdiscrepancyradio}.

As the faint-end slope of the measured RLF is constrained primarily by the lowest redshift bin, we compare our $0.2<z<0.4$ RLF to the NVSS-based local 1.4\,GHz RLF of \cite{Mauch_2007}. We find a similar exponential drop beyond the knee of the RLF for SFGs, however, we find a significantly steeper faint-end slope, with \cite{Mauch_2007} finding $\delta = 1.02$ compared to our value of $\delta = 1.48$. Although consistent at the 2$\sigma$ level, the difference is also likely due to the relatively bright $K-$band magnitude limit coupled with the strong correlation between radio luminosity and $K-$band magnitude which leads to some incompleteness at the lowest radio luminosities in the local study of \cite{Mauch_2007}.

The other noticeable aspect in our measured luminosity functions is the upturn towards higher luminosities that is due to the contribution of the AGN, i.e. the total RLF is a combination of two underlying distributions of two different populations, where the radio emission is related to different physical process. Indeed, this is the reason we choose to model the total RLF as described in Section~\ref{sec:local_rlf}.

The observed higher normalisation of the low-luminosity end of the RLF clearly leads to a different best-fit functional form for the local radio luminosity function of SFGs, compared to \cite{Novak_2017}, which is the closest study in terms of depth and frequency, albeit with smaller area. We find a value of $\delta = 1.48^{+0.20}_{-0.26}$ for the faint-end slope of the SFG luminosity function. This value is steeper than the faint-end slope measured for local RLF of SFGs found by \cite{Novak_2017} of $\delta = 1.22$, although we note that they are formally consistent within the uncertainties. However, such a steep slope is consistent with the faint-end slope of the luminosity function at other wavelengths. For example, the ultraviolet luminosity function at low-redshift \citep{Arnouts2005} has a slope of $-1.5 < \delta < -1.2$.
At far-infrared wavelengths \cite{Koprowski2017} find a similar value of $\delta = -1.4$ (converting their definition of $\delta$ to that used from Eq.~\ref{sflocal}). We also note that in Fig.~3 of \cite{Novak_2017} the data points at low luminosity generally sit above their model faint-end slope and it is only the final points in each sample, where the completeness corrections are the most significant, that force the faint-end slope to be flatter than what we find here.

However, this steeper slope is also coupled with a significantly higher normalisation term $\log_{10}(\phi_{*}^{\rm SF} /$ Mpc$^{-3}$\,dex$^{-1}) = -2.00^{+0.10}_{-0.11}$, compared to $\log_{10}(\phi_{*}^{\rm SF} /$ Mpc$^{-3}$\,dex$^{-1}) = -2.45$ in \cite{Novak_2017}, even with a log-normal prior set around the latter value. Together, these provide a best-fit model that suggests there is a far greater number of SFGs in deep radio surveys than inferred from previous surveys that cannot detect the low-surface brightness diffuse emission. However, although the best-fit local luminosity function parameters for the SFG population are robust, Fig.~\ref{fig:pdf_corner_plot} shows that $\delta$, $\phi_{*}^{\rm SF}$ and $L_{*}^{\rm SF}$ are highly degenerate. This is a well known aspect of the functional form adopted, and emphasises the requirement to have a robust measurement of the luminosity function both above and below the knee. Indeed, this is the reason we fix the AGN-related part of the luminosity function to the local values.

\subsection{The evolution of the SFG and AGN luminosity function}

We next turn to the best-fit to the evolutionary parameters: $\alpha_{L}^{\mathrm{SF}}$ and $\beta_{L}^{\mathrm{SF}}$ for the pure luminosity evolution of the SFGs and $\alpha_{D}^{\mathrm{AGN}}$ and $\beta_{D}^{\mathrm{AGN}}$ for the pure density evolution of the AGN.
We find evidence for strong evolution of the SFG luminosity function, with $\alpha_L^{\mathrm{SF}} = 4.0$ with a shallow but significant turnover at high redshifts with $\beta_L^{\mathrm{SF}} = -0.42$ (for both cases of redistributing the non-XID radio sources). The strength of this evolution means that the comoving space density of SFGs increases by a factor of $\sim 10$ from $z\sim 0\rightarrow 2$. This evolution is similar in magnitude to that found at other wavelengths for SFGs, providing independent evidence that the combined SFG and AGN luminosity functions used in this work are behaving as one would expect in terms of their evolution. However, due to the fact that the normalisation of the local luminosity function in this work has a much higher value, coupled with a steep faint-end slope, we also obtain a significantly higher comoving space density at high redshift. This is largely due to the adoption of pure luminosity evolution and a faint-end slope and normalisation fixed by the low-redshift data. However, we also note that a pure-density evolution model produces a similar degree of evolution due to the need to fit the bright end of the SFG luminosity function at higher redshifts. We return to the implications of this for the evolution of the cosmic star-formation rate density in Section~\ref{subsec:sfrd}.

For AGN, the PDE fit ($\alpha_{D}^{\rm AGN}$, $\beta_{D}^{\rm AGN}$) produces a vertical shift in normalisation while keeping the RLF shape fixed. The comoving space density of AGN also rises from $z=0$ to $z\sim 1-2$ and then flattens, in agreement with previous deep radio surveys \citep{McAlpine2013, Smolcic2017AGN}. The posterior distributions (Fig.~\ref{fig:pdf_corner_plot}) show the expected anti–correlations within $\alpha_{L}^{\rm SF}$, $\beta_{L}^{\rm SF}$ and $\alpha_{D}^{\rm AGN}$, $\beta_{D}^{\rm AGN}$, with relatively weak cross–component couplings between the AGN and SFG luminosity functions, indicating that PLE (SFG) and PDE (AGN) are independently constrained.

These results support the adopted modelling choice of PLE for SFG and PDE for AGN. Our  data do not bracket the break in the RLF for AGN ($L_{*}^{\rm AGN}$) in most redshift bins, so the AGN LF is effectively single power–law over the luminosity range we probe; allowing the AGN shape to shift would be weakly constrained and largely degenerate with normalisation. Fixing the AGN shape and evolving only its density thus captures what the data measure most directly. In contrast, the SFG constraints are strongest near the bright-end of the SFG RLF, so a horizontal shift in $L_{*}^{SF}$ is well measured and physically interpretable as the movement of the knee in the SFG RLF to higher luminosities with increasing redshift.

The single–redshift analysis (Fig.~\ref{fig:rlf_single_appendix}), which assigns each source a single best redshift, likewise agrees with the $z>1$ results in overall shape and amplitude. Differences arise in the sparsest, highest–redshift bins and at the brightest luminosities, where the absence of smoothing over the photometric redshift PDFs can accentuate Poisson spikes. Refitting under the different treatments produces small, compensating shifts in local RLF parameters (e.g.\ $\phi_{*}^{\rm SF}$, $L_{*}^{\rm SF}$). Quantitatively, inferred evolutionary trends are statistically unchanged and the SFG/AGN separation remains consistent across redshift; any deviations are limited to minor shifts in a few $1/V_{\max}$ points at the highest luminosities.

\subsection{Star Formation Rate Density}
\label{subsec:sfrd}

\begin{figure*}
    \includegraphics[width=\textwidth]{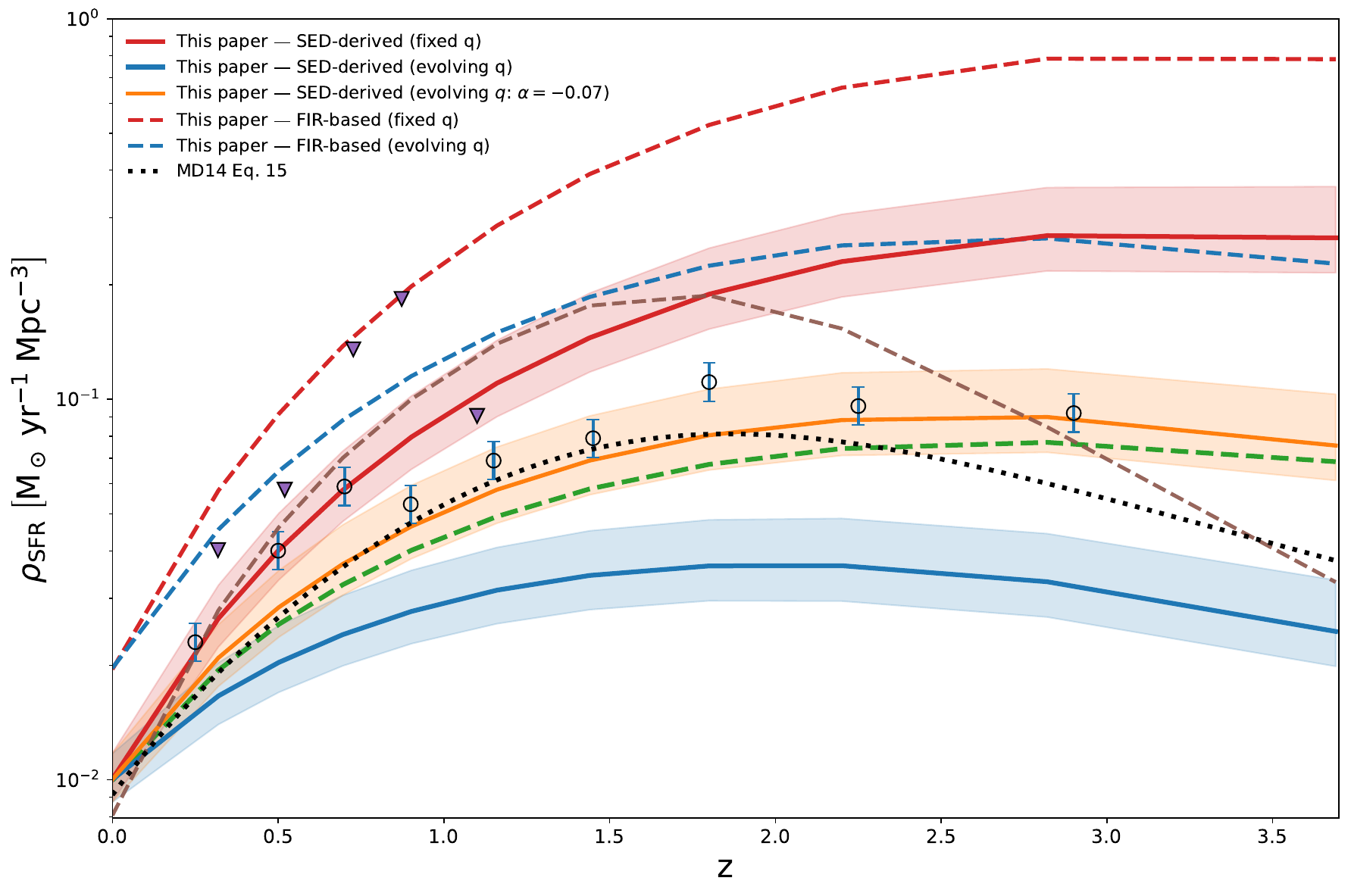}
    \centering
    \caption{
    Star-formation rate density, $\rho_{\mathrm{SFR}}$, as a function of redshift. Solid lines show the evolution assuming the SED-derived conversion to SFR from radio luminosity \citep{Cook2024}. The fixed-$q$ track is denoted by the red curve (with 68\% confidence band) and the evolving-$q$ track by the blue line (with 68\% confidence band).
    We also show an alternative line with $q_{\rm tot} \propto (1+z)^{-0.07}$ (orange). 
    For comparison, the dashed lines show the SFRD based on the FIR-based conversion from radio luminosity to SFR with fixed-$q$ (red) and evolving-$q$ (blue). The brown dashed curve shows the radio-only model from  \citet{Matthews_2021}. We also show the radio derived SFRD from \citet{Novak_2017} (green dashed line),  \citet{Cochrane2023} (blue errorbars) and \citet{matthews2024confirmationsubstantialdiscrepancyradio} (purple inverted triangles). The black dotted line is the UV+IR compilation fit of \citet{Madau_2014}. We only show the confidence region for our baseline model for clarity.
    }
    \label{fig:sfrd_pdf}
\end{figure*}

The functional form of the evolving SFG luminosity function allows us to measure the evolution in the comoving star-formation rate density of the Universe, assuming the radio luminosity for these sources closely traces the star-formation rate. The link between SFR and radio luminosity has been extensively studied at both low and high redshifts. Traditionally this has been done through the use of the radio--far-infrared correlation \citep[e.g.][]{Helou1985,Yun2001,Bell2003,Jarvis2010,Delhaize2017}. However more recent studies have used full spectral energy distribution modelling to determine the SFR from all components of a galaxy SED \citep[e.g.][]{Davies2017,Gurkan2018,Smith2021,Cook2024}.

Here we derive the SFRD by integrating our RLFs in redshift slices. For the radio–to–SFR conversion we use two tracks, an evolving FIR–radio correlation (evolving–\(q\)) following \cite{Delhaize2017} and \cite{Novak_2017}, and a non–evolving (fixed–\(q\)) control (with a constant FIR–radio correlation) from \citet{Murphy2009}.
Using two separate evolution models is valuable as they give an indication of the range of plausible SFRs derived from the far-infrared -- radio correlation for the SFGs in our sample and the associated RLF.

For our evolving-\(q\) track we adopt the relation from \cite{Novak_2017}:
\begin{equation}\label{eq:qtir}
q_{\rm TIR}(z) = 2.78\,(1+z)^{-0.14},
\end{equation}
\begin{equation}\label{eq:SFRqevol}
{\rm SFR_{\rm evolving\text{-}q}} = f_{\rm IMF}\,10^{-24}\,10^{q_{\rm TIR}(z)}\,\frac{L_{1.4}}{\rm 
W\,Hz^{-1}},
\end{equation}
where \( f_{\text{IMF}} \) is the IMF normalisation factor (set to 1 for a Chabrier IMF), and $L_{1.4}$ is the rest–frame 1.4\,GHz luminosity. This evolving $q_{\rm TIR}$ allows for the observed decline of \(q_{\rm IR}\) with redshift, plausibly tied to a stellar mass dependence coupled with a flux-limited sample at near-infrared wavelengths \citep{Delvecchio2021,Smith2021}.

The cosmic SFRD is then given by:
\begin{equation}
{\rm SFRD}(z)\;=\;\int_{L_{\min}}^{L_{\max}}
\phi_{\rm SF}(L,z)\,{\rm SFR}(L,z)\,{\rm d}\log L   ,
\end{equation}
where $\phi_{\rm SF}(L,z)$ is the SFG RLF. For consistency across redshift bins, we integrate in log-space from $\log({L_{\min)}}=20.0$ to $\log({L_{\max}})=26.1$, noting that the bulk of the luminosity density arises from near the knee of the RLF and the exact choice of these limits does not affect the overall results.

Fig.~\ref{fig:sfrd_pdf} shows both the fixed-$q$ (red-dashed) and evolving-$q$ (blue-dashed) curves derived using the same RLF fit,  their separation reflects only the calibration used to convert to SFR. Both calibrations are informative, the evolving-$q$ curve provides a baseline that accounts for the observed decline of $q_{\rm TIR}(z)$, whereas the fixed-$q$ curve offers continuity with much of the historical radio literature and serves as a transparent control. Together they bracket the radio-to-SFR calibration uncertainty based on the far-infrared--radio correlation. In both cases the radio-derived SFRD rises to $z\!\sim\!2$ and then declines at earlier epochs as expected from a range of multi-wavelength studies \citep[e.g.][]{Madau_2014}. However, the radio–derived SFRDs from our measured RLF for SFGs are consistently higher than UV+IR compilations, especially below $z\!\sim\!1.3$, where we have the strongest constraints.
Given the very high spectroscopic completeness at the lowest redshifts in our sample (see Table~\ref{tab:combined_vmax_counts}) this cannot be attributed to photometric redshift uncertainties. Indeed, we have almost 100 per cent spectroscopic redshifts for our lowest redshift bin in the COSMOS field and find that the normalisation is consistent between COSMOS and the XMM-LSS field. Therefore, we are finding a genuinely higher luminosity function normalisation. Once this form of the luminosity function is set by the low-redshift points, then it is inevitable that the SFRD remains consistently high across all redshifts, as the overall shape of the RLF is retained across all redshifts.

This higher normalisation has also been found by a comparable study using MeerKAT data over the DEEP2 field by \cite{matthews2024confirmationsubstantialdiscrepancyradio} and to a lesser degree in other recent deep radio survey \citep[e.g.][]{Enia2022,Cochrane2023}. Indeed, our analysis confirms the results of \cite{matthews2024confirmationsubstantialdiscrepancyradio}, where they acknowledge the limitations of their single-deep-field survey, which increases the uncertainty in the normalisation due to cosmic variance, which are of the order of 20~per cent in the lowest redshift bins. Given we have two separate fields, which we would not expect to be at all correlated in the large-scale structure they contain, the cosmic variance across the two fields (noting the larger area in XMM-LSS) should therefore be reduced by a factor of $\sim 2$ \citep[see Eq~9 in][]{Moster2011}. Thus, cosmic variance cannot explain the factor of $\sim 2$ higher SFRD in the radio compared to the other multi-wavelength tracers.

We overplot the radio–only SFRD model of \citet{Matthews_2021} in Fig.~\ref{fig:sfrd_pdf} The model constrains the redshift evolution of the SFRD statistically by fitting the 1.4\,GHz source counts, without requiring individual redshifts. The resulting track lies between our fixed–$q$ and evolving–$q$ curves and shares the same broad rise and decline with redshift, supporting the view that our photometric-redshift PDF–based RLF modelling is fully consistent with earlier counts–based constraints on the radio SFRD from MeerKAT observations, and that the elevated radio
normalisation is unlikely to be driven by redshift incompleteness or photometric–redshift systematics.

The most likely explanation is the conversion used to determine the galaxy SFR from the radio luminosity. Although we use the standard relation from \cite{Murphy_2011} this relation is derived solely from far-infrared data. Using the MIGHTEE survey data, \cite{Cook2024} derived SFRs for all the MIGHTEE radio continuum galaxies using full SED modelling of the UV-far-infrared data. They found a slightly shallower relation between radio luminosity and SFR of the form:
\begin{equation}\label{eq:cook}
{\rm SFR} / {\rm M}_{\odot}\,{\rm yr}^{-1} = 10^{1.014\pm0.003}\left(\frac{L_{1.4}}{5\times10^{22}~{\rm W Hz}^{-1}}\right)^{0.868\pm 0.005}.
\end{equation}
For a similar evolving $q$ to that described by \cite{Delhaize2017} and \cite{Novak_2017}, this leads to
\begin{equation}
q_{\rm tot}(z) = 5.31(1+z)^{-0.14},
\end{equation}
where, in this case $q_{\rm tot}$ denotes the evolving relation between SFR and $1.4$\,GHz luminosity defined using the \cite{Cook2024} relation at $z=0$. This then leads to an alternative evolutionary form:
\begin{equation}\label{eq:SFRqtot}
{\rm SFR}_{q_{\rm tot}} / {\rm M}_{\odot}\,{\rm yr}^{-1}  = f_{\rm IMF}\,10^{-24}\,10^{q_{\rm tot}(z)}\,\left(\frac{L_{1.4}}{\rm 
W\,Hz^{-1}}\right)^{0.868}.
\end{equation}
The evolution of the SFRD shown by the lines labelled ``SED-derived'' in Fig.~\ref{fig:sfrd_pdf} use this
calibration. This relation results in lower overall SFRs for a given radio luminosity compared to Eqns.~\ref{eq:qtir} and \ref{eq:SFRqevol}. Indeed, the difference in SFR at the knee of the luminosity function at $L_{*}^{\rm SF} = 10^{21.32}$\,W\,Hz$^{-1}$ where the bulk of the SFRD arises,
is a factor of 1.8, which can account for the factor of $\sim 2$ higher normalisation in the low-redshift SFRD. Fig.~\ref{fig:sfrd_pdf} demonstrates this, and also largely reconciles the apparent discrepancy between previous work on the SFRD derived from deep radio data compared to UV and far-infrared data. 

We also show how adopting the relationship between SFR and radio luminosity from \cite{Cook2024}, alongside a non-evolving dependence of this relation with redshift (red line in Fig~\ref{fig:sfrd_pdf}), which shows excellent agreement with the similar study by \cite{Cochrane2023} using deep LOFAR data, at least out to z $\sim 1$.  

However, it should also be noted that the closer agreement between the SFRD based on the \cite{Cook2024} relation and the UV+FIR derived SFRD from \cite{Madau_2014}, is inevitable to a certain degree, and not based on a deeper understanding of the physical processes involved. By insisting that the SFR derived from radio luminosity is tied to SFR derived from the full SED fitting, we should obtain the same normalisation in the SFRD as one would obtain from using the UV--far-infrared data, assuming that both samples are complete to a given SFR. This highlights the importance of using the correct relation between radio luminosity and SFR, but also highlights that when enforcing consistency at the lowest redshifts then we find that the radio-derived evolving SFRD closely matches that of the dust-corrected SFRD \citep{Madau_2014}.

Beyond $z\sim 1$, the non-evolving model using Eq.~\ref{eq:cook} generally over-predicts the SFRD compared to the UV- and far-infrared tracers. On the other hand incorporating a similar redshift dependence to that of \cite{Novak_2017} (Eq.~\ref{eq:SFRqtot}) appears to under-predict the SFRD, although field-to-field variations and cosmic variance may account for some of this discrepancy. 
However, this may also be resolved by considering the mass-dependence of the SFR--radio luminosity relation \citep[e.g.][]{Gurkan2018, Delvecchio2021, Smith2021} or by adopting a shallower redshift evolution in Eqn.~\ref{eq:qtir}, consistent with the lack of significant evolution in the radio–SFR relation at $1.5<z<3.5$ reported by \citet{Tabatabaei_2025}, also using  MIGHTEE data. For illustration, in Fig.~\ref{fig:sfrd_pdf}, we also show a line for $q_{\rm tot} \propto (1+z)^{-0.07}$ highlighting the sensitivity of the high-redshift SFRD to the form of the evolution, especially for a non-linear relationship between SFR and radio luminosity.
 
Fig.~\ref{fig:sfrd_uniform} shows the SFRD obtained from our uniform-$z$ approach. When compared with our $z>1$ approach, both methods recover the same overall shape a rise to a broad maximum at $z\!\sim\!2$ followed by a gradual decline and statistically consistent evolution parameters. Refitting the RLF leads to small, compensating shifts in the local SFG parameters, keeping the overall SFRD offset modest. The two confidence envelopes overlap across the full redshift range, confirming that the SFRD shape and peak position remain unchanged within the combined uncertainties. Fig.~\ref{fig:single_sfrd_appendix} shows the SFRD obtained derived using the results of using just the single best-fit redshift analysis. The two curves (evolving-$q$ and fixed-$q$) are nearly identical across the full redshift range, with modest differences. Table~\ref{tab:best_fit_params} shows the best-fit parameters obtained using the single-redshift approach, consistent with the PDF–based solutions within uncertainties.

\subsection{Field-to-field variations}
Although we are relatively limited in terms of assessing the level of cosmic variance between the two fields. We can compare the best-fit RLF for each field independently, the results are shown in Table~\ref{tab:combined_bestfit_cosmos_xmm_rows} with the RLFs shown in Figs.~\ref{fig:rlf_single_cosmos} and \ref{fig:rlf_single_xmmlss}. We find that the fits are formally consistent across all the key parameters, albeit with some interesting small differences. The normalisation for the SFGs in the COSMOS field is a factor of $\sim 0.2$~dex higher at the knee of the SFG luminosity function. This is coupled with a higher degree of evolution, marginally increasing the difference at $z\sim 0.2$, our lowest redshift bin.  This suggests that COSMOS has 60 percent higher density at $\sim 0.2$ than the XMM-LSS field. It would be difficult to attribute this to poor photometric redshifts in the XMM-LSS field, given that the performance of the photometric redshift algorithm used in \cite{Hatfield_2022} is most accurate at $0.2<z<0.5$, due to the wealth of training data for the GPz algorithm. As such, it is likely due to a genuinely higher comoving space density in the COSMOS field. However, such an "overdensity" does not necessarily have to be at low redshift. Due to the method of inferring the best-fit radio luminosity parameters across all redshift bins, then an overdensity at moderately high redshift can also lead to an elevated $\phi_{*}^{\rm SF}$. Indeed, COSMOS is well known to have a significant overdensity at $z\sim 1$ and this can also be seen in Fig.~\ref{fig:histogram}, which in turn can drive up the evolution term. This highlights how just studying a single, relatively narrow field, across a significant redshift range can lead to differing results and conclusions. Importantly, moving forward we will have both wide and deep radio data coupled with wide and deep optical data over the full MIGHTEE fields to address this issue fully.

\begin{table*}
\caption{Best-fit parameters for SF and AGN for the COSMOS and XMM-LSS fields. We fit a PLE model for SF galaxies ($\alpha_{L}^{\mathrm{SF}}$ and $\beta_{L}^{\mathrm{SF}}$) and a PDE model for AGN ($\alpha_{D}^{\mathrm{AGN}}$ and $\beta_{D}^{\mathrm{AGN}}$). The local AGN LF parameters are held fixed in all fits (see Table~\ref{tab:best_fit_params}).}
\label{tab:combined_bestfit_cosmos_xmm_rows}
\begin{tabular}{|l|c|c|c|c|c|c|c|c|}
\hline
Field &
$\log_{10}(\phi_*^{\rm SF}/\mathrm{Mpc^{-3}\,dex^{-1})}$ &
$\log_{10}(L_*^{\rm SF}/\mathrm{W\,Hz^{-1})}$ &
$\delta$ &
$\sigma$ &
$\alpha_L^{\mathrm{SF}}$ &
$\beta_L^{\mathrm{SF}}$ &
$\alpha_D^{\mathrm{AGN}}$ &
$\beta_D^{\mathrm{AGN}}$ \\
\hline\hline

COSMOS &
$-1.79_{-0.15}^{+0.08}$ &
$21.26_{-0.20}^{+0.18}$ &
$1.55_{-0.43}^{+0.33}$ &
$0.46_{-0.04}^{+0.04}$ &
$4.56_{-0.34}^{+0.35}$ &
$-0.73_{-0.15}^{+0.14}$ &
$1.41_{-0.28}^{+0.27}$ &
$-0.40_{-0.11}^{+0.11}$ \\
\hline
XMM-LSS &
$-2.15_{-0.13}^{+0.10}$ &
$21.40_{-0.21}^{+0.18}$ &
$1.49_{-0.26}^{+0.19}$ &
$0.50_{-0.03}^{+0.03}$ &
$3.91_{-0.11}^{+0.12}$ &
$-0.42_{-0.04}^{+0.03}$ &
$2.79_{-0.10}^{+0.10}$ &
$-0.67_{-0.05}^{+0.04}$ \\
\hline
\end{tabular}
\end{table*}

\section{Summary and Conclusions}\label{sec:methods}

We have measured the evolution of the RLF for SFGs and AGN using deep 1.28\,GHz MeerKAT observations from the MIGHTEE survery, combining the COSMOS and XMM-LSS fields. The total RLF was decomposed into SFG and AGN components and modelled with PLE for SFGs and PDE for AGN. 

We adopt a new approach to account for photometric redshift uncertainties and incompleteness due to some radio sources not having an optical identification and therefore redshift. 
Spectroscopic redshifts are adopted where available and, for sources without spectroscopy, we propagate full photometric redshift PDFs through 100 Monte-Carlo resamples to generate robust uncertainties on \(V_{\max}\) and the binned LFs. This Monte-Carlo PDF propagation both broadens the uncertainties to include redshift error and reduces Poisson-driven spikes at the bright-end of the RLF by redistributing redshift probability across bins.

To assess sensitivity to optical/NIR identification incompleteness we employ three different approaches. Our primary analysis assigns all non-identified sources to $z>1$, providing a clear upper bound on their high–$z$ impact. As alternative cross-checks we distribute the same sources uniformly in $z$ and also adopt a single–redshift analysis that assigns one best redshift per source and then redetermine the RLF. All three analyses produce consistent RLFs within the combined uncertainties; differences are confined to modest, localized shifts.

We find a much higher normalisation for the SFG luminosity function compared to previous studies at 3\,GHz over the COSMOS field by \cite{Novak_2017}, however, our results are consistent with more recent studies with both MeerKAT \citep{matthews2024confirmationsubstantialdiscrepancyradio} and LOFAR \citep{Cochrane2023}. We attribute this to the much better surface brightness sensitivity of the MeerKAT (and LOFAR) telescopes, compared to the A+C configuration used for the VLA-3GHz COSMOS survey of \cite{VLACOSMOS3GHz}. However, even with this higher normalisation, coupled with a steeper faint-end slope, we still find evidence for strong evolution for the SFG population to $z\sim 2$.

From the radio luminosity function, we infer the evolution in the cosmic star-formation rate density (SFRD). We determine the SFRD using two radio–SFR calibrations: one where we adopt a fixed linear relationship between radio luminosity and star-formation rate from \cite{Murphy2009} and one that evolves from this $z=0$ calibration, with redshift following the work of \cite{Delhaize2017} and \cite{Novak_2017}. 

Both calibrations show an increase in the SFRDs that exceed the SFRD from UV+IR compilations and also the radio-based work of \cite{Novak_2017}, but are consistent with the \cite{matthews2024confirmationsubstantialdiscrepancyradio} radio-based measurements. 
However, we find that adopting more recent determinations of the correlation between radio luminosity and star-formation rates from \cite{Cook2024}, based on full spectral energy distribution modelling of the MIGHTEE continuum data, can reconcile the SFRD with UV and infrared tracers of the SFRD.

Given the interplay between the calibration of the radio luminosity -- star-formation rate calibration and the SFRD through the luminosity function, it is critical to adopt consistent calibrations and assumptions when combining these. Indeed, if there is also a mass-dependent aspect to the relationship between the star-formation rate and radio luminosity then this also needs to be included as a factor in the determination of the SFRD based on radio continuum data.

However, if the radio continuum emission can be properly calibrated against robust measurements of the star-formation rate across all stellar masses, overcoming the various Malmquist bias effects, then it is clear that the new generation of radio telescopes are a tremendously powerful probe of the evolution in the total, both obscured and unobscured, SFRD. Moreover, \cite{Malefahlo2026} have recently shown that utilising new Bayesian source extraction methods based on the multi-wavelength catalogues, enables robust flux-density measurement much closer to the confusion noise, opening up the potential to determine the RLF to significantly fainter flux-densities.
Thus, the full MIGHTEE radio continuum data when fully cross-matched with both current and forthcoming data from Rubin and {\em Euclid},  should provide the most robust measurement of the evolution in both the SFRD, where depth is critical, and the evolution of the radio-AGN activity, where areal coverage becomes increasingly important. 

Furthermore, this study shows that these different components can be modelled separately in a combined radio luminosity function, thus mitigating against issues surrounding classifying sources as AGN or star-formation dominated galaxies. Indeed, the statistical measurement of both populations may provide more robust and unbiased insights as it naturally accounts for hybrid sources.

\section*{Acknowledgements}
MJJ, CLH and IHW acknowledge support from the Hintze Family Charitable Foundation through the Oxford Hintze Centre for Astrophysical Surveys. MJJ, IH, RGV and NS acknowledge the support from a UKRI Frontiers Research Grant [EP/X026639/1]. CLH also acknowledges support from the Science and Technology Facilities Council (STFC) through grant ST/Y000951/1. CP acknowledges support via the RAL Space In House Research programme funded by the Science and Technology Facilities Council of the UK Research and Innovation (award ST/M001083/1). MV acknowledges financial support from the Inter-University Institute for Data Intensive Astronomy (IDIA),
a partnership of the University of Cape Town, the University of Pretoria and the University of the Western Cape, and from the South African Department of Science and Innovation's National Research Foundation under the
ISARP RADIOMAP Joint Research Scheme (DSI-NRF Grant Number 150551) and the CPRR HIPPO Project (DSI-NRF Grant Number SRUG22031677).

The MeerKAT telescope is operated by the South African Radio Astronomy Observatory, which is a facility of the National Research Foundation, an agency of the Department of Science and Innovation. We acknowledge the use of the ilifu cloud computing facility – www.ilifu.ac.za, a partnership between the University of Cape Town, the University of the Western Cape, Stellenbosch University, Sol Plaatje University and the Cape Peninsula University of Technology. The Ilifu facility is supported by contributions from the Inter-University Institute for Data Intensive Astronomy (IDIA – a partnership between the University of Cape Town, the University of Pretoria and the University of the Western Cape, the Computational Biology division at UCT and the Data Intensive Research Initiative of South Africa (DIRISA). The authors acknowledge the Centre for High Performance Computing (CHPC), South Africa, for providing computational resources to this research project. 

\section*{Data Availability}
The MIGHTEE ES data products are available from the SARAO archive at \blue{https://doi.org/10.48479/emmd-kf31}.



\bibliographystyle{mnras}
\bibliography{example} 

@ARTICLE{Malefahlo2026,
       author = {{Malefahlo}, Eliab D. and {Jarvis}, Matt J. and {Santos}, Mario G. and {Cress}, Catherine and {Smith}, Daniel J.~B. and {Hale}, Catherine and {Afonso}, Jos{\'e} and {Whittam}, Imogen H. and {Vaccari}, Mattia and {Heywood}, Ian and {Jin}, Shuowen and {An}, Fangxia},
        title = "{Deblending the MIGHTEE-COSMOS survey with XID+: The resolved radio source counts to $S_{1.4}\approx 5μ$Jy}",
      journal = {arXiv e-prints},
         year = 2025,
        month = dec,
          eid = {arXiv:2512.09290},
        pages = {arXiv:2512.09290},
          doi = {10.48550/arXiv.2512.09290},
archivePrefix = {arXiv},
       eprint = {2512.09290},
 primaryClass = {astro-ph.GA},
       adsurl = {https://ui.adsabs.harvard.edu/abs/2025arXiv251209290M}
}

@ARTICLE{Varadaraj2026,
       author = {{Varadaraj}, R.~G. and {Bowler}, R.~A.~A. and {Jarvis}, M.~J. and {Weaver}, J.~R. and {Ba{\~n}ados}, E. and {Holloway}, P. and {Caputi}, K.~I. and {Wilkins}, S.~M. and {Yang}, D. and {Milvang-Jensen}, B. and {Gabarra}, L. and {Oesch}, P.~A. and {Amara}, A. and {Andreon}, S. and {Auricchio}, N. and {Baccigalupi}, C. and {Baldi}, M. and {Bardelli}, S. and {Biviano}, A. and {Branchini}, E. and {Brescia}, M. and {Camera}, S. and {Ca{\~n}as-Herrera}, G. and {Capobianco}, V. and {Carbone}, C. and {Carretero}, J. and {Castellano}, M. and {Castignani}, G. and {Cavuoti}, S. and {Chambers}, K.~C. and {Cimatti}, A. and {Colodro-Conde}, C. and {Congedo}, G. and {Conselice}, C.~J. and {Conversi}, L. and {Copin}, Y. and {Courbin}, F. and {Courtois}, H.~M. and {Cropper}, M. and {Da Silva}, A. and {Degaudenzi}, H. and {De Lucia}, G. and {Dole}, H. and {Dubath}, F. and {Duncan}, C.~A.~J. and {Dupac}, X. and {Dusini}, S. and {Escoffier}, S. and {Farina}, M. and {Farinelli}, R. and {Faustini}, F. and {Ferriol}, S. and {Finelli}, F. and {Fosalba}, P. and {Fourmanoit}, N. and {Frailis}, M. and {Franceschi}, E. and {Fumana}, M. and {Galeotta}, S. and {George}, K. and {Gillis}, B. and {Giocoli}, C. and {Gracia-Carpio}, J. and {Grazian}, A. and {Grupp}, F. and {Guzzo}, L. and {Haugan}, S.~V.~H. and {Hoar}, J. and {Hoekstra}, H. and {Holmes}, W. and {Hook}, I.~M. and {Hormuth}, F. and {Hornstrup}, A. and {Jahnke}, K. and {Jhabvala}, M. and {Joachimi}, B. and {Keih{\"a}nen}, E. and {Kermiche}, S. and {Kiessling}, A. and {Kilbinger}, M. and {Kubik}, B. and {K{\"u}mmel}, M. and {Kunz}, M. and {Kurki-Suonio}, H. and {Le Brun}, A.~M.~C. and {Ligori}, S. and {Lilje}, P.~B. and {Lindholm}, V. and {Lloro}, I. and {Mainetti}, G. and {Maino}, D. and {Maiorano}, E. and {Mansutti}, O. and {Marggraf}, O. and {Martinelli}, M. and {Martinet}, N. and {Marulli}, F. and {Massey}, R.~J. and {Medinaceli}, E. and {Mei}, S. and {Melchior}, M. and {Mellier}, Y. and {Meneghetti}, M. and {Merlin}, E. and {Meylan}, G. and {Mora}, A. and {Moresco}, M. and {Moscardini}, L. and {Nakajima}, R. and {Neissner}, C. and {Niemi}, S. -M. and {Padilla}, C. and {Paltani}, S. and {Pasian}, F. and {Pedersen}, K. and {Percival}, W.~J. and {Pettorino}, V. and {Pires}, S. and {Polenta}, G. and {Poncet}, M. and {Popa}, L.~A. and {Pozzetti}, L. and {Raison}, F. and {Renzi}, A. and {Rhodes}, J. and {Riccio}, G. and {Romelli}, E. and {Roncarelli}, M. and {Rossetti}, E. and {Saglia}, R. and {Sakr}, Z. and {Sapone}, D. and {Sartoris}, B. and {Schirmer}, M. and {Schneider}, P. and {Schrabback}, T. and {Secroun}, A. and {Seidel}, G. and {Serrano}, S. and {Simon}, P. and {Sirignano}, C. and {Sirri}, G. and {Stanco}, L. and {Starck}, J. -L. and {Steinwagner}, J. and {Tallada-Cresp{\'\i}}, P. and {Taylor}, A.~N. and {Teplitz}, H.~I. and {Tereno}, I. and {Tessore}, N. and {Toft}, S. and {Toledo-Moreo}, R. and {Torradeflot}, F. and {Tutusaus}, I. and {Valenziano}, L. and {Valiviita}, J. and {Vassallo}, T. and {Veropalumbo}, A. and {Wang}, Y. and {Weller}, J. and {Zamorani}, G. and {Zerbi}, F.~M. and {Zucca}, E. and {Mart{\'\i}n-Fleitas}, J. and {Scottez}, V. and {Viel}, M.},
        title = "{Euclid: Discovery of bright $z\simeq7$ Lyman-break galaxies in UltraVISTA and Euclid COSMOS}",
      journal = {arXiv e-prints},
         year = 2025,
        month = oct,
          eid = {arXiv:2510.00945},
        pages = {arXiv:2510.00945},
          doi = {10.48550/arXiv.2510.00945},
archivePrefix = {arXiv},
       eprint = {2510.00945},
 primaryClass = {astro-ph.GA},
       adsurl = {https://ui.adsabs.harvard.edu/abs/2025arXiv251000945V}
}

@ARTICLE{Rigby2011,
       author = {{Rigby}, E.~E. and {Best}, P.~N. and {Brookes}, M.~H. and {Peacock}, J.~A. and {Dunlop}, J.~S. and {R{\"o}ttgering}, H.~J.~A. and {Wall}, J.~V. and {Ker}, L.},
        title = "{The luminosity-dependent high-redshift turnover in the steep spectrum radio luminosity function: clear evidence for downsizing in the radio-AGN population}",
      journal = {\mnras},
         year = 2011,
        month = sep,
       volume = {416},
       number = {3},
        pages = {1900-1915},
          doi = {10.1111/j.1365-2966.2011.19167.x},
archivePrefix = {arXiv},
       eprint = {1104.5020},
 primaryClass = {astro-ph.CO},
       adsurl = {https://ui.adsabs.harvard.edu/abs/2011MNRAS.416.1900R}
}

@ARTICLE{Willott2001,
       author = {{Willott}, Chris J. and {Rawlings}, Steve and {Blundell}, Katherine M. and {Lacy}, Mark and {Eales}, Stephen A.},
        title = "{The radio luminosity function from the low-frequency 3CRR, 6CE and 7CRS complete samples}",
      journal = {\mnras},
         year = 2001,
        month = apr,
       volume = {322},
       number = {3},
        pages = {536-552},
          doi = {10.1046/j.1365-8711.2001.04101.x},
archivePrefix = {arXiv},
       eprint = {astro-ph/0010419},
 primaryClass = {astro-ph},
       adsurl = {https://ui.adsabs.harvard.edu/abs/2001MNRAS.322..536W}
}

@ARTICLE{Jarvis2001RLF,
       author = {{Jarvis}, Matt J. and {Rawlings}, Steve and {Willott}, Chris J. and {Blundell}, Katherine M. and {Eales}, Steve and {Lacy}, Mark},
        title = "{On the redshift cut-off for steep-spectrum radio sources}",
      journal = {\mnras},
         year = 2001,
        month = nov,
       volume = {327},
       number = {3},
        pages = {907-917},
          doi = {10.1046/j.1365-8711.2001.04778.x},
archivePrefix = {arXiv},
       eprint = {astro-ph/0106473},
 primaryClass = {astro-ph},
       adsurl = {https://ui.adsabs.harvard.edu/abs/2001MNRAS.327..907J}
}

@ARTICLE{DP90,
       author = {{Dunlop}, J.~S. and {Peacock}, J.~A.},
        title = "{The redshift cut-off in the luminosity function of radio galaxies and quasars.}",
      journal = {\mnras},
         year = 1990,
        month = nov,
       volume = {247},
        pages = {19},
       adsurl = {https://ui.adsabs.harvard.edu/abs/1990MNRAS.247...19D}
}

@ARTICLE{VLACOSMOS3GHz,
       author = {{Smol{\v{c}}i{\'c}}, V. and {Novak}, M. and {Bondi}, M. and {Ciliegi}, P. and {Mooley}, K.~P. and {Schinnerer}, E. and {Zamorani}, G. and {Navarrete}, F. and {Bourke}, S. and {Karim}, A. and {Vardoulaki}, E. and {Leslie}, S. and {Delhaize}, J. and {Carilli}, C.~L. and {Myers}, S.~T. and {Baran}, N. and {Delvecchio}, I. and {Miettinen}, O. and {Banfield}, J. and {Balokovi{\'c}}, M. and {Bertoldi}, F. and {Capak}, P. and {Frail}, D.~A. and {Hallinan}, G. and {Hao}, H. and {Herrera Ruiz}, N. and {Horesh}, A. and {Ilbert}, O. and {Intema}, H. and {Jeli{\'c}}, V. and {Kl{\"o}ckner}, H.-R. and {Krpan}, J. and {Kulkarni}, S.~R. and {McCracken}, H. and {Laigle}, C. and {Middleberg}, E. and {Murphy}, E.~J. and {Sargent}, M. and {Scoville}, N.~Z. and {Sheth}, K.},
        title = "{The VLA-COSMOS 3 GHz Large Project: Continuum data and source catalog release}",
      journal = {\aap},
         year = 2017,
        month = jun,
       volume = {602},
          eid = {A1},
        pages = {A1},
          doi = {10.1051/0004-6361/201628704},
archivePrefix = {arXiv},
       eprint = {1703.09713},
 primaryClass = {astro-ph.GA},
       adsurl = {https://ui.adsabs.harvard.edu/abs/2017A&A...602A...1S}
}

@ARTICLE{Khostovan2025,
       author = {{Khostovan}, Ali Ahmad and {Kartaltepe}, Jeyhan S. and {Salvato}, Mara and {Ilbert}, Olivier and {Casey}, Caitlin M. and {Algera}, Hiddo and {Antwi-Danso}, Jacqueline and {Battisti}, Andrew and {Brinch}, Malte and {Brusa}, Marcella and {Calabr{\`o}}, Antonello and {Capak}, Peter L. and {Chartab}, Nima and {Cooper}, Olivia R. and {Cox}, Isa G. and {Darvish}, Behnam and {Drakos}, Nicole E. and {Faisst}, Andreas L. and {George}, Matthew R. and {Gozaliasl}, Ghassem and {Harish}, Santosh and {Hasinger}, G{\"u}nther and {Hatamnia}, Hossein and {Iovino}, Angela and {Jin}, Shuowen and {Kashino}, Daichi and {Koekemoer}, Anton M. and {Laishram}, Ronaldo and {Lee}, Khee-Gan and {Lertprasertpong}, Jitrapon and {Lilly}, Simon J. and {Liu}, Daizhong and {Masters}, Daniel C. and {Mobasher}, Bahram and {Nagao}, Tohru and {Onodera}, Masato and {Peng}, Yingjie and {Sanders}, David B. and {Sanders}, Ryan L. and {Sattari}, Zahra and {Scoville}, Nick and {Shah}, Ekta A. and {Silverman}, John D. and {Suzuki}, Nao and {Taamoli}, Sina and {Tanaka}, Masayuki and {Tasca}, Lidia A.~M. and {Toft}, Sune and {Toni}, Greta and {Trakhtenbrot}, Benny and {Trump}, Jonathan R. and {Vaccari}, Mattia and {Valentino}, Francesco and {Vanderhoof}, Brittany N. and {Weaver}, John R. and {Yun}, Min S. and {Zavala}, Jorge A.},
        title = "{COSMOS Spectroscopic Redshift Compilation (First Data Release): 488,000 Redshifts Encompassing Two Decades of Spectroscopy}",
      journal = {\apjs},
         year = 2026,
        month = jan,
       volume = {282},
       number = {1},
          eid = {6},
        pages = {6},
          doi = {10.3847/1538-4365/ae1cb9},
archivePrefix = {arXiv},
       eprint = {2503.00120},
 primaryClass = {astro-ph.GA},
       adsurl = {https://ui.adsabs.harvard.edu/abs/2026ApJS..282....6K}
}

@ARTICLE{Moster2011,
       author = {{Moster}, Benjamin P. and {Somerville}, Rachel S. and {Newman}, Jeffrey A. and {Rix}, Hans-Walter},
        title = "{A Cosmic Variance Cookbook}",
      journal = {\apj},
         year = 2011,
        month = apr,
       volume = {731},
       number = {2},
          eid = {113},
        pages = {113},
          doi = {10.1088/0004-637X/731/2/113},
archivePrefix = {arXiv},
       eprint = {1001.1737},
 primaryClass = {astro-ph.CO},
       adsurl = {https://ui.adsabs.harvard.edu/abs/2011ApJ...731..113M}
}

@ARTICLE{Enia2022,
       author = {{Enia}, Andrea and {Talia}, Margherita and {Pozzi}, Francesca and {Cimatti}, Andrea and {Delvecchio}, Ivan and {Zamorani}, Gianni and {D'Amato}, Quirino and {Bisigello}, Laura and {Gruppioni}, Carlotta and {Rodighiero}, Giulia and {Calura}, Francesco and {Dallacasa}, Daniele and {Giulietti}, Marika and {Barchiesi}, Luigi and {Behiri}, Meriem and {Romano}, Michael},
        title = "{A New Estimate of the Cosmic Star Formation Density from a Radio-selected Sample, and the Contribution of H-dark Galaxies at z {\ensuremath{\geq}} 3}",
      journal = {\apj},
         year = 2022,
        month = mar,
       volume = {927},
       number = {2},
          eid = {204},
        pages = {204},
          doi = {10.3847/1538-4357/ac51ca},
archivePrefix = {arXiv},
       eprint = {2202.00019},
 primaryClass = {astro-ph.CO},
       adsurl = {https://ui.adsabs.harvard.edu/abs/2022ApJ...927..204E}
}

@ARTICLE{Delhaize2017,
       author = {{Delhaize}, J. and {Smol{\v{c}}i{\'c}}, V. and {Delvecchio}, I. and {Novak}, M. and {Sargent}, M. and {Baran}, N. and {Magnelli}, B. and {Zamorani}, G. and {Schinnerer}, E. and {Murphy}, E.~J. and {Aravena}, M. and {Berta}, S. and {Bondi}, M. and {Capak}, P. and {Carilli}, C. and {Ciliegi}, P. and {Civano}, F. and {Ilbert}, O. and {Karim}, A. and {Laigle}, C. and {Le F{\`e}vre}, O. and {Marchesi}, S. and {McCracken}, H.~J. and {Salvato}, M. and {Seymour}, N. and {Tasca}, L.},
        title = "{The VLA-COSMOS 3 GHz Large Project: The infrared-radio correlation of star-forming galaxies and AGN to z {\ensuremath{\lesssim}} 6}",
      journal = {\aap},
         year = 2017,
        month = jun,
       volume = {602},
          eid = {A4},
        pages = {A4},
          doi = {10.1051/0004-6361/201629430},
archivePrefix = {arXiv},
       eprint = {1703.09723},
 primaryClass = {astro-ph.GA},
       adsurl = {https://ui.adsabs.harvard.edu/abs/2017A&A...602A...4D}
}

@ARTICLE{Davies2017,
       author = {{Davies}, L.~J.~M. and {Huynh}, M.~T. and {Hopkins}, A.~M. and {Seymour}, N. and {Driver}, S.~P. and {Robotham}, A.~G.~R. and {Baldry}, I.~K. and {Bland-Hawthorn}, J. and {Bourne}, N. and {Bremer}, M.~N. and {Brown}, M.~J.~I. and {Brough}, S. and {Cluver}, M. and {Grootes}, M.~W. and {Jarvis}, M. and {Loveday}, J. and {Moffet}, A. and {Owers}, M. and {Phillipps}, S. and {Sadler}, E. and {Wang}, L. and {Wilkins}, S. and {Wright}, A.},
        title = "{Galaxy And Mass Assembly: the 1.4 GHz SFR indicator, SFR-M$_{*}$ relation and predictions for ASKAP-GAMA}",
      journal = {\mnras},
         year = 2017,
        month = apr,
       volume = {466},
       number = {2},
        pages = {2312-2324},
          doi = {10.1093/mnras/stw3080},
archivePrefix = {arXiv},
       eprint = {1701.06242},
 primaryClass = {astro-ph.GA},
       adsurl = {https://ui.adsabs.harvard.edu/abs/2017MNRAS.466.2312D}
}

@ARTICLE{Gurkan2018,
       author = {{G{\"u}rkan}, G. and {Hardcastle}, M.~J. and {Smith}, D.~J.~B. and {Best}, P.~N. and {Bourne}, N. and {Calistro-Rivera}, G. and {Heald}, G. and {Jarvis}, M.~J. and {Prandoni}, I. and {R{\"o}ttgering}, H.~J.~A. and {Sabater}, J. and {Shimwell}, T. and {Tasse}, C. and {Williams}, W.~L.},
        title = "{LOFAR/H-ATLAS: the low-frequency radio luminosity-star formation rate relation}",
      journal = {\mnras},
         year = 2018,
        month = apr,
       volume = {475},
       number = {3},
        pages = {3010-3028},
          doi = {10.1093/mnras/sty016},
archivePrefix = {arXiv},
       eprint = {1801.02629},
 primaryClass = {astro-ph.GA},
       adsurl = {https://ui.adsabs.harvard.edu/abs/2018MNRAS.475.3010G}
}

@ARTICLE{Smith2021,
       author = {{Smith}, D.~J.~B. and {Haskell}, P. and {G{\"u}rkan}, G. and {Best}, P.~N. and {Hardcastle}, M.~J. and {Kondapally}, R. and {Williams}, W. and {Duncan}, K.~J. and {Cochrane}, R.~K. and {McCheyne}, I. and {R{\"o}ttgering}, H.~J.~A. and {Sabater}, J. and {Shimwell}, T.~W. and {Tasse}, C. and {Bonato}, M. and {Bondi}, M. and {Jarvis}, M.~J. and {Leslie}, S.~K. and {Prandoni}, I. and {Wang}, L.},
        title = "{The LOFAR Two-metre Sky Survey Deep Fields. The star-formation rate-radio luminosity relation at low frequencies}",
      journal = {\aap},
         year = 2021,
        month = apr,
       volume = {648},
          eid = {A6},
        pages = {A6},
          doi = {10.1051/0004-6361/202039343},
archivePrefix = {arXiv},
       eprint = {2011.08196},
 primaryClass = {astro-ph.GA},
       adsurl = {https://ui.adsabs.harvard.edu/abs/2021A&A...648A...6S}
}

@ARTICLE{Helou1985,
       author = {{Helou}, G. and {Soifer}, B.~T. and {Rowan-Robinson}, M.},
        title = "{Thermal infrared and nonthermal radio : remarkable correlation in disks of galaxies.}",
      journal = {\apjl},
         year = 1985,
        month = nov,
       volume = {298},
        pages = {L7-L11},
          doi = {10.1086/184556},
       adsurl = {https://ui.adsabs.harvard.edu/abs/1985ApJ...298L...7H}
}

@ARTICLE{Smolcic2017AGN,
       author = {{Smol{\v{c}}i{\'c}}, V. and {Novak}, M. and {Delvecchio}, I. and {Ceraj}, L. and {Bondi}, M. and {Delhaize}, J. and {Marchesi}, S. and {Murphy}, E. and {Schinnerer}, E. and {Vardoulaki}, E. and {Zamorani}, G.},
        title = "{The VLA-COSMOS 3 GHz Large Project: Cosmic evolution of radio AGN and implications for radio-mode feedback since z   5}",
      journal = {\aap},
         year = 2017,
        month = jun,
       volume = {602},
          eid = {A6},
        pages = {A6},
          doi = {10.1051/0004-6361/201730685},
archivePrefix = {arXiv},
       eprint = {1705.07090},
 primaryClass = {astro-ph.GA},
       adsurl = {https://ui.adsabs.harvard.edu/abs/2017A&A...602A...6S}
}

@ARTICLE{Koprowski2017,
       author = {{Koprowski}, M.~P. and {Dunlop}, J.~S. and {Micha{\l}owski}, M.~J. and {Coppin}, K.~E.~K. and {Geach}, J.~E. and {McLure}, R.~J. and {Scott}, D. and {van der Werf}, P.~P.},
        title = "{The evolving far-IR galaxy luminosity function and dust-obscured star formation rate density out to z≃5.}",
      journal = {\mnras},
         year = 2017,
        month = nov,
       volume = {471},
       number = {4},
        pages = {4155-4169},
          doi = {10.1093/mnras/stx1843},
archivePrefix = {arXiv},
       eprint = {1706.00426},
 primaryClass = {astro-ph.GA},
       adsurl = {https://ui.adsabs.harvard.edu/abs/2017MNRAS.471.4155K}
}

@ARTICLE{Arnouts2005,
       author = {{Arnouts}, S. and {Schiminovich}, D. and {Ilbert}, O. and {Tresse}, L. and {Milliard}, B. and {Treyer}, M. and {Bardelli}, S. and {Budavari}, T. and {Wyder}, T.~K. and {Zucca}, E. and {Le F{\`e}vre}, O. and {Martin}, D.~C. and {Vettolani}, G. and {Adami}, C. and {Arnaboldi}, M. and {Barlow}, T. and {Bianchi}, L. and {Bolzonella}, M. and {Bottini}, D. and {Byun}, Y.-I. and {Cappi}, A. and {Charlot}, S. and {Contini}, T. and {Donas}, J. and {Forster}, K. and {Foucaud}, S. and {Franzetti}, P. and {Friedman}, P.~G. and {Garilli}, B. and {Gavignaud}, I. and {Guzzo}, L. and {Heckman}, T.~M. and {Hoopes}, C. and {Iovino}, A. and {Jelinsky}, P. and {Le Brun}, V. and {Lee}, Y.-W. and {Maccagni}, D. and {Madore}, B.~F. and {Malina}, R. and {Marano}, B. and {Marinoni}, C. and {McCracken}, H.~J. and {Mazure}, A. and {Meneux}, B. and {Merighi}, R. and {Morrissey}, P. and {Neff}, S. and {Paltani}, S. and {Pell{\`o}}, R. and {Picat}, J.~P. and {Pollo}, A. and {Pozzetti}, L. and {Radovich}, M. and {Rich}, R.~M. and {Scaramella}, R. and {Scodeggio}, M. and {Seibert}, M. and {Siegmund}, O. and {Small}, T. and {Szalay}, A.~S. and {Welsh}, B. and {Xu}, C.~K. and {Zamorani}, G. and {Zanichelli}, A.},
        title = "{The GALEX VIMOS-VLT Deep Survey Measurement of the Evolution of the 1500 {\r{A}} Luminosity Function}",
      journal = {\apjl},
         year = 2005,
        month = jan,
       volume = {619},
       number = {1},
        pages = {L43-L46},
          doi = {10.1086/426733},
archivePrefix = {arXiv},
       eprint = {astro-ph/0411391},
 primaryClass = {astro-ph},
       adsurl = {https://ui.adsabs.harvard.edu/abs/2005ApJ...619L..43A}
}

@ARTICLE{Gentile2024,
       author = {{Gentile}, Fabrizio and {Talia}, Margherita and {Behiri}, Meriem and {Zamorani}, Giovanni and {Barchiesi}, Luigi and {Vignali}, Cristian and {Pozzi}, Francesca and {Bethermin}, Matthieu and {Enia}, Andrea and {Faisst}, Andreas L. and {Giulietti}, Marika and {Gruppioni}, Carlotta and {Lapi}, Andrea and {Massardi}, Marcella and {Smol{\v{c}}i{\'c}}, Vernesa and {Vaccari}, Mattia and {Cimatti}, Andrea},
        title = "{Illuminating the Dark Side of Cosmic Star Formation. III. Building the Largest Homogeneous Sample of Radio-selected Dusty Star-forming Galaxies in COSMOS with PhoEBO}",
      journal = {\apj},
         year = 2024,
        month = feb,
       volume = {962},
       number = {1},
          eid = {26},
        pages = {26},
          doi = {10.3847/1538-4357/ad1519},
archivePrefix = {arXiv},
       eprint = {2312.05305},
 primaryClass = {astro-ph.GA},
       adsurl = {https://ui.adsabs.harvard.edu/abs/2024ApJ...962...26G}
}

@ARTICLE{Gentile2025,
       author = {{Gentile}, Fabrizio and {Talia}, Margherita and {Enia}, Andrea and {Pozzi}, Francesca and {Traina}, Alberto and {Zamorani}, Giovanni and {Andika}, Irham T. and {Behiri}, Meriem and {Barrufet}, Laia and {Casey}, Caitlin M. and {Cimatti}, Andrea and {Drakos}, Nicole E. and {Faisst}, Andreas L. and {Franco}, Maximilien and {Gillman}, Steven and {Giulietti}, Marika and {Gottumukkala}, Rashmi and {Hayward}, Christopher C. and {Ilbert}, Olivier and {Jin}, Shuowen and {Lapi}, Andrea and {McKinney}, Jed and {Shuntov}, Marko and {Vaccari}, Mattia and {Vignali}, Cristian and {Akins}, Hollis B. and {Allen}, Natalie and {Harish}, Santosh and {McCracken}, Henry Joy and {Kartaltepe}, Jeyhan S. and {Koekemoer}, Anton M. and {Liu}, Daizhong and {Paquereau}, Louise and {Rhodes}, Jason and {Rich}, Micheal R. and {Robertson}, Brant E. and {Toft}, Sune},
        title = "{Going deeper into the dark with COSMOS-Web: JWST unveils the total contribution of radio-selected NIR-faint galaxies to the cosmic star formation rate density}",
      journal = {\aap},
         year = 2025,
        month = may,
       volume = {697},
          eid = {A46},
        pages = {A46},
          doi = {10.1051/0004-6361/202452461},
archivePrefix = {arXiv},
       eprint = {2503.00112},
 primaryClass = {astro-ph.GA},
       adsurl = {https://ui.adsabs.harvard.edu/abs/2025A&A...697A..46G}
}

@INPROCEEDINGS{Cuillandre2012,
       author = {{Cuillandre}, Jean-Charles J. and {Withington}, Kanoa and {Hudelot}, Patrick and {Goranova}, Yuliana and {McCracken}, Henry and {Magnard}, Fr{\'e}d{\'e}ric and {Mellier}, Yannick and {Regnault}, Nicolas and {B{\'e}toule}, Marc and {Aussel}, Herv{\'e} and {Kavelaars}, J.~J. and {Fernique}, Pierre and {Bonnarel}, Fran{\c{c}}ois and {Ochsenbein}, Francois and {Ilbert}, Olivier},
        title = "{Introduction to the CFHT Legacy Survey final release (CFHTLS T0007)}",
    booktitle = {Observatory Operations: Strategies, Processes, and Systems IV},
         year = 2012,
       editor = {{Peck}, Alison B. and {Seaman}, Robert L. and {Comeron}, Fernando},
       series = {Society of Photo-Optical Instrumentation Engineers (SPIE) Conference Series},
       volume = {8448},
        month = sep,
          eid = {84480M},
        pages = {84480M},
          doi = {10.1117/12.925584},
       adsurl = {https://ui.adsabs.harvard.edu/abs/2012SPIE.8448E..0MC}
}

@ARTICLE{Ilbert2009,
       author = {{Ilbert}, O. and {Capak}, P. and {Salvato}, M. and {Aussel}, H. and {McCracken}, H.~J. and {Sanders}, D.~B. and {Scoville}, N. and {Kartaltepe}, J. and {Arnouts}, S. and {Le Floc'h}, E. and {Mobasher}, B. and {Taniguchi}, Y. and {Lamareille}, F. and {Leauthaud}, A. and {Sasaki}, S. and {Thompson}, D. and {Zamojski}, M. and {Zamorani}, G. and {Bardelli}, S. and {Bolzonella}, M. and {Bongiorno}, A. and {Brusa}, M. and {Caputi}, K.~I. and {Carollo}, C.~M. and {Contini}, T. and {Cook}, R. and {Coppa}, G. and {Cucciati}, O. and {de la Torre}, S. and {de Ravel}, L. and {Franzetti}, P. and {Garilli}, B. and {Hasinger}, G. and {Iovino}, A. and {Kampczyk}, P. and {Kneib}, J.-P. and {Knobel}, C. and {Kovac}, K. and {Le Borgne}, J.~F. and {Le Brun}, V. and {Le F{\`e}vre}, O. and {Lilly}, S. and {Looper}, D. and {Maier}, C. and {Mainieri}, V. and {Mellier}, Y. and {Mignoli}, M. and {Murayama}, T. and {Pell{\`o}}, R. and {Peng}, Y. and {P{\'e}rez-Montero}, E. and {Renzini}, A. and {Ricciardelli}, E. and {Schiminovich}, D. and {Scodeggio}, M. and {Shioya}, Y. and {Silverman}, J. and {Surace}, J. and {Tanaka}, M. and {Tasca}, L. and {Tresse}, L. and {Vergani}, D. and {Zucca}, E.},
        title = "{Cosmos Photometric Redshifts with 30-Bands for 2-deg$^{2}$}",
      journal = {\apj},
         year = 2009,
        month = jan,
       volume = {690},
       number = {2},
        pages = {1236-1249},
          doi = {10.1088/0004-637X/690/2/1236},
archivePrefix = {arXiv},
       eprint = {0809.2101},
 primaryClass = {astro-ph},
       adsurl = {https://ui.adsabs.harvard.edu/abs/2009ApJ...690.1236I}
}

@ARTICLE{Varadaraj2023,
       author = {{Varadaraj}, R.~G. and {Bowler}, R.~A.~A. and {Jarvis}, M.~J. and {Adams}, N.~J. and {H{\"a}u{\ss}ler}, B.},
        title = "{The bright end of the galaxy luminosity function at z ≃ 7 from the VISTA VIDEO survey}",
      journal = {\mnras},
         year = 2023,
        month = sep,
       volume = {524},
       number = {3},
        pages = {4586-4613},
          doi = {10.1093/mnras/stad2081},
archivePrefix = {arXiv},
       eprint = {2304.02494},
 primaryClass = {astro-ph.GA},
       adsurl = {https://ui.adsabs.harvard.edu/abs/2023MNRAS.524.4586V}
}

@ARTICLE{Bowler2021,
       author = {{Bowler}, R.~A.~A. and {Adams}, N.~J. and {Jarvis}, M.~J. and {H{\"a}u{\ss}ler}, B.},
        title = "{The rapid transition from star formation to AGN-dominated rest-frame ultraviolet light at z ≃ 4}",
      journal = {\mnras},
         year = 2021,
        month = mar,
       volume = {502},
       number = {1},
        pages = {662-677},
          doi = {10.1093/mnras/stab038},
archivePrefix = {arXiv},
       eprint = {2101.01195},
 primaryClass = {astro-ph.GA},
       adsurl = {https://ui.adsabs.harvard.edu/abs/2021MNRAS.502..662B}
}

@ARTICLE{Jarvis2013,
author = {{Jarvis}, Matt J. and {Bonfield}, D.~G. and {Bruce}, V.~A. and {Geach}, J.~E. and {McAlpine}, K. and {McLure}, R.~J. and {Gonz{\'a}lez-Solares}, E. and {Irwin}, M. and {Lewis}, J. and {Yoldas}, A. Kupcu and {Andreon}, S. and {Cross}, N.~J.~G. and {Emerson}, J.~P. and {Dalton}, G. and {Dunlop}, J.~S. and {Hodgkin}, S.~T. and {Le}, F{\`e}vre O. and {Karouzos}, M. and {Meisenheimer}, K. and {Oliver}, S. and {Rawlings}, S. and {Simpson}, C. and {Smail}, I. and {Smith}, D.~J.~B. and {Sullivan}, M. and {Sutherland}, W. and {White}, S.~V. and {Zwart}, J.~T.~L.},
        title = "{The VISTA Deep Extragalactic Observations (VIDEO) survey}",
      journal = {\mnras},
         year = 2013,
        month = jan,
       volume = {428},
       number = {2},
        pages = {1281-1295},
          doi = {10.1093/mnras/sts118},
archivePrefix = {arXiv},
       eprint = {1206.4263},
 primaryClass = {astro-ph.CO},
       adsurl = {https://ui.adsabs.harvard.edu/abs/2013MNRAS.428.1281J}
}

@ARTICLE{Aihara2019,
       author = {{Aihara}, Hiroaki and {AlSayyad}, Yusra and {Ando}, Makoto and {Armstrong}, Robert and {Bosch}, James and {Egami}, Eiichi and {Furusawa}, Hisanori and {Furusawa}, Junko and {Goulding}, Andy and {Harikane}, Yuichi and {Hikage}, Chiaki and {Ho}, Paul T.~P. and {Hsieh}, Bau-Ching and {Huang}, Song and {Ikeda}, Hiroyuki and {Imanishi}, Masatoshi and {Ito}, Kei and {Iwata}, Ikuru and {Jaelani}, Anton T. and {Kakuma}, Ryota and {Kawana}, Kojiro and {Kikuta}, Satoshi and {Kobayashi}, Umi and {Koike}, Michitaro and {Komiyama}, Yutaka and {Li}, Xiangchong and {Liang}, Yongming and {Lin}, Yen-Ting and {Luo}, Wentao and {Lupton}, Robert and {Lust}, Nate B. and {MacArthur}, Lauren A. and {Matsuoka}, Yoshiki and {Mineo}, Sogo and {Miyatake}, Hironao and {Miyazaki}, Satoshi and {More}, Surhud and {Murata}, Ryoma and {Namiki}, Shigeru V. and {Nishizawa}, Atsushi J. and {Oguri}, Masamune and {Okabe}, Nobuhiro and {Okamoto}, Sakurako and {Okura}, Yuki and {Ono}, Yoshiaki and {Onodera}, Masato and {Onoue}, Masafusa and {Osato}, Ken and {Ouchi}, Masami and {Shibuya}, Takatoshi and {Strauss}, Michael A. and {Sugiyama}, Naoshi and {Suto}, Yasushi and {Takada}, Masahiro and {Takagi}, Yuhei and {Takata}, Tadafumi and {Takita}, Satoshi and {Tanaka}, Masayuki and {Terai}, Tsuyoshi and {Toba}, Yoshiki and {Uchiyama}, Hisakazu and {Utsumi}, Yousuke and {Wang}, Shiang-Yu and {Wang}, Wenting and {Yamada}, Yoshihiko},
        title = "{Second data release of the Hyper Suprime-Cam Subaru Strategic Program}",
      journal = {\pasj},
         year = 2019,
        month = dec,
       volume = {71},
       number = {6},
          eid = {114},
        pages = {114},
          doi = {10.1093/pasj/psz103},
archivePrefix = {arXiv},
       eprint = {1905.12221},
 primaryClass = {astro-ph.IM},
       adsurl = {https://ui.adsabs.harvard.edu/abs/2019PASJ...71..114A}
}

@ARTICLE{Aihara2018b,
       author = {{Aihara}, Hiroaki and {Armstrong}, Robert and {Bickerton}, Steven and {Bosch}, James and {Coupon}, Jean and {Furusawa}, Hisanori and {Hayashi}, Yusuke and {Ikeda}, Hiroyuki and {Kamata}, Yukiko and {Karoji}, Hiroshi and {Kawanomoto}, Satoshi and {Koike}, Michitaro and {Komiyama}, Yutaka and {Lang}, Dustin and {Lupton}, Robert H. and {Mineo}, Sogo and {Miyatake}, Hironao and {Miyazaki}, Satoshi and {Morokuma}, Tomoki and {Obuchi}, Yoshiyuki and {Oishi}, Yukie and {Okura}, Yuki and {Price}, Paul A. and {Takata}, Tadafumi and {Tanaka}, Manobu M. and {Tanaka}, Masayuki and {Tanaka}, Yoko and {Uchida}, Tomohisa and {Uraguchi}, Fumihiro and {Utsumi}, Yousuke and {Wang}, Shiang-Yu and {Yamada}, Yoshihiko and {Yamanoi}, Hitomi and {Yasuda}, Naoki and {Arimoto}, Nobuo and {Chiba}, Masashi and {Finet}, Francois and {Fujimori}, Hiroki and {Fujimoto}, Seiji and {Furusawa}, Junko and {Goto}, Tomotsugu and {Goulding}, Andy and {Gunn}, James E. and {Harikane}, Yuichi and {Hattori}, Takashi and {Hayashi}, Masao and {He{\l}miniak}, Krzysztof G. and {Higuchi}, Ryo and {Hikage}, Chiaki and {Ho}, Paul T.~P. and {Hsieh}, Bau-Ching and {Huang}, Kuiyun and {Huang}, Song and {Imanishi}, Masatoshi and {Iwata}, Ikuru and {Jaelani}, Anton T. and {Jian}, Hung-Yu and {Kashikawa}, Nobunari and {Katayama}, Nobuhiko and {Kojima}, Takashi and {Konno}, Akira and {Koshida}, Shintaro and {Kusakabe}, Haruka and {Leauthaud}, Alexie and {Lee}, Chien-Hsiu and {Lin}, Lihwai and {Lin}, Yen-Ting and {Mandelbaum}, Rachel and {Matsuoka}, Yoshiki and {Medezinski}, Elinor and {Miyama}, Shoken and {Momose}, Rieko and {More}, Anupreeta and {More}, Surhud and {Mukae}, Shiro and {Murata}, Ryoma and {Murayama}, Hitoshi and {Nagao}, Tohru and {Nakata}, Fumiaki and {Niida}, Mana and {Niikura}, Hiroko and {Nishizawa}, Atsushi J. and {Oguri}, Masamune and {Okabe}, Nobuhiro and {Ono}, Yoshiaki and {Onodera}, Masato and {Onoue}, Masafusa and {Ouchi}, Masami and {Pyo}, Tae-Soo and {Shibuya}, Takatoshi and {Shimasaku}, Kazuhiro and {Simet}, Melanie and {Speagle}, Joshua and {Spergel}, David N. and {Strauss}, Michael A. and {Sugahara}, Yuma and {Sugiyama}, Naoshi and {Suto}, Yasushi and {Suzuki}, Nao and {Tait}, Philip J. and {Takada}, Masahiro and {Terai}, Tsuyoshi and {Toba}, Yoshiki and {Turner}, Edwin L. and {Uchiyama}, Hisakazu and {Umetsu}, Keiichi and {Urata}, Yuji and {Usuda}, Tomonori and {Yeh}, Sherry and {Yuma}, Suraphong},
        title = "{First data release of the Hyper Suprime-Cam Subaru Strategic Program}",
      journal = {\pasj},
         year = 2018,
        month = jan,
       volume = {70},
          eid = {S8},
        pages = {S8},
          doi = {10.1093/pasj/psx081},
archivePrefix = {arXiv},
       eprint = {1702.08449},
 primaryClass = {astro-ph.IM},
       adsurl = {https://ui.adsabs.harvard.edu/abs/2018PASJ...70S...8A}
}

@ARTICLE{Aihara2018a,
       author = {{Aihara}, Hiroaki and {Arimoto}, Nobuo and {Armstrong}, Robert and {Arnouts}, St{\'e}phane and {Bahcall}, Neta A. and {Bickerton}, Steven and {Bosch}, James and {Bundy}, Kevin and {Capak}, Peter L. and {Chan}, James H.~H. and {Chiba}, Masashi and {Coupon}, Jean and {Egami}, Eiichi and {Enoki}, Motohiro and {Finet}, Francois and {Fujimori}, Hiroki and {Fujimoto}, Seiji and {Furusawa}, Hisanori and {Furusawa}, Junko and {Goto}, Tomotsugu and {Goulding}, Andy and {Greco}, Johnny P. and {Greene}, Jenny E. and {Gunn}, James E. and {Hamana}, Takashi and {Harikane}, Yuichi and {Hashimoto}, Yasuhiro and {Hattori}, Takashi and {Hayashi}, Masao and {Hayashi}, Yusuke and {He{\l}miniak}, Krzysztof G. and {Higuchi}, Ryo and {Hikage}, Chiaki and {Ho}, Paul T.~P. and {Hsieh}, Bau-Ching and {Huang}, Kuiyun and {Huang}, Song and {Ikeda}, Hiroyuki and {Imanishi}, Masatoshi and {Inoue}, Akio K. and {Iwasawa}, Kazushi and {Iwata}, Ikuru and {Jaelani}, Anton T. and {Jian}, Hung-Yu and {Kamata}, Yukiko and {Karoji}, Hiroshi and {Kashikawa}, Nobunari and {Katayama}, Nobuhiko and {Kawanomoto}, Satoshi and {Kayo}, Issha and {Koda}, Jin and {Koike}, Michitaro and {Kojima}, Takashi and {Komiyama}, Yutaka and {Konno}, Akira and {Koshida}, Shintaro and {Koyama}, Yusei and {Kusakabe}, Haruka and {Leauthaud}, Alexie and {Lee}, Chien-Hsiu and {Lin}, Lihwai and {Lin}, Yen-Ting and {Lupton}, Robert H. and {Mandelbaum}, Rachel and {Matsuoka}, Yoshiki and {Medezinski}, Elinor and {Mineo}, Sogo and {Miyama}, Shoken and {Miyatake}, Hironao and {Miyazaki}, Satoshi and {Momose}, Rieko and {More}, Anupreeta and {More}, Surhud and {Moritani}, Yuki and {Moriya}, Takashi J. and {Morokuma}, Tomoki and {Mukae}, Shiro and {Murata}, Ryoma and {Murayama}, Hitoshi and {Nagao}, Tohru and {Nakata}, Fumiaki and {Niida}, Mana and {Niikura}, Hiroko and {Nishizawa}, Atsushi J. and {Obuchi}, Yoshiyuki and {Oguri}, Masamune and {Oishi}, Yukie and {Okabe}, Nobuhiro and {Okamoto}, Sakurako and {Okura}, Yuki and {Ono}, Yoshiaki and {Onodera}, Masato and {Onoue}, Masafusa and {Osato}, Ken and {Ouchi}, Masami and {Price}, Paul A. and {Pyo}, Tae-Soo and {Sako}, Masao and {Sawicki}, Marcin and {Shibuya}, Takatoshi and {Shimasaku}, Kazuhiro and {Shimono}, Atsushi and {Shirasaki}, Masato and {Silverman}, John D. and {Simet}, Melanie and {Speagle}, Joshua and {Spergel}, David N. and {Strauss}, Michael A. and {Sugahara}, Yuma and {Sugiyama}, Naoshi and {Suto}, Yasushi and {Suyu}, Sherry H. and {Suzuki}, Nao and {Tait}, Philip J. and {Takada}, Masahiro and {Takata}, Tadafumi and {Tamura}, Naoyuki and {Tanaka}, Manobu M. and {Tanaka}, Masaomi and {Tanaka}, Masayuki and {Tanaka}, Yoko and {Terai}, Tsuyoshi and {Terashima}, Yuichi and {Toba}, Yoshiki and {Tominaga}, Nozomu and {Toshikawa}, Jun and {Turner}, Edwin L. and {Uchida}, Tomohisa and {Uchiyama}, Hisakazu and {Umetsu}, Keiichi and {Uraguchi}, Fumihiro and {Urata}, Yuji and {Usuda}, Tomonori and {Utsumi}, Yousuke and {Wang}, Shiang-Yu and {Wang}, Wei-Hao and {Wong}, Kenneth C. and {Yabe}, Kiyoto and {Yamada}, Yoshihiko and {Yamanoi}, Hitomi and {Yasuda}, Naoki and {Yeh}, Sherry and {Yonehara}, Atsunori and {Yuma}, Suraphong},
        title = "{The Hyper Suprime-Cam SSP Survey: Overview and survey design}",
      journal = {\pasj},
         year = 2018,
        month = jan,
       volume = {70},
          eid = {S4},
        pages = {S4},
          doi = {10.1093/pasj/psx066},
archivePrefix = {arXiv},
       eprint = {1704.05858},
 primaryClass = {astro-ph.IM},
       adsurl = {https://ui.adsabs.harvard.edu/abs/2018PASJ...70S...4A}
}

@ARTICLE{McCracken2012,
       author = {{McCracken}, H.~J. and {Milvang-Jensen}, B. and {Dunlop}, J. and {Franx}, M. and {Fynbo}, J.~P.~U. and {Le F{\`e}vre}, O. and {Holt}, J. and {Caputi}, K.~I. and {Goranova}, Y. and {Buitrago}, F. and {Emerson}, J.~P. and {Freudling}, W. and {Hudelot}, P. and {L{\'o}pez-Sanjuan}, C. and {Magnard}, F. and {Mellier}, Y. and {M{\o}ller}, P. and {Nilsson}, K.~K. and {Sutherland}, W. and {Tasca}, L. and {Zabl}, J.},
        title = "{UltraVISTA: a new ultra-deep near-infrared survey in COSMOS}",
      journal = {\aap},
         year = 2012,
        month = aug,
       volume = {544},
          eid = {A156},
        pages = {A156},
          doi = {10.1051/0004-6361/201219507},
archivePrefix = {arXiv},
       eprint = {1204.6586},
 primaryClass = {astro-ph.CO},
       adsurl = {https://ui.adsabs.harvard.edu/abs/2012A&A...544A.156M}
}

@INPROCEEDINGS{Jarvis2016,
       author = {{Jarvis}, M. and {Taylor}, R. and {Agudo}, I. and {Allison}, J.~R. and {Deane}, R.~P. and {Frank}, B. and {Gupta}, N. and {Heywood}, I. and {Maddox}, N. and {McAlpine}, K. and {Santos}, M. and {Scaife}, A.~M.~M. and {Vaccari}, M. and {Zwart}, J.~T.~L. and {Adams}, E. and {Bacon}, D.~J. and {Baker}, A.~J. and {Bassett}, B.~A. and {Best}, P.~N. and {Beswick}, R. and {Blyth}, S. and {Brown}, M.~L. and {Bruggen}, M. and {Cluver}, M. and {Colafrancesco}, S. and {Cotter}, G. and {Cress}, C. and {Dav{\'e}}, R. and {Ferrari}, C. and {Hardcastle}, M.~J. and {Hale}, C.~L. and {Harrison}, I. and {Hatfield}, P.~W. and {Klockner}, H.~R. and {Kolwa}, S. and {Malefahlo}, E. and {Marubini}, T. and {Mauch}, T. and {Moodley}, K. and {Morganti}, R. and {Norris}, R.~P. and {Peters}, J.~A. and {Prandoni}, I. and {Prescott}, M. and {Oliver}, S. and {Oozeer}, N. and {Rottgering}, H.~J.~A. and {Seymour}, N. and {Simpson}, C. and {Smirnov}, O. and {Smith}, D.~J.~B.},
        title = "{The MeerKAT International GHz Tiered Extragalactic Exploration (MIGHTEE) Survey}",
    booktitle = {MeerKAT Science: On the Pathway to the SKA},
         year = 2016,
        month = jan,
          eid = {6},
        pages = {6},
          doi = {10.22323/1.277.0006},
archivePrefix = {arXiv},
       eprint = {1709.01901},
 primaryClass = {astro-ph.GA},
       adsurl = {https://ui.adsabs.harvard.edu/abs/2016mks..confE...6J}
}

@ARTICLE{Adams2023,
       author = {{Adams}, N.~J. and {Bowler}, R.~A.~A. and {Jarvis}, M.~J. and {Varadaraj}, R.~G. and {H{\"a}u{\ss}ler}, B.},
        title = "{The total rest-frame UV luminosity function from 3 < z < 5: a simultaneous study of AGN and galaxies from -28 < M$_{UV}$ < -16}",
      journal = {\mnras},
         year = 2023,
        month = jul,
       volume = {523},
       number = {1},
        pages = {327-346},
          doi = {10.1093/mnras/stad1333},
archivePrefix = {arXiv},
       eprint = {2207.09342},
 primaryClass = {astro-ph.GA},
       adsurl = {https://ui.adsabs.harvard.edu/abs/2023MNRAS.523..327A}
}

@ARTICLE{Tasse2021,
       author = {{Tasse}, C. and {Shimwell}, T. and {Hardcastle}, M.~J. and {O'Sullivan}, S.~P. and {van Weeren}, R. and {Best}, P.~N. and {Bester}, L. and {Hugo}, B. and {Smirnov}, O. and {Sabater}, J. and {Calistro-Rivera}, G. and {de Gasperin}, F. and {Morabito}, L.~K. and {R{\"o}ttgering}, H. and {Williams}, W.~L. and {Bonato}, M. and {Bondi}, M. and {Botteon}, A. and {Br{\"u}ggen}, M. and {Brunetti}, G. and {Chy{\.z}y}, K.~T. and {Garrett}, M.~A. and {G{\"u}rkan}, G. and {Jarvis}, M.~J. and {Kondapally}, R. and {Mandal}, S. and {Prandoni}, I. and {Repetti}, A. and {Retana-Montenegro}, E. and {Schwarz}, D.~J. and {Shulevski}, A. and {Wiaux}, Y.},
        title = "{The LOFAR Two-meter Sky Survey: Deep Fields Data Release 1. I. Direction-dependent calibration and imaging}",
      journal = {\aap},
         year = 2021,
        month = apr,
       volume = {648},
          eid = {A1},
        pages = {A1},
          doi = {10.1051/0004-6361/202038804},
archivePrefix = {arXiv},
       eprint = {2011.08328},
 primaryClass = {astro-ph.IM},
       adsurl = {https://ui.adsabs.harvard.edu/abs/2021A&A...648A...1T}
}

@ARTICLE{Sabater2021,
       author = {{Sabater}, J. and {Best}, P.~N. and {Tasse}, C. and {Hardcastle}, M.~J. and {Shimwell}, T.~W. and {Nisbet}, D. and {Jelic}, V. and {Callingham}, J.~R. and {R{\"o}ttgering}, H.~J.~A. and {Bonato}, M. and {Bondi}, M. and {Ciardi}, B. and {Cochrane}, R.~K. and {Jarvis}, M.~J. and {Kondapally}, R. and {Koopmans}, L.~V.~E. and {O'Sullivan}, S.~P. and {Prandoni}, I. and {Schwarz}, D.~J. and {Smith}, D.~J.~B. and {Wang}, L. and {Williams}, W.~L. and {Zaroubi}, S.},
        title = "{The LOFAR Two-meter Sky Survey: Deep Fields Data Release 1. II. The ELAIS-N1 LOFAR deep field}",
      journal = {\aap},
         year = 2021,
        month = apr,
       volume = {648},
          eid = {A2},
        pages = {A2},
          doi = {10.1051/0004-6361/202038828},
archivePrefix = {arXiv},
       eprint = {2011.08211},
 primaryClass = {astro-ph.GA},
       adsurl = {https://ui.adsabs.harvard.edu/abs/2021A&A...648A...2S}
}

@ARTICLE{DESI,
       author = {{DESI Collaboration} and {Adame}, A.~G. and {Aguilar}, J. and {Ahlen}, S. and {Alam}, S. and {Aldering}, G. and {Alexander}, D.~M. and {Alfarsy}, R. and {Allende Prieto}, C. and {Alvarez}, M. and {Alves}, O. and {Anand}, A. and {Andrade-Oliveira}, F. and {Armengaud}, E. and {Asorey}, J. and {Avila}, S. and {Aviles}, A. and {Bailey}, S. and {Balaguera-Antol{\'\i}nez}, A. and {Ballester}, O. and {Baltay}, C. and {Bault}, A. and {Bautista}, J. and {Behera}, J. and {Beltran}, S.~F. and {BenZvi}, S. and {Beraldo e Silva}, L. and {Bermejo-Climent}, J.~R. and {Berti}, A. and {Besuner}, R. and {Beutler}, F. and {Bianchi}, D. and {Blake}, C. and {Blum}, R. and {Bolton}, A.~S. and {Brieden}, S. and {Brodzeller}, A. and {Brooks}, D. and {Brown}, Z. and {Buckley-Geer}, E. and {Burtin}, E. and {Cabayol-Garcia}, L. and {Cai}, Z. and {Canning}, R. and {Cardiel-Sas}, L. and {Carnero Rosell}, A. and {Castander}, F.~J. and {Cervantes-Cota}, J.~L. and {Chabanier}, S. and {Chaussidon}, E. and {Chaves-Montero}, J. and {Chen}, S. and {Chen}, X. and {Chuang}, C. and {Claybaugh}, T. and {Cole}, S. and {Cooper}, A.~P. and {Cuceu}, A. and {Davis}, T.~M. and {Dawson}, K. and {de Belsunce}, R. and {de la Cruz}, R. and {de la Macorra}, A. and {Della Costa}, J. and {de Mattia}, A. and {Demina}, R. and {Demirbozan}, U. and {DeRose}, J. and {Dey}, A. and {Dey}, B. and {Dhungana}, G. and {Ding}, J. and {Ding}, Z. and {Doel}, P. and {Doshi}, R. and {Douglass}, K. and {Edge}, A. and {Eftekharzadeh}, S. and {Eisenstein}, D.~J. and {Elliott}, A. and {Ereza}, J. and {Escoffier}, S. and {Fagrelius}, P. and {Fan}, X. and {Fanning}, K. and {Fawcett}, V.~A. and {Ferraro}, S. and {Flaugher}, B. and {Font-Ribera}, A. and {Forero-Romero}, J.~E. and {Forero-S{\'a}nchez}, D. and {Frenk}, C.~S. and {G{\"a}nsicke}, B.~T. and {Garc{\'\i}a}, L. {\'A}. and {Garc{\'\i}a-Bellido}, J. and {Garcia-Quintero}, C. and {Garrison}, L.~H. and {Gil-Mar{\'\i}n}, H. and {Golden-Marx}, J. and {Gontcho A Gontcho}, S. and {Gonzalez-Morales}, A.~X. and {Gonzalez-Perez}, V. and {Gordon}, C. and {Graur}, O. and {Green}, D. and {Gruen}, D. and {Guy}, J. and {Hadzhiyska}, B. and {Hahn}, C. and {Han}, J.~J. and {Hanif}, M.~M.~S. and {Herrera-Alcantar}, H.~K. and {Honscheid}, K. and {Hou}, J. and {Howlett}, C. and {Huterer}, D. and {Ir{\v{s}}i{\v{c}}}, V. and {Ishak}, M. and {Jacques}, A. and {Jana}, A. and {Jiang}, L. and {Jimenez}, J. and {Jing}, Y.~P. and {Joudaki}, S. and {Joyce}, R. and {Jullo}, E. and {Juneau}, S. and {Kara{\c{c}}ayl{\i}}, N.~G. and {Karim}, T. and {Kehoe}, R. and {Kent}, S. and {Khederlarian}, A. and {Kim}, S. and {Kirkby}, D. and {Kisner}, T. and {Kitaura}, F. and {Kizhuprakkat}, N. and {Kneib}, J. and {Koposov}, S.~E. and {Kov{\'a}cs}, A. and {Kremin}, A. and {Krolewski}, A. and {L'Huillier}, B. and {Lahav}, O. and {Lambert}, A. and {Lamman}, C. and {Lan}, T. -W. and {Landriau}, M. and {Lang}, D. and {Lange}, J.~U. and {Lasker}, J. and {Leauthaud}, A. and {Le Guillou}, L. and {Levi}, M.~E. and {Li}, T.~S. and {Linder}, E. and {Lyons}, A. and {Magneville}, C. and {Manera}, M. and {Manser}, C.~J. and {Margala}, D. and {Martini}, P. and {McDonald}, P. and {Medina}, G.~E. and {Medina-Varela}, L. and {Meisner}, A. and {Mena-Fern{\'a}ndez}, J. and {Meneses-Rizo}, J. and {Mezcua}, M. and {Miquel}, R. and {Montero-Camacho}, P. and {Moon}, J. and {Moore}, S. and {Moustakas}, J. and {Mueller}, E. and {Mundet}, J. and {Mu{\~n}oz-Guti{\'e}rrez}, A. and {Myers}, A.~D. and {Nadathur}, S. and {Napolitano}, L. and {Neveux}, R. and {Newman}, J.~A. and {Nie}, J. and {Nikutta}, R. and {Niz}, G. and {Norberg}, P. and {Noriega}, H.~E. and {Paillas}, E. and {Palanque-Delabrouille}, N. and {Palmese}, A. and {Pan}, Z. and {Parkinson}, D. and {Penmetsa}, S. and {Percival}, W.~J. and {P{\'e}rez-Fern{\'a}ndez}, A. and {P{\'e}rez-R{\`a}fols}, I. and {Pieri}, M. and {Poppett}, C. and {Porredon}, A. and {Pothier}, S.},
        title = "{The Early Data Release of the Dark Energy Spectroscopic Instrument}",
      journal = {\aj},
         year = 2024,
        month = aug,
       volume = {168},
       number = {2},
          eid = {58},
        pages = {58},
          doi = {10.3847/1538-3881/ad3217},
archivePrefix = {arXiv},
       eprint = {2306.06308},
 primaryClass = {astro-ph.CO},
       adsurl = {https://ui.adsabs.harvard.edu/abs/2024AJ....168...58D}
}

@ARTICLE{Arnaudova2025,
       author = {{Arnaudova}, M.~I. and {Smith}, D.~J.~B. and {Hardcastle}, M.~J. and {Best}, P.~N. and {Das}, S. and {Shenoy}, S. and {Duncan}, K.~J. and {Holden}, L.~R. and {Kondapally}, R. and {Morabito}, L.~K. and {R{\"o}ttgering}, H.~J.~A.},
        title = "{The LOFAR Two-metre Sky Survey Deep Fields: new probabilistic spectroscopic classifications and the accretion rates of radio galaxies}",
      journal = {\mnras},
         year = 2025,
        month = sep,
       volume = {542},
       number = {3},
        pages = {2245-2268},
          doi = {10.1093/mnras/staf1347},
archivePrefix = {arXiv},
       eprint = {2508.18347},
 primaryClass = {astro-ph.GA},
       adsurl = {https://ui.adsabs.harvard.edu/abs/2025MNRAS.542.2245A}
}

@ARTICLE{Malefahlo2020,
       author = {{Malefahlo}, Eliab and {Santos}, Mario G. and {Jarvis}, Matt J. and {White}, Sarah V. and {Zwart}, Jonathan T.~L.},
        title = "{The optically selected 1.4-GHz quasar luminosity function below 1 mJy}",
      journal = {\mnras},
         year = 2020,
        month = mar,
       volume = {492},
       number = {4},
        pages = {5297-5312},
          doi = {10.1093/mnras/staa112},
archivePrefix = {arXiv},
       eprint = {1908.05316},
 primaryClass = {astro-ph.GA},
       adsurl = {https://ui.adsabs.harvard.edu/abs/2020MNRAS.492.5297M}
}

@ARTICLE{Kondapally2022,
       author = {{Kondapally}, Rohit and {Best}, Philip N. and {Cochrane}, Rachel K. and {Sabater}, Jos{\'e} and {Duncan}, Kenneth J. and {Hardcastle}, Martin J. and {Haskell}, Paul and {Mingo}, Beatriz and {R{\"o}ttgering}, Huub J.~A. and {Smith}, Daniel J.~B. and {Williams}, Wendy L. and {Bonato}, Matteo and {Calistro Rivera}, Gabriela and {Gao}, Fangyou and {Hale}, Catherine L. and {Ma{\l}ek}, Katarzyna and {Miley}, George K. and {Prandoni}, Isabella and {Wang}, Lingyu},
        title = "{Cosmic evolution of low-excitation radio galaxies in the LOFAR two-metre sky survey deep fields}",
      journal = {\mnras},
         year = 2022,
        month = jul,
       volume = {513},
       number = {3},
        pages = {3742-3767},
          doi = {10.1093/mnras/stac1128},
archivePrefix = {arXiv},
       eprint = {2204.07588},
 primaryClass = {astro-ph.GA},
       adsurl = {https://ui.adsabs.harvard.edu/abs/2022MNRAS.513.3742K}
}

@ARTICLE{Heckman2024,
       author = {{Heckman}, Timothy M. and {Roy}, Namrata and {Best}, Philip N. and {Kondapally}, Rohit},
        title = "{Mergers, Radio Jets, and Quenching Star Formation in Massive Galaxies: Quantifying Their Synchronized Cosmic Evolution and Assessing the Energetics}",
      journal = {\apj},
         year = 2024,
        month = dec,
       volume = {977},
       number = {1},
          eid = {125},
        pages = {125},
          doi = {10.3847/1538-4357/ad8f3e},
archivePrefix = {arXiv},
       eprint = {2410.09157},
 primaryClass = {astro-ph.GA},
       adsurl = {https://ui.adsabs.harvard.edu/abs/2024ApJ...977..125H}
}

@ARTICLE{Kondapally2023,
       author = {{Kondapally}, Rohit and {Best}, Philip N. and {Raouf}, Mojtaba and {Thomas}, Nicole L. and {Dav{\'e}}, Romeel and {Shabala}, Stanislav S. and {R{\"o}ttgering}, Huub J.~A. and {Hardcastle}, Martin J. and {Bonato}, Matteo and {Cochrane}, Rachel K. and {Ma{\l}ek}, Katarzyna and {Morabito}, Leah K. and {Prandoni}, Isabella and {Smith}, Daniel J.~B.},
        title = "{Cosmic evolution of radio-AGN feedback: confronting models with data}",
      journal = {\mnras},
         year = 2023,
        month = aug,
       volume = {523},
       number = {4},
        pages = {5292-5305},
          doi = {10.1093/mnras/stad1813},
archivePrefix = {arXiv},
       eprint = {2306.11795},
 primaryClass = {astro-ph.GA},
       adsurl = {https://ui.adsabs.harvard.edu/abs/2023MNRAS.523.5292K}
}

@ARTICLE{Yue2025,
       author = {{Yue}, B. -H. and {Duncan}, K.~J. and {Best}, P.~N. and {Arnaudova}, M.~I. and {Morabito}, L.~K. and {Petley}, J.~W. and {R{\"o}ttgering}, H.~J.~A. and {Shenoy}, S. and {Smith}, D.~J.~B.},
        title = "{A novel Bayesian approach for decomposing the radio emission of quasars - II. Link between quasar radio emission and black hole mass}",
      journal = {\mnras},
         year = 2025,
        month = feb,
       volume = {537},
       number = {2},
        pages = {858-875},
          doi = {10.1093/mnras/staf077},
archivePrefix = {arXiv},
       eprint = {2501.07629},
 primaryClass = {astro-ph.GA},
       adsurl = {https://ui.adsabs.harvard.edu/abs/2025MNRAS.537..858Y}
}

@ARTICLE{Macfarlane2021,
       author = {{Macfarlane}, C. and {Best}, P.~N. and {Sabater}, J. and {G{\"u}rkan}, G. and {Jarvis}, M.~J. and {R{\"o}ttgering}, H.~J.~A. and {Baldi}, R.~D. and {Calistro Rivera}, G. and {Duncan}, K.~J. and {Morabito}, L.~K. and {Prandoni}, I. and {Retana-Montenegro}, E.},
        title = "{The radio loudness of SDSS quasars from the LOFAR Two-metre Sky Survey: ubiquitous jet activity and constraints on star formation}",
      journal = {\mnras},
         year = 2021,
        month = oct,
       volume = {506},
       number = {4},
        pages = {5888-5907},
          doi = {10.1093/mnras/stab1998},
archivePrefix = {arXiv},
       eprint = {2107.09141},
 primaryClass = {astro-ph.GA},
       adsurl = {https://ui.adsabs.harvard.edu/abs/2021MNRAS.506.5888M}
}

@ARTICLE{Condon2013,
       author = {{Condon}, J.~J. and {Kellermann}, K.~I. and {Kimball}, Amy E. and {Ivezi{\'c}}, {\v{Z}}eljko and {Perley}, R.~A.},
        title = "{Active Galactic Nucleus and Starburst Radio Emission from Optically Selected Quasi-stellar Objects}",
      journal = {\apj},
         year = 2013,
        month = may,
       volume = {768},
       number = {1},
          eid = {37},
        pages = {37},
          doi = {10.1088/0004-637X/768/1/37},
archivePrefix = {arXiv},
       eprint = {1303.3448},
 primaryClass = {astro-ph.CO},
       adsurl = {https://ui.adsabs.harvard.edu/abs/2013ApJ...768...37C}
}

@ARTICLE{Kimball2011,
       author = {{Kimball}, Amy E. and {Kellermann}, K.~I. and {Condon}, J.~J. and {Ivezi{\'c}}, {\v{Z}}eljko and {Perley}, Richard A.},
        title = "{The Two-component Radio Luminosity Function of Quasi-stellar Objects: Star Formation and Active Galactic Nucleus}",
      journal = {\apjl},
         year = 2011,
        month = sep,
       volume = {739},
       number = {1},
          eid = {L29},
        pages = {L29},
          doi = {10.1088/2041-8205/739/1/L29},
archivePrefix = {arXiv},
       eprint = {1107.3551},
 primaryClass = {astro-ph.CO},
       adsurl = {https://ui.adsabs.harvard.edu/abs/2011ApJ...739L..29K}
}

@ARTICLE{White2015,
       author = {{White}, Sarah V. and {Jarvis}, Matt J. and {H{\"a}u{\ss}ler}, Boris and {Maddox}, Natasha},
        title = "{Radio-quiet quasars in the VIDEO survey: evidence for AGN-powered radio emission at S\_\{1.4 GHz < 1\} mJy}",
      journal = {\mnras},
         year = 2015,
        month = apr,
       volume = {448},
       number = {3},
        pages = {2665-2686},
          doi = {10.1093/mnras/stv134},
archivePrefix = {arXiv},
       eprint = {1410.3892},
 primaryClass = {astro-ph.GA},
       adsurl = {https://ui.adsabs.harvard.edu/abs/2015MNRAS.448.2665W}
}

@ARTICLE{White2017,
       author = {{White}, Sarah V. and {Jarvis}, Matt J. and {Kalfountzou}, Eleni and {Hardcastle}, Martin J. and {Verma}, Aprajita and {Cao Orjales}, Jos{\'e} M. and {Stevens}, Jason},
        title = "{Evidence that the AGN dominates the radio emission in z {\ensuremath{\sim}} 1 radio-quiet quasars}",
      journal = {\mnras},
         year = 2017,
        month = jun,
       volume = {468},
       number = {1},
        pages = {217-238},
          doi = {10.1093/mnras/stx284},
archivePrefix = {arXiv},
       eprint = {1702.00904},
 primaryClass = {astro-ph.GA},
       adsurl = {https://ui.adsabs.harvard.edu/abs/2017MNRAS.468..217W}
}

@ARTICLE{Morabito2025,
       author = {{Morabito}, Leah K. and {Kondapally}, R. and {Best}, P.~N. and {Yue}, B. -H. and {de Jong}, J.~M.~G.~H.~J. and {Sweijen}, F. and {Bondi}, Marco and {Schwarz}, Dominik J. and {Smith}, D.~J.~B. and {van Weeren}, R.~J. and {R{\"o}ttgering}, H.~J.~A. and {Shimwell}, T.~W. and {Prandoni}, Isabella},
        title = "{A hidden active galactic nucleus population: the first radio luminosity functions constructed by physical process}",
      journal = {\mnras},
         year = 2025,
        month = jan,
       volume = {536},
       number = {1},
        pages = {L32-L37},
          doi = {10.1093/mnrasl/slae104},
archivePrefix = {arXiv},
       eprint = {2411.05069},
 primaryClass = {astro-ph.GA},
       adsurl = {https://ui.adsabs.harvard.edu/abs/2025MNRAS.536L..32M}
}

@ARTICLE{Muxlow2020,
       author = {{Muxlow}, T.~W.~B. and {Thomson}, A.~P. and {Radcliffe}, J.~F. and {Wrigley}, N.~H. and {Beswick}, R.~J. and {Smail}, Ian and {McHardy}, I.~M. and {Garrington}, S.~T. and {Ivison}, R.~J. and {Jarvis}, M.~J. and {Prandoni}, I. and {Bondi}, M. and {Guidetti}, D. and {Argo}, M.~K. and {Bacon}, David and {Best}, P.~N. and {Biggs}, A.~D. and {Chapman}, S.~C. and {Coppin}, K. and {Chen}, H. and {Garratt}, T.~K. and {Garrett}, M.~A. and {Ibar}, E. and {Kneib}, Jean-Paul and {Knudsen}, Kirsten K. and {Koopmans}, L.~V.~E. and {Morabito}, L.~K. and {Murphy}, E.~J. and {Njeri}, A. and {Pearson}, Chris and {P{\'e}rez-Torres}, M.~A. and {Richards}, A.~M.~S. and {R{\"o}ttgering}, H.~J.~A. and {Sargent}, M.~T. and {Serjeant}, Stephen and {Simpson}, C. and {Simpson}, J.~M. and {Swinbank}, A.~M. and {Varenius}, E. and {Venturi}, T.},
        title = "{The e-MERGE Survey (e-MERLIN Galaxy Evolution Survey): overview and survey description}",
      journal = {\mnras},
         year = 2020,
        month = jun,
       volume = {495},
       number = {1},
        pages = {1188-1208},
          doi = {10.1093/mnras/staa1279},
archivePrefix = {arXiv},
       eprint = {2005.02407},
 primaryClass = {astro-ph.GA},
       adsurl = {https://ui.adsabs.harvard.edu/abs/2020MNRAS.495.1188M}
}

@ARTICLE{Radcliffe2021,
       author = {{Radcliffe}, J.~F. and {Barthel}, P.~D. and {Thomson}, A.~P. and {Garrett}, M.~A. and {Beswick}, R.~J. and {Muxlow}, T.~W.~B.},
        title = "{Nowhere to hide: Radio-faint AGN in the GOODS-N field. II. Multi-wavelength AGN selection techniques and host galaxy properties}",
      journal = {\aap},
         year = 2021,
        month = may,
       volume = {649},
          eid = {A27},
        pages = {A27},
          doi = {10.1051/0004-6361/202038591},
archivePrefix = {arXiv},
       eprint = {2103.08575},
 primaryClass = {astro-ph.GA},
       adsurl = {https://ui.adsabs.harvard.edu/abs/2021A&A...649A..27R}
}

@ARTICLE{Murphy2009,
       author = {{Murphy}, Eric J.},
        title = "{The Far-Infrared-Radio Correlation at High Redshifts: Physical Considerations and Prospects for the Square Kilometer Array}",
      journal = {\apj},
         year = 2009,
        month = nov,
       volume = {706},
       number = {1},
        pages = {482-496},
          doi = {10.1088/0004-637X/706/1/482},
archivePrefix = {arXiv},
       eprint = {0910.0011},
 primaryClass = {astro-ph.CO},
       adsurl = {https://ui.adsabs.harvard.edu/abs/2009ApJ...706..482M}
}

@ARTICLE{Whittam2025,
       author = {{Whittam}, I.~H. and {Jarvis}, M.~J. and {Murphy}, Eric J. and {Adams}, N.~J. and {Bowler}, R.~A.~A. and {Matthews}, A. and {Varadaraj}, R.~G. and {Hale}, C.~L. and {Heywood}, I. and {Knowles}, K. and {Marchetti}, L. and {Seymour}, N. and {Tabatabaei}, F. and {Taylor}, A.~R. and {Vaccari}, M. and {Verma}, A.},
        title = "{Evidence for inverse Compton scattering in high-redshift Lyman-break galaxies}",
      journal = {\mnras},
         year = 2025,
        month = oct,
       volume = {543},
       number = {1},
        pages = {507-517},
          doi = {10.1093/mnras/staf1505},
archivePrefix = {arXiv},
       eprint = {2509.06677},
 primaryClass = {astro-ph.GA},
       adsurl = {https://ui.adsabs.harvard.edu/abs/2025MNRAS.543..507W}
}

@ARTICLE{Read2018,
       author = {{Read}, S.~C. and {Smith}, D.~J.~B. and {G{\"u}rkan}, G. and {Hardcastle}, M.~J. and {Williams}, W.~L. and {Best}, P.~N. and {Brinks}, E. and {Calistro-Rivera}, G. and {Chy{\.Z}y}, K.~T. and {Duncan}, K. and {Dunne}, L. and {Jarvis}, M.~J. and {Morabito}, L.~K. and {Prandoni}, I. and {R{\"o}ttgering}, H.~J.~A. and {Sabater}, J. and {Viaene}, S.},
        title = "{The Far-Infrared Radio Correlation at low radio frequency with LOFAR/H-ATLAS}",
      journal = {\mnras},
         year = 2018,
        month = nov,
       volume = {480},
       number = {4},
        pages = {5625-5644},
          doi = {10.1093/mnras/sty2198},
archivePrefix = {arXiv},
       eprint = {1808.10452},
 primaryClass = {astro-ph.GA},
       adsurl = {https://ui.adsabs.harvard.edu/abs/2018MNRAS.480.5625R}
}

@ARTICLE{Wang2019,
       author = {{Wang}, L. and {Gao}, F. and {Duncan}, K.~J. and {Williams}, W.~L. and {Rowan-Robinson}, M. and {Sabater}, J. and {Shimwell}, T.~W. and {Bonato}, M. and {Calistro-Rivera}, G. and {Chy{\.z}y}, K.~T. and {Farrah}, D. and {G{\"u}rkan}, G. and {Hardcastle}, M.~J. and {McCheyne}, I. and {Prandoni}, I. and {Read}, S.~C. and {R{\"o}ttgering}, H.~J.~A. and {Smith}, D.~J.~B.},
        title = "{A LOFAR-IRAS cross-match study: the far-infrared radio correlation and the 150 MHz luminosity as a star-formation rate tracer}",
      journal = {\aap},
         year = 2019,
        month = nov,
       volume = {631},
          eid = {A109},
        pages = {A109},
          doi = {10.1051/0004-6361/201935913},
archivePrefix = {arXiv},
       eprint = {1909.04489},
 primaryClass = {astro-ph.GA},
       adsurl = {https://ui.adsabs.harvard.edu/abs/2019A&A...631A.109W}
}

@ARTICLE{Algera2020,
       author = {{Algera}, H.~S.~B. and {Smail}, I. and {Dudzevi{\v{c}}i{\={u}}t{\.{e}}}, U. and {Swinbank}, A.~M. and {Stach}, S. and {Hodge}, J.~A. and {Thomson}, A.~P. and {Almaini}, O. and {Arumugam}, V. and {Blain}, A.~W. and {Calistro-Rivera}, G. and {Chapman}, S.~C. and {Chen}, C. -C. and {da Cunha}, E. and {Farrah}, D. and {Leslie}, S. and {Scott}, D. and {van der Vlugt}, D. and {Wardlow}, J.~L. and {van der Werf}, P.},
        title = "{An ALMA Survey of the SCUBA-2 Cosmology Legacy Survey UKIDSS/UDS Field: The Far-infrared/Radio Correlation for High-redshift Dusty Star-forming Galaxies}",
      journal = {\apj},
         year = 2020,
        month = nov,
       volume = {903},
       number = {2},
          eid = {138},
        pages = {138},
          doi = {10.3847/1538-4357/abb77b},
archivePrefix = {arXiv},
       eprint = {2009.06647},
 primaryClass = {astro-ph.GA},
       adsurl = {https://ui.adsabs.harvard.edu/abs/2020ApJ...903..138A}
}

@ARTICLE{Gurkan2021,
       author = {{Smith}, D.~J.~B. and {Haskell}, P. and {G{\"u}rkan}, G. and {Best}, P.~N. and {Hardcastle}, M.~J. and {Kondapally}, R. and {Williams}, W. and {Duncan}, K.~J. and {Cochrane}, R.~K. and {McCheyne}, I. and {R{\"o}ttgering}, H.~J.~A. and {Sabater}, J. and {Shimwell}, T.~W. and {Tasse}, C. and {Bonato}, M. and {Bondi}, M. and {Jarvis}, M.~J. and {Leslie}, S.~K. and {Prandoni}, I. and {Wang}, L.},
        title = "{The LOFAR Two-metre Sky Survey Deep Fields. The star-formation rate-radio luminosity relation at low frequencies}",
      journal = {\aap},
         year = 2021,
        month = apr,
       volume = {648},
          eid = {A6},
        pages = {A6},
          doi = {10.1051/0004-6361/202039343},
archivePrefix = {arXiv},
       eprint = {2011.08196},
 primaryClass = {astro-ph.GA},
       adsurl = {https://ui.adsabs.harvard.edu/abs/2021A&A...648A...6S}
}

@ARTICLE{Cook2024,
       author = {{Cook}, Robin H.~W. and {Davies}, Luke J.~M. and {Rhee}, Jonghwan and {Hale}, Catherine L. and {Bellstedt}, Sabine and {Thorne}, Jessica E. and {Delvecchio}, Ivan and {Collier}, Jordan D. and {Dodson}, Richard and {Driver}, Simon P. and {Holwerda}, Benne W. and {Jarvis}, Matt J. and {Knowles}, Kenda and {Lagos}, Claudia and {Maddox}, Natasha and {Meyer}, Martin and {Robotham}, Aaron S.~G. and {Roychowdhury}, Sambit and {Rozgonyi}, Kristof and {Seymour}, Nicholas and {Siudek}, Malgorzata and {Whiting}, Matthew and {Whittam}, Imogen},
        title = "{DEVILS/MIGHTEE/GAMA/DINGO: the impact of SFR time-scales on the SFR-radio luminosity correlation}",
      journal = {\mnras},
         year = 2024,
        month = jun,
       volume = {531},
       number = {1},
        pages = {708-727},
          doi = {10.1093/mnras/stae1215},
archivePrefix = {arXiv},
       eprint = {2405.00337},
 primaryClass = {astro-ph.GA},
       adsurl = {https://ui.adsabs.harvard.edu/abs/2024MNRAS.531..708C}
}

@ARTICLE{Jarvis2010,
       author = {{Jarvis}, Matt J. and {Smith}, D.~J.~B. and {Bonfield}, D.~G. and {Hardcastle}, M.~J. and {Falder}, J.~T. and {Stevens}, J.~A. and {Ivison}, R.~J. and {Auld}, R. and {Baes}, M. and {Baldry}, I.~K. and {Bamford}, S.~P. and {Bourne}, N. and {Buttiglione}, S. and {Cava}, A. and {Cooray}, A. and {Dariush}, A. and {de Zotti}, G. and {Dunlop}, J.~S. and {Dunne}, L. and {Dye}, S. and {Eales}, S. and {Fritz}, J. and {Hill}, D.~T. and {Hopwood}, R. and {Hughes}, D.~H. and {Ibar}, E. and {Jones}, D.~H. and {Kelvin}, L. and {Lawrence}, A. and {Leeuw}, L. and {Loveday}, J. and {Maddox}, S.~J. and {Micha{\l}owski}, M.~J. and {Negrello}, M. and {Norberg}, P. and {Pohlen}, M. and {Prescott}, M. and {Rigby}, E.~E. and {Robotham}, A. and {Rodighiero}, G. and {Scott}, D. and {Sharp}, R. and {Temi}, P. and {Thompson}, M.~A. and {van der Werf}, P. and {van Kampen}, E. and {Vlahakis}, C. and {White}, G.},
        title = "{Herschel-ATLAS: the far-infrared-radio correlation at z < 0.5}",
      journal = {\mnras},
         year = 2010,
        month = nov,
       volume = {409},
       number = {1},
        pages = {92-101},
          doi = {10.1111/j.1365-2966.2010.17772.x},
archivePrefix = {arXiv},
       eprint = {1009.5390},
 primaryClass = {astro-ph.CO},
       adsurl = {https://ui.adsabs.harvard.edu/abs/2010MNRAS.409...92J}
}

@ARTICLE{Yun2001,
       author = {{Yun}, Min S. and {Reddy}, Naveen A. and {Condon}, J.~J.},
        title = "{Radio Properties of Infrared-selected Galaxies in the IRAS 2 Jy Sample}",
      journal = {\apj},
         year = 2001,
        month = jun,
       volume = {554},
       number = {2},
        pages = {803-822},
          doi = {10.1086/323145},
archivePrefix = {arXiv},
       eprint = {astro-ph/0102154},
 primaryClass = {astro-ph},
       adsurl = {https://ui.adsabs.harvard.edu/abs/2001ApJ...554..803Y}
}

@ARTICLE{Karim2011,
       author = {{Karim}, A. and {Schinnerer}, E. and {Mart{\'\i}nez-Sansigre}, A. and {Sargent}, M.~T. and {van der Wel}, A. and {Rix}, H. -W. and {Ilbert}, O. and {Smol{\v{c}}i{\'c}}, V. and {Carilli}, C. and {Pannella}, M. and {Koekemoer}, A.~M. and {Bell}, E.~F. and {Salvato}, M.},
        title = "{The Star Formation History of Mass-selected Galaxies in the COSMOS Field}",
      journal = {\apj},
         year = 2011,
        month = apr,
       volume = {730},
       number = {2},
          eid = {61},
        pages = {61},
          doi = {10.1088/0004-637X/730/2/61},
archivePrefix = {arXiv},
       eprint = {1011.6370},
 primaryClass = {astro-ph.CO},
       adsurl = {https://ui.adsabs.harvard.edu/abs/2011ApJ...730...61K}
}

@ARTICLE{Bell2003,
       author = {{Bell}, Eric F.},
        title = "{Estimating Star Formation Rates from Infrared and Radio Luminosities: The Origin of the Radio-Infrared Correlation}",
      journal = {\apj},
         year = 2003,
        month = apr,
       volume = {586},
       number = {2},
        pages = {794-813},
          doi = {10.1086/367829},
archivePrefix = {arXiv},
       eprint = {astro-ph/0212121},
 primaryClass = {astro-ph},
       adsurl = {https://ui.adsabs.harvard.edu/abs/2003ApJ...586..794B}
}

@ARTICLE{Delvecchio2021,
       author = {{Delvecchio}, I. and {Daddi}, E. and {Sargent}, M.~T. and {Jarvis}, M.~J. and {Elbaz}, D. and {Jin}, S. and {Liu}, D. and {Whittam}, I.~H. and {Algera}, H. and {Carraro}, R. and {D'Eugenio}, C. and {Delhaize}, J. and {Kalita}, B.~S. and {Leslie}, S. and {Moln{\'a}r}, D. Cs. and {Novak}, M. and {Prandoni}, I. and {Smol{\v{c}}i{\'c}}, V. and {Ao}, Y. and {Aravena}, M. and {Bournaud}, F. and {Collier}, J.~D. and {Randriamampandry}, S.~M. and {Randriamanakoto}, Z. and {Rodighiero}, G. and {Schober}, J. and {White}, S.~V. and {Zamorani}, G.},
        title = "{The infrared-radio correlation of star-forming galaxies is strongly M$_{{\ensuremath{\star}}}$-dependent but nearly redshift-invariant since z {\ensuremath{\sim}} 4}",
      journal = {\aap},
         year = 2021,
        month = mar,
       volume = {647},
          eid = {A123},
        pages = {A123},
          doi = {10.1051/0004-6361/202039647},
archivePrefix = {arXiv},
       eprint = {2010.05510},
 primaryClass = {astro-ph.GA},
       adsurl = {https://ui.adsabs.harvard.edu/abs/2021A&A...647A.123D}
}

@ARTICLE{Novak_2017,
       author = {{Novak}, M. and {Smol{\v{c}}i{\'c}}, V. and {Delhaize}, J. and {Delvecchio}, I. and {Zamorani}, G. and {Baran}, N. and {Bondi}, M. and {Capak}, P. and {Carilli}, C.~L. and {Ciliegi}, P. and {Civano}, F. and {Ilbert}, O. and {Karim}, A. and {Laigle}, C. and {Le F{\`e}vre}, O. and {Marchesi}, S. and {McCracken}, H. and {Miettinen}, O. and {Salvato}, M. and {Sargent}, M. and {Schinnerer}, E. and {Tasca}, L.},
        title = "{The VLA-COSMOS 3 GHz Large Project: Cosmic star formation history since z   5}",
      journal = {\aap},
         year = 2017,
        month = jun,
       volume = {602},
          eid = {A5},
        pages = {A5},
          doi = {10.1051/0004-6361/201629436},
archivePrefix = {arXiv},
       eprint = {1703.09724},
 primaryClass = {astro-ph.GA},
       adsurl = {https://ui.adsabs.harvard.edu/abs/2017A&A...602A...5N}
}

@ARTICLE{Schmidt_1968,
       author = {{Schmidt}, Maarten},
        title = "{Space Distribution and Luminosity Functions of Quasi-Stellar Radio Sources}",
      journal = {\apj},
         year = 1968,
        month = feb,
       volume = {151},
        pages = {393},
          doi = {10.1086/149446},
       adsurl = {https://ui.adsabs.harvard.edu/abs/1968ApJ...151..393S}
}

@article{Novak_2018,
    author = {{Novak}, M. and {Smol{\v{c}}i{\'c}}, V. and {Schinnerer}, E. and {Zamorani}, G. and {Delvecchio}, I. and {Bondi}, M. and {Delhaize}, J.},
    title = "{Constraints on submicrojansky radio number counts based on evolving VLA-COSMOS luminosity functions}",
    journal = {\aap},
    year = 2018,
    month = jun,
    volume = {614}, 
    eid = {A47},
    pages = {A47},
    doi = {10.1051/0004-6361/201731635},
    archivePrefix = {arXiv},
    eprint = {1803.01569},
    primaryClass = {astro-ph.GA},
    adsurl = {https://ui.adsabs.harvard.edu/abs/2018A&A...614A..47N}
}

@software{Mohan2015PyBDSFPB,
       author = {{Mohan}, Niruj and {Rafferty}, David},
        title = "{PyBDSF: Python Blob Detection and Source Finder}",
 howpublished = {Astrophysics Source Code Library, record ascl:1502.007},
         year = 2015,
        month = feb,
          eid = {ascl:1502.007},
archivePrefix = {ascl},
       eprint = {1502.007},
       adsurl = {https://ui.adsabs.harvard.edu/abs/2015ascl.soft02007M}
}

@ARTICLE{Mauch_2007,
       author = {{Mauch}, Tom and {Sadler}, Elaine M.},
        title = "{Radio sources in the 6dFGS: local luminosity functions at 1.4 GHz for star-forming galaxies and radio-loud AGN}",
      journal = {\mnras},
         year = 2007,
        month = mar,
       volume = {375},
       number = {3},
        pages = {931-950},
          doi = {10.1111/j.1365-2966.2006.11353.x},
archivePrefix = {arXiv},
       eprint = {astro-ph/0612018},
 primaryClass = {astro-ph},
       adsurl = {https://ui.adsabs.harvard.edu/abs/2007MNRAS.375..931M}
}

@article{2008MNRAS.384..449F,
       author = {{Feroz}, F. and {Hobson}, M.~P.},
        title = "{Multimodal nested sampling: an efficient and robust alternative to Markov Chain Monte Carlo methods for astronomical data analyses}",
      journal = {\mnras},
         year = 2008,
        month = feb,
       volume = {384},
       number = {2},
        pages = {449-463},
          doi = {10.1111/j.1365-2966.2007.12353.x},
archivePrefix = {arXiv},
       eprint = {0704.3704},
 primaryClass = {astro-ph},
       adsurl = {https://ui.adsabs.harvard.edu/abs/2008MNRAS.384..449F}
}

@INPROCEEDINGS{Vaccari2022,
       author = {{Vaccari}, Mattia},
        title = "{The Spitzer Spectroscopic Data Fusion - Merged Spectroscopic Redshift Catalogs in Spitzer Fields}",
    booktitle = {Zenodo dataset},
         year = 2022,
       volume = {63},
        month = mar,
    publisher = {Zenodo},
          eid = {6368348},
        pages = {6368348},
          doi = {10.5281/zenodo.6368347},
       adsurl = {https://ui.adsabs.harvard.edu/abs/2022zndo...6368348V}
}

@ARTICLE{Almosallam_2016b,
       author = {{Almosallam}, Ibrahim A. and {Jarvis}, Matt J. and {Roberts}, Stephen J.},
        title = "{GPZ: non-stationary sparse Gaussian processes for heteroscedastic uncertainty estimation in photometric redshifts}",
      journal = {\mnras},
         year = 2016,
        month = oct,
       volume = {462},
       number = {1},
        pages = {726-739},
          doi = {10.1093/mnras/stw1618},
archivePrefix = {arXiv},
       eprint = {1604.03593},
 primaryClass = {astro-ph.IM},
       adsurl = {https://ui.adsabs.harvard.edu/abs/2016MNRAS.462..726A}
}

@ARTICLE{Kennicutt_1998,
       author = {{Kennicutt}, Jr., Robert C.},
        title = "{Star Formation in Galaxies Along the Hubble Sequence}",
      journal = {\araa},
         year = 1998,
        month = jan,
       volume = {36},
        pages = {189-232},
          doi = {10.1146/annurev.astro.36.1.189},
archivePrefix = {arXiv},
       eprint = {astro-ph/9807187},
 primaryClass = {astro-ph},
       adsurl = {https://ui.adsabs.harvard.edu/abs/1998ARA&A..36..189K}
}

@ARTICLE{1992ARA&A..30..575C,
       author = {{Condon}, J.~J.},
        title = "{Radio emission from normal galaxies.}",
      journal = {\araa},
         year = 1992,
        month = jan,
       volume = {30},
        pages = {575-611},
          doi = {10.1146/annurev.aa.30.090192.003043},
       url = {, http://dx.doi.org/10.1146/annurev.aa.30.090192.00304}
}

@ARTICLE{Madau_2014,
       author = {{Madau}, Piero and {Dickinson}, Mark},
        title = "{Cosmic Star-Formation History}",
      journal = {\araa},
         year = 2014,
        month = aug,
       volume = {52},
        pages = {415-486},
          doi = {10.1146/annurev-astro-081811-125615},
archivePrefix = {arXiv},
       eprint = {1403.0007},
 primaryClass = {astro-ph.CO},
       adsurl = {https://ui.adsabs.harvard.edu/abs/2014ARA&A..52..415M}
}

@ARTICLE{Almosallam2016a,
       author = {{Almosallam}, Ibrahim A. and {Lindsay}, Sam N. and {Jarvis}, Matt J. and {Roberts}, Stephen J.},
        title = "{A sparse Gaussian process framework for photometric redshift estimation}",
      journal = {\mnras},
         year = 2016,
        month = jan,
       volume = {455},
       number = {3},
        pages = {2387-2401},
          doi = {10.1093/mnras/stv2425},
archivePrefix = {arXiv},
       eprint = {1505.05489},
 primaryClass = {astro-ph.IM},
       adsurl = {https://ui.adsabs.harvard.edu/abs/2016MNRAS.455.2387A}
}

@ARTICLE{Duncan2021,
       author = {{Duncan}, K.~J. and {Kondapally}, R. and {Brown}, M.~J.~I. and {Bonato}, M. and {Best}, P.~N. and {R{\"o}ttgering}, H.~J.~A. and {Bondi}, M. and {Bowler}, R.~A.~A. and {Cochrane}, R.~K. and {G{\"u}rkan}, G. and {Hardcastle}, M.~J. and {Jarvis}, M.~J. and {Kunert-Bajraszewska}, M. and {Leslie}, S.~K. and {Ma{\l}ek}, K. and {Morabito}, L.~K. and {O'Sullivan}, S.~P. and {Prandoni}, I. and {Sabater}, J. and {Shimwell}, T.~W. and {Smith}, D.~J.~B. and {Wang}, L. and {Wo{\l}owska}, A. and {Tasse}, C.},
        title = "{The LOFAR Two-meter Sky Survey: Deep Fields Data Release 1. IV. Photometric redshifts and stellar masses}",
      journal = {\aap},
         year = 2021,
        month = apr,
       volume = {648},
          eid = {A4},
        pages = {A4},
          doi = {10.1051/0004-6361/202038809},
archivePrefix = {arXiv},
       eprint = {2011.08204},
 primaryClass = {astro-ph.GA},
       adsurl = {https://ui.adsabs.harvard.edu/abs/2021A&A...648A...4D}
}

@ARTICLE{Duncan2018,
       author = {{Duncan}, Kenneth J. and {Jarvis}, Matt J. and {Brown}, Michael J.~I. and {R{\"o}ttgering}, Huub J.~A.},
        title = "{Photometric redshifts for the next generation of deep radio continuum surveys - II. Gaussian processes and hybrid estimates}",
      journal = {\mnras},
         year = 2018,
        month = jul,
       volume = {477},
       number = {4},
        pages = {5177-5190},
          doi = {10.1093/mnras/sty940},
archivePrefix = {arXiv},
       eprint = {1712.04476},
 primaryClass = {astro-ph.GA},
       adsurl = {https://ui.adsabs.harvard.edu/abs/2018MNRAS.477.5177D}
}

@ARTICLE{Cochrane2023,
       author = {{Cochrane}, R.~K. and {Kondapally}, R. and {Best}, P.~N. and {Sabater}, J. and {Duncan}, K.~J. and {Smith}, D.~J.~B. and {Hardcastle}, M.~J. and {R{\"o}ttgering}, H.~J.~A. and {Prandoni}, I. and {Haskell}, P. and {G{\"u}rkan}, G. and {Miley}, G.~K.},
        title = "{The LOFAR Two-metre Sky Survey: the radio view of the cosmic star formation history}",
      journal = {\mnras},
         year = 2023,
        month = aug,
       volume = {523},
       number = {4},
        pages = {6082-6102},
          doi = {10.1093/mnras/stad1602},
archivePrefix = {arXiv},
       eprint = {2305.15510},
 primaryClass = {astro-ph.GA},
       adsurl = {https://ui.adsabs.harvard.edu/abs/2023MNRAS.523.6082C}
}

@ARTICLE{McAlpine2013,
       author = {{McAlpine}, K. and {Jarvis}, M.~J. and {Bonfield}, D.~G.},
        title = "{Evolution of faint radio sources in the VIDEO-XMM3 field}",
      journal = {\mnras},
         year = 2013,
        month = dec,
       volume = {436},
       number = {2},
        pages = {1084-1095},
          doi = {10.1093/mnras/stt1638},
archivePrefix = {arXiv},
       eprint = {1309.0358},
 primaryClass = {astro-ph.CO},
       adsurl = {https://ui.adsabs.harvard.edu/abs/2013MNRAS.436.1084M}
}

@ARTICLE{1985ApJ...299..109M,
       author = {{Marshall}, H.~L.},
        title = "{The evolution of optically selected quasars with Z < 2.2 and B <20.}",
      journal = {\apj},
         year = 1985,
        month = dec,
       volume = {299},
        pages = {109-121},
          doi = {10.1086/163685},
       adsurl = {https://ui.adsabs.harvard.edu/abs/1985ApJ...299..109M}
}

@ARTICLE{Saunders1990The6A,
       author = {{Saunders}, W. and {Rowan-Robinson}, M. and {Lawrence}, A. and {Efstathiou}, G. and {Kaiser}, N. and {Ellis}, R.~S. and {Frenk}, C.~S.},
        title = "{The 60-mu.m and far-infrared luminosity functions of IRAS galaxies.}",
      journal = {\mnras},
         year = 1990,
        month = jan,
       volume = {242},
        pages = {318-337},
          doi = {10.1093/mnras/242.3.318},
       adsurl = {https://ui.adsabs.harvard.edu/abs/1990MNRAS.242..318S}
}

@ARTICLE{Malefahlo_2021,
       author = {{Malefahlo}, Eliab D. and {Jarvis}, Matt J. and {Santos}, Mario G. and {White}, Sarah V. and {Adams}, Nathan J. and {Bowler}, Rebecca A.~A.},
        title = "{A deep radio view of the evolution of the cosmic star formation rate density from a stellar-mass-selected sample in VLA-COSMOS}",
      journal = {\mnras},
         year = 2022,
        month = jan,
       volume = {509},
       number = {3},
        pages = {4291-4307},
          doi = {10.1093/mnras/stab3242},
archivePrefix = {arXiv},
       eprint = {2012.09797},
 primaryClass = {astro-ph.GA},
       adsurl = {https://ui.adsabs.harvard.edu/abs/2022MNRAS.509.4291M}
}

@ARTICLE{Best_2012,
       author = {{Best}, P.~N. and {Heckman}, T.~M.},
        title = "{On the fundamental dichotomy in the local radio-AGN population: accretion, evolution and host galaxy properties}",
      journal = {\mnras},
         year = 2012,
        month = apr,
       volume = {421},
       number = {2},
        pages = {1569-1582},
          doi = {10.1111/j.1365-2966.2012.20414.x},
archivePrefix = {arXiv},
       eprint = {1201.2397},
 primaryClass = {astro-ph.CO},
       adsurl = {https://ui.adsabs.harvard.edu/abs/2012MNRAS.421.1569B}
}

@ARTICLE{2014ARA&A..52..589H,
       author = {{Heckman}, Timothy M. and {Best}, Philip N.},
        title = "{The Coevolution of Galaxies and Supermassive Black Holes: Insights from Surveys of the Contemporary Universe}",
      journal = {\araa},
         year = 2014,
        month = aug,
       volume = {52},
        pages = {589-660},
          doi = {10.1146/annurev-astro-081913-035722},
archivePrefix = {arXiv},
       eprint = {1403.4620},
 primaryClass = {astro-ph.GA},
       adsurl = {https://ui.adsabs.harvard.edu/abs/2014ARA&A..52..589H}
}

@ARTICLE{1983ApJ...266..713O,
       author = {{Oke}, J.~B. and {Gunn}, J.~E.},
        title = "{Secondary standard stars for absolute spectrophotometry.}",
      journal = {\apj},
         year = 1983,
        month = mar,
       volume = {266},
        pages = {713-717},
          doi = {10.1086/160817},
       adsurl = {https://ui.adsabs.harvard.edu/abs/1983ApJ...266..713O}
}

@ARTICLE{Clewley_2004,
       author = {{Clewley}, L. and {Jarvis}, Matt J.},
        title = "{The cosmic evolution of low-luminosity radio sources from the Sloan Digital Sky Survey Data Release 1}",
      journal = {\mnras},
         year = 2004,
        month = aug,
       volume = {352},
       number = {3},
        pages = {909-914},
          doi = {10.1111/j.1365-2966.2004.07981.x},
archivePrefix = {arXiv},
       eprint = {astro-ph/0405080},
 primaryClass = {astro-ph},
       adsurl = {https://ui.adsabs.harvard.edu/abs/2004MNRAS.352..909C}
}

@ARTICLE{Fabian_2012,
       author = {{Fabian}, A.~C.},
        title = "{Observational Evidence of Active Galactic Nuclei Feedback}",
      journal = {\araa},
         year = 2012,
        month = sep,
       volume = {50},
        pages = {455-489},
          doi = {10.1146/annurev-astro-081811-125521},
archivePrefix = {arXiv},
       eprint = {1204.4114},
 primaryClass = {astro-ph.CO},
       adsurl = {https://ui.adsabs.harvard.edu/abs/2012ARA&A..50..455F}
}

@ARTICLE{2009Natur.460..213C,
       author = {{Cattaneo}, A. and {Faber}, S.~M. and {Binney}, J. and {Dekel}, A. and {Kormendy}, J. and {Mushotzky}, R. and {Babul}, A. and {Best}, P.~N. and {Br{\"u}ggen}, M. and {Fabian}, A.~C. and {Frenk}, C.~S. and {Khalatyan}, A. and {Netzer}, H. and {Mahdavi}, A. and {Silk}, J. and {Steinmetz}, M. and {Wisotzki}, L.},
        title = "{The role of black holes in galaxy formation and evolution}",
      journal = {\nat},
         year = 2009,
        month = jul,
       volume = {460},
       number = {7252},
        pages = {213-219},
          doi = {10.1038/nature08135},
archivePrefix = {arXiv},
       eprint = {0907.1608},
 primaryClass = {astro-ph.CO},
       adsurl = {https://ui.adsabs.harvard.edu/abs/2009Natur.460..213C}
}

@ARTICLE{2f494631963a431cb58dfea47f6ad8e5,
       author = {{McAlpine}, Kim and {Jarvis}, Matt J.},
        title = "{The evolution of radio sources in the UKIDSS-DXS-XMM-LSS field}",
      journal = {\mnras},
         year = 2011,
        month = may,
       volume = {413},
       number = {2},
        pages = {1054-1060},
          doi = {10.1111/j.1365-2966.2010.18191.x},
archivePrefix = {arXiv},
       eprint = {1012.3020},
 primaryClass = {astro-ph.CO},
       adsurl = {https://ui.adsabs.harvard.edu/abs/2011MNRAS.413.1054M}
}

@ARTICLE{2009ApJ...696...24S,
       author = {{Smol{\v{c}}i{\'c}}, V. and {Zamorani}, G. and {Schinnerer}, E. and {Bardelli}, S. and {Bondi}, M. and {B{\^\i}rzan}, L. and {Carilli}, C.~L. and {Ciliegi}, P. and {Elvis}, M. and {Impey}, C.~D. and {Koekemoer}, A.~M. and {Merloni}, A. and {Paglione}, T. and {Salvato}, M. and {Scodeggio}, M. and {Scoville}, N. and {Trump}, J.~R.},
        title = "{Cosmic Evolution of Radio Selected Active Galactic Nuclei in the Cosmos Field}",
      journal = {\apj},
         year = 2009,
        month = may,
       volume = {696},
       number = {1},
        pages = {24-39},
          doi = {10.1088/0004-637X/696/1/24},
archivePrefix = {arXiv},
       eprint = {0901.3372},
 primaryClass = {astro-ph.CO},
       adsurl = {https://ui.adsabs.harvard.edu/abs/2009ApJ...696...24S}
}

@ARTICLE{Hale_2022,
       author = {{Hale}, C.~L. and {Whittam}, I.~H. and {Jarvis}, M.~J. and {Best}, P.~N. and {Thomas}, N.~L. and {Heywood}, I. and {Prescott}, M. and {Adams}, N. and {Afonso}, J. and {An}, Fangxia and {Bowler}, R.~A.~A. and {Collier}, J.~D. and {Cook}, R.~H.~W. and {Dav{\'e}}, R. and {Frank}, B.~S. and {Glowacki}, M. and {Hatfield}, P.~W. and {Kolwa}, S. and {Lovell}, C.~C. and {Maddox}, N. and {Marchetti}, L. and {Morabito}, L.~K. and {Murphy}, E. and {Prandoni}, I. and {Randriamanakoto}, Z. and {Taylor}, A.~R.},
        title = "{MIGHTEE: deep 1.4 GHz source counts and the sky temperature contribution of star-forming galaxies and active galactic nuclei}",
      journal = {\mnras},
         year = 2023,
        month = apr,
       volume = {520},
       number = {2},
        pages = {2668-2691},
          doi = {10.1093/mnras/stac3320},
archivePrefix = {arXiv},
       eprint = {2211.05741},
 primaryClass = {astro-ph.GA},
       adsurl = {https://ui.adsabs.harvard.edu/abs/2023MNRAS.520.2668H}
}

@ARTICLE{Buchner_2014,
       author = {{Buchner}, J. and {Georgakakis}, A. and {Nandra}, K. and {Hsu}, L. and {Rangel}, C. and {Brightman}, M. and {Merloni}, A. and {Salvato}, M. and {Donley}, J. and {Kocevski}, D.},
        title = "{X-ray spectral modelling of the AGN obscuring region in the CDFS: Bayesian model selection and catalogue}",
      journal = {\aap},
         year = 2014,
        month = apr,
       volume = {564},
          eid = {A125},
        pages = {A125},
          doi = {10.1051/0004-6361/201322971},
archivePrefix = {arXiv},
       eprint = {1402.0004},
 primaryClass = {astro-ph.HE},
       adsurl = {https://ui.adsabs.harvard.edu/abs/2014A&A...564A.125B}
}

@ARTICLE{Zhu_2023,
       author = {{Zhu}, Shifu and {Brandt}, W.~N. and {Zou}, Fan and {Luo}, Bin and {Ni}, Qingling and {Xue}, Yongquan and {Yan}, Wei},
        title = "{Radio AGN selection and characterization in three Deep-Drilling Fields of the Vera C. Rubin Observatory Legacy Survey of Space and Time}",
      journal = {\mnras},
         year = 2023,
        month = jul,
       volume = {522},
       number = {3},
        pages = {3506-3528},
          doi = {10.1093/mnras/stad1178},
archivePrefix = {arXiv},
       eprint = {2304.07864},
 primaryClass = {astro-ph.GA},
       adsurl = {https://ui.adsabs.harvard.edu/abs/2023MNRAS.522.3506Z}
}

@ARTICLE{matthews2024confirmationsubstantialdiscrepancyradio,
       author = {{Matthews}, A.~M. and {Kelson}, D.~D. and {Newman}, A.~B. and {Camilo}, F. and {Condon}, J.~J. and {Cotton}, W.~D. and {Dickinson}, M. and {Jarrett}, T.~H. and {Lacy}, M.},
        title = "{Confirmation of a Substantial Discrepancy between Radio and UV{\textendash}IR Measures of the Star Formation Rate Density at 0.2 < z < 1.3}",
      journal = {\apj},
         year = 2024,
        month = may,
       volume = {966},
       number = {2},
          eid = {194},
        pages = {194},
          doi = {10.3847/1538-4357/ad3912},
archivePrefix = {arXiv},
       eprint = {2310.17701},
 primaryClass = {astro-ph.GA},
       adsurl = {https://ui.adsabs.harvard.edu/abs/2024ApJ...966..194M}
}

@ARTICLE{whittam2023mighteemultiwavelengthcounterpartscosmos,
       author = {{Whittam}, I.~H. and {Prescott}, M. and {Hale}, C.~L. and {Jarvis}, M.~J. and {Heywood}, I. and {An}, Fangxia and {Glowacki}, M. and {Maddox}, N. and {Marchetti}, L. and {Morabito}, L.~K. and {Adams}, N.~J. and {Bowler}, R.~A.~A. and {Hatfield}, P.~W. and {Varadaraj}, R.~G. and {Collier}, J. and {Frank}, B. and {Taylor}, A.~R. and {Santos}, M.~G. and {Vaccari}, M. and {Afonso}, J. and {Ao}, Y. and {Delhaize}, J. and {Knowles}, K. and {Kolwa}, S. and {Randriamampandry}, S.~M. and {Randriamanakoto}, Z. and {Smirnov}, O. and {Smith}, D.~J.~B. and {White}, S.~V.},
        title = "{MIGHTEE: Multi-wavelength counterparts in the COSMOS field}",
      journal = {\mnras},
         year = 2024,
        month = jan,
       volume = {527},
       number = {2},
        pages = {3231-3245},
          doi = {10.1093/mnras/stad3307},
archivePrefix = {arXiv},
       eprint = {2310.17409},
 primaryClass = {astro-ph.GA},
       adsurl = {https://ui.adsabs.harvard.edu/abs/2024MNRAS.527.3231W}
}

@ARTICLE{Hatfield_2022,
       author = {{Hatfield}, P.~W. and {Jarvis}, M.~J. and {Adams}, N. and {Bowler}, R.~A.~A. and {H{\"a}u{\ss}ler}, B. and {Duncan}, K.~J.},
        title = "{Hybrid photometric redshifts for sources in the COSMOS and XMM-LSS fields}",
      journal = {\mnras},
         year = 2022,
        month = jul,
       volume = {513},
       number = {3},
        pages = {3719-3733},
          doi = {10.1093/mnras/stac1042},
archivePrefix = {arXiv},
       eprint = {2206.00748},
 primaryClass = {astro-ph.GA},
       adsurl = {https://ui.adsabs.harvard.edu/abs/2022MNRAS.513.3719H}
}

@ARTICLE{Matthews_2021,
       author = {{Matthews}, A.~M. and {Condon}, J.~J. and {Cotton}, W.~D. and {Mauch}, T.},
        title = "{Cosmic Star Formation History Measured at 1.4 GHz}",
      journal = {\apj},
         year = 2021,
        month = jun,
       volume = {914},
       number = {2},
          eid = {126},
        pages = {126},
          doi = {10.3847/1538-4357/abfaf6},
archivePrefix = {arXiv},
       eprint = {2104.11756},
 primaryClass = {astro-ph.GA},
       adsurl = {https://ui.adsabs.harvard.edu/abs/2021ApJ...914..126M}
}

@ARTICLE{Heywood_2021,
       author = {{Heywood}, I. and {Jarvis}, M.~J. and {Hale}, C.~L. and {Whittam}, I.~H. and {Bester}, H.~L. and {Hugo}, B. and {Kenyon}, J.~S. and {Prescott}, M. and {Smirnov}, O.~M. and {Tasse}, C. and {Afonso}, J.~M. and {Best}, P.~N. and {Collier}, J.~D. and {Deane}, R.~P. and {Frank}, B.~S. and {Hardcastle}, M.~J. and {Knowles}, K. and {Maddox}, N. and {Murphy}, E.~J. and {Prandoni}, I. and {Randriamampandry}, S.~M. and {Santos}, M.~G. and {Sekhar}, S. and {Tabatabaei}, F. and {Taylor}, A.~R. and {Thorat}, K.},
        title = "{MIGHTEE: total intensity radio continuum imaging and the COSMOS/XMM-LSS Early Science fields}",
      journal = {\mnras},
         year = 2022,
        month = jan,
       volume = {509},
       number = {2},
        pages = {2150-2168},
          doi = {10.1093/mnras/stab3021},
archivePrefix = {arXiv},
       eprint = {2110.00347},
 primaryClass = {astro-ph.GA},
       adsurl = {https://ui.adsabs.harvard.edu/abs/2022MNRAS.509.2150H}
}

@ARTICLE{Murphy_2011,
       author = {{Murphy}, E.~J. and {Condon}, J.~J. and {Schinnerer}, E. and {Kennicutt}, R.~C. and {Calzetti}, D. and {Armus}, L. and {Helou}, G. and {Turner}, J.~L. and {Aniano}, G. and {Beir{\~a}o}, P. and {Bolatto}, A.~D. and {Brandl}, B.~R. and {Croxall}, K.~V. and {Dale}, D.~A. and {Donovan Meyer}, J.~L. and {Draine}, B.~T. and {Engelbracht}, C. and {Hunt}, L.~K. and {Hao}, C.-N. and {Koda}, J. and {Roussel}, H. and {Skibba}, R. and {Smith}, J.-D.~T.},
        title = "{Calibrating Extinction-free Star Formation Rate Diagnostics with 33 GHz Free-free Emission in NGC 6946}",
      journal = {\apj},
         year = 2011,
        month = aug,
       volume = {737},
       number = {2},
          eid = {67},
        pages = {67},
          doi = {10.1088/0004-637X/737/2/67},
archivePrefix = {arXiv},
       eprint = {1105.4877},
 primaryClass = {astro-ph.CO},
       adsurl = {https://ui.adsabs.harvard.edu/abs/2011ApJ...737...67M}
}

@ARTICLE{Kennicutt_2012,
       author = {{Kennicutt}, Robert C. and {Evans}, Neal J.},
        title = "{Star Formation in the Milky Way and Nearby Galaxies}",
      journal = {\araa},
         year = 2012,
        month = sep,
       volume = {50},
        pages = {531-608},
          doi = {10.1146/annurev-astro-081811-125610},
archivePrefix = {arXiv},
       eprint = {1204.3552},
 primaryClass = {astro-ph.GA},
       adsurl = {https://ui.adsabs.harvard.edu/abs/2012ARA&A..50..531K}
}

@ARTICLE{Hardcastle_2020,
       author = {{Hardcastle}, M.~J. and {Croston}, J.~H.},
        title = "{Radio galaxies and feedback from AGN jets}",
      journal = {\nar},
         year = 2020,
        month = jun,
       volume = {88},
          eid = {101539},
        pages = {101539},
          doi = {10.1016/j.newar.2020.101539},
archivePrefix = {arXiv},
       eprint = {2003.06137},
 primaryClass = {astro-ph.HE},
       adsurl = {https://ui.adsabs.harvard.edu/abs/2020NewAR..8801539H}
}

@ARTICLE{Tabatabaei_2025,
       author = {{Tabatabaei}, Fatemeh and {Khademi}, Maryam and {Jarvis}, Matt J. and {Taylor}, Russ and {Whittam}, Imogen H. and {An}, Fangxia and {Javadi}, Reihaneh and {Murphy}, Eric J. and {Vaccari}, Mattia},
        title = "{The Radio Spectral Energy Distribution and Star Formation Calibration in MIGHTEE-COSMOS Highly Star-forming Galaxies at 1.5 < z < 3.5}",
      journal = {\apj},
         year = 2025,
        month = aug,
       volume = {989},
       number = {1},
          eid = {44},
        pages = {44},
          doi = {10.3847/1538-4357/ade233},
archivePrefix = {arXiv},
       eprint = {2506.16275},
 primaryClass = {astro-ph.GA},
       adsurl = {https://ui.adsabs.harvard.edu/abs/2025ApJ...989...44T}
}

@ARTICLE{Wilman_2008,
       author = {{Wilman}, R.~J. and {Miller}, L. and {Jarvis}, M.~J. and {Mauch}, T. and {Levrier}, F. and {Abdalla}, F.~B. and {Rawlings}, S. and {Kl{\"o}ckner}, H.-R. and {Obreschkow}, D. and {Olteanu}, D. and {Young}, S.},
        title = "{A semi-empirical simulation of the extragalactic radio continuum sky for next generation radio telescopes}",
      journal = {\mnras},
         year = 2008,
        month = aug,
       volume = {388},
       number = {3},
        pages = {1335-1348},
          doi = {10.1111/j.1365-2966.2008.13486.x},
archivePrefix = {arXiv},
       eprint = {0805.3413},
 primaryClass = {astro-ph},
       adsurl = {https://ui.adsabs.harvard.edu/abs/2008MNRAS.388.1335W}
}

@ARTICLE{Dave_2019,
       author = {{Dav{\'e}}, Romeel and {Angl{\'e}s-Alc{\'a}zar}, Daniel and {Narayanan}, Desika and {Li}, Qi and {Rafieferantsoa}, Mika H. and {Appleby}, Sarah},
        title = "{SIMBA: Cosmological simulations with black hole growth and feedback}",
      journal = {\mnras},
         year = 2019,
        month = jun,
       volume = {486},
       number = {2},
        pages = {2827-2849},
          doi = {10.1093/mnras/stz937},
archivePrefix = {arXiv},
       eprint = {1901.10203},
 primaryClass = {astro-ph.GA},
       adsurl = {https://ui.adsabs.harvard.edu/abs/2019MNRAS.486.2827D}
}

@ARTICLE{Condon_1992,
       author = {{Condon}, J.~J.},
        title = "{Radio emission from normal galaxies.}",
      journal = {\araa},
         year = 1992,
        month = jan,
       volume = {30},
        pages = {575-611},
          doi = {10.1146/annurev.aa.30.090192.003043},
       adsurl = {https://ui.adsabs.harvard.edu/abs/1992ARA&A..30..575C}
}


\appendix
\section{Source Counts and Luminosity Function}
\label{appendix:source-counts}

A summary of the redshift-bin source counts (spectroscopic, photometric, and non-XID) used in the $V_{\mathrm{max}}$ calculation for the luminosity function analysis.

\clearpage

\setlength\LTcapwidth{\textwidth} 

\captionsetup{width=\LTcapwidth,
              justification=raggedright,
              singlelinecheck=false}

\begingroup
\scriptsize
\small
\setlength{\tabcolsep}{3.5pt}
\onecolumn
\begin{longtable}{@{}l c c r r r r r r@{}}
\caption{Luminosity functions of the total radio-selected sample obtained with the $V_{\mathrm{max}}$ method for the $z>1$ analysis. Source counts per redshift bin for the COSMOS and XMM-LSS fields, including spectroscopic redshifts ($N_{\mathrm{spec}}$), photometric redshifts ($N_{\mathrm{phot}}$, counted only when $N_{\mathrm{spec}}$ is not available), and sources without optical counterparts ($N_{\mathrm{noXID}}$), assigned statistical redshifts at $z>1$.}
\label{tab:combined_vmax_counts} \\

\toprule
Redshift & $\log(L_{1.4\mathrm{GHz}} / \mathrm{W\,Hz}^{-1})$ & $\log(\phi / \mathrm{Mpc}^{-3}\,\mathrm{dex}^{-1})$ & COSMOS N$_{\mathrm{spec}}$ & COSMOS N$_{\mathrm{phot}}$ & COSMOS N$_{\mathrm{noXID}}$ & XMM N$_{\mathrm{spec}}$ & XMM N$_{\mathrm{phot}}$ & XMM N$_{\mathrm{noXID}}$ \\
\midrule
\endfirsthead

\multicolumn{9}{@{}l}{{\tablename\ \thetable{} -- continued from previous page}} \\
\toprule
Redshift & $\log(L_{1.4\mathrm{GHz}} / \mathrm{W\,Hz}^{-1})$ & $\log(\phi / \mathrm{Mpc}^{-3}\,\mathrm{dex}^{-1})$ & COSMOS N$_{\mathrm{spec}}$ & COSMOS N$_{\mathrm{phot}}$ & COSMOS N$_{\mathrm{noXID}}$ & XMM N$_{\mathrm{spec}}$ & XMM N$_{\mathrm{phot}}$ & XMM N$_{\mathrm{noXID}}$ \\
\midrule
\endhead

\midrule
\multicolumn{9}{r}{{Continued on next page}} \\
\endfoot

\bottomrule
\endlastfoot
0.2 \textless z \textless 0.4 & 21.88 & $-2.15 \pm 0.02$ & 71 & 1 & 0 & 116 & 116 & 0 \\
  & 22.12 & $-2.34 \pm 0.02$ & 159 & 1 & 0 & 195 & 212 & 0 \\
  & 22.38 & $-2.73 \pm 0.02$ & 152 & 1 & 0 & 179 & 168 & 0 \\
  & 22.62 & $-3.14 \pm 0.03$ & 86 & 1 & 0 & 86 & 89 & 0 \\
  & 22.88 & $-3.45 \pm 0.04$ & 32 & 1 & 0 & 37 & 45 & 0 \\
  & 23.12 & $-3.91 \pm 0.07$ & 12 & 1 & 0 & 13 & 17 & 0 \\
  & 23.38 & $-4.10 \pm 0.08$ & 8 & 1 & 0 & 13 & 6 & 0 \\
  & 23.62 & $-4.73 \pm 0.18$ & 1 & 0 & 0 & 2 & 3 & 0 \\
  & 23.88 & $-4.55 \pm 0.14$ & 1 & 0 & 0 & 7 & 1 & 0 \\
  & 24.12 & $-4.90 \pm 0.22$ & 1 & 0 & 0 & 2 & 1 & 0 \\
  & 24.38 & $-4.81 \pm 0.18$ & 2 & 0 & 0 & 3 & 1 & 0 \\
  & 24.62 & $-4.46 \pm 0.13$ & 0 & 0 & 0 & 10 & 1 & 0 \\
  & 24.88 & $-5.52 \pm 0.43$ & 0 & 0 & 0 & 0 & 1 & 0 \\
  & 25.12 & $-5.52 \pm 0.43$ & 0 & 0 & 0 & 0 & 1 & 0 \\
  & 25.38 & $-5.52 \pm 0.43$ & 0 & 0 & 0 & 0 & 1 & 0 \\
0.4 \textless z \textless 0.6 & 22.62 & $-2.60 \pm 0.05$ & 135 & 5 & 0 & 324 & 394 & 0 \\
  & 22.88 & $-3.15 \pm 0.02$ & 95 & 3 & 0 & 195 & 223 & 0 \\
  & 23.12 & $-3.62 \pm 0.03$ & 34 & 3 & 0 & 61 & 78 & 0 \\
  & 23.38 & $-3.93 \pm 0.05$ & 10 & 1 & 0 & 36 & 39 & 0 \\
  & 23.62 & $-4.43 \pm 0.08$ & 2 & 1 & 0 & 12 & 13 & 0 \\
  & 23.88 & $-4.70 \pm 0.11$ & 3 & 1 & 0 & 8 & 3 & 0 \\
  & 24.12 & $-5.00 \pm 0.16$ & 0 & 1 & 0 & 5 & 2 & 0 \\
  & 24.38 & $-5.06 \pm 0.18$ & 1 & 1 & 0 & 3 & 2 & 0 \\
  & 24.62 & $-5.54 \pm 0.31$ & 0 & 0 & 0 & 2 & 1 & 0 \\
  & 24.88 & $-5.25 \pm 0.22$ & 1 & 0 & 0 & 3 & 1 & 0 \\
  & 25.12 & $-5.85 \pm 0.43$ & 0 & 0 & 0 & 0 & 1 & 0 \\
  & 25.38 & $-5.86 \pm 0.43$ & 0 & 0 & 0 & 0 & 1 & 0 \\
  & 25.62 & $-5.86 \pm 0.43$ & 0 & 0 & 0 & 0 & 1 & 0 \\
  & 25.88 & $-5.86 \pm 0.43$ & 0 & 0 & 0 & 0 & 1 & 0 \\
0.6 \textless z \textless 0.8 & 23.12 & $-3.21 \pm 0.02$ & 115 & 9 & 0 & 241 & 282 & 0 \\
  & 23.38 & $-3.60 \pm 0.02$ & 60 & 7 & 0 & 109 & 117 & 0 \\
  & 23.62 & $-4.00 \pm 0.04$ & 33 & 2 & 0 & 39 & 42 & 0 \\
  & 23.88 & $-4.35 \pm 0.05$ & 9 & 1 & 0 & 21 & 21 & 0 \\
  & 24.12 & $-4.65 \pm 0.07$ & 9 & 1 & 0 & 7 & 11 & 0 \\
  & 24.38 & $-4.81 \pm 0.10$ & 5 & 1 & 0 & 7 & 5 & 0 \\
  & 24.62 & $-4.69 \pm 0.09$ & 1 & 1 & 0 & 15 & 6 & 0 \\
  & 24.88 & $-5.19 \pm 0.18$ & 0 & 1 & 0 & 5 & 2 & 0 \\
  & 25.12 & $-5.17 \pm 0.12$ & 0 & 0 & 0 & 7 & 1 & 0 \\
  & 25.38 & $-5.57 \pm 0.19$ & 0 & 0 & 0 & 2 & 1 & 0 \\
  & 25.62 & $-6.05 \pm 0.31$ & 0 & 0 & 0 & 1 & 1 & 0 \\
  & 25.88 & $-6.05 \pm 0.69$ & 0 & 0 & 0 & 0 & 1 & 0 \\
  & 26.20 & $-5.35 \pm 0.40$ & 0 & 0 & 0 & 8 & 1 & 0 \\
0.8 \textless z \textless 1.0 & 23.38 & $-3.26 \pm 0.02$ & 143 & 18 & 0 & 198 & 442 & 0 \\
  & 23.62 & $-3.65 \pm 0.02$ & 59 & 12 & 0 & 81 & 190 & 0 \\
  & 23.88 & $-4.02 \pm 0.04$ & 18 & 4 & 0 & 39 & 79 & 0 \\
  & 24.12 & $-4.35 \pm 0.05$ & 13 & 1 & 0 & 22 & 33 & 0 \\
  & 24.38 & $-4.62 \pm 0.07$ & 9 & 2 & 0 & 11 & 14 & 0 \\
  & 24.62 & $-4.86 \pm 0.10$ & 4 & 1 & 0 & 7 & 9 & 0 \\
  & 24.88 & $-4.79 \pm 0.09$ & 3 & 1 & 0 & 13 & 6 & 0 \\
  & 25.12 & $-5.38 \pm 0.18$ & 2 & 0 & 0 & 2 & 2 & 0 \\
  & 25.38 & $-5.05 \pm 0.12$ & 3 & 0 & 0 & 6 & 4 & 0 \\
  & 25.62 & $-5.46 \pm 0.19$ & 0 & 0 & 0 & 3 & 2 & 0 \\
  & 25.88 & $-5.87 \pm 0.31$ & 0 & 0 & 0 & 0 & 2 & 0 \\
  & 26.20 & $-6.38 \pm 0.69$ & 0 & 0 & 0 & 0 & 1 & 0 \\
  & 26.60 & $-5.90 \pm 0.40$ & 0 & 0 & 0 & 3 & 1 & 0 \\
1.0 \textless z \textless 1.3 & 23.62 & $-3.69 \pm 0.02$ & 86 & 14 & 61 & 82 & 367 & 46 \\
  & 23.88 & $-4.06 \pm 0.03$ & 39 & 8 & 45 & 48 & 158 & 29 \\
  & 24.12 & $-4.43 \pm 0.04$ & 21 & 3 & 19 & 19 & 64 & 38 \\
  & 24.38 & $-4.73 \pm 0.06$ & 9 & 3 & 6 & 13 & 30 & 23 \\
  & 24.62 & $-5.18 \pm 0.10$ & 0 & 1 & 3 & 6 & 12 & 27 \\
  & 24.88 & $-5.28 \pm 0.12$ & 1 & 2 & 1 & 4 & 7 & 6 \\
  & 25.12 & $-5.22 \pm 0.11$ & 7 & 1 & 0 & 6 & 3 & 2 \\
  & 25.38 & $-5.42 \pm 0.14$ & 0 & 0 & 0 & 4 & 7 & 1 \\
  & 25.62 & $-5.58 \pm 0.16$ & 0 & 0 & 0 & 1 & 6 & 2 \\
  & 25.88 & $-6.42 \pm 0.43$ & 0 & 0 & 0 & 0 & 1 & 2 \\
  & 26.20 & $-6.64 \pm 0.69$ & 0 & 0 & 0 & 0 & 1 & 0 \\
  & 26.60 & $-6.64 \pm 0.69$ & 0 & 0 & 0 & 0 & 1 & 0 \\
1.3 \textless z \textless 1.6 & 23.88 & $-3.86 \pm 0.01$ & 55 & 17 & 69 & 35 & 340 & 29 \\
  & 24.12 & $-4.20 \pm 0.02$ & 26 & 7 & 26 & 17 & 147 & 45 \\
  & 24.38 & $-4.62 \pm 0.04$ & 12 & 2 & 4 & 6 & 58 & 24 \\
  & 24.62 & $-5.02 \pm 0.06$ & 4 & 1 & 6 & 0 & 26 & 27 \\
  & 24.88 & $-5.20 \pm 0.10$ & 2 & 1 & 1 & 1 & 17 & 21 \\
  & 25.12 & $-5.31 \pm 0.11$ & 1 & 1 & 0 & 2 & 12 & 5 \\
  & 25.38 & $-5.64 \pm 0.15$ & 0 & 1 & 1 & 0 & 6 & 3 \\
  & 25.62 & $-5.79 \pm 0.15$ & 0 & 0 & 0 & 0 & 5 & 4 \\
  & 25.88 & $-5.80 \pm 0.25$ & 0 & 0 & 0 & 0 & 5 & 2 \\
  & 26.20 & $-5.79 \pm 0.49$ & 0 & 0 & 0 & 5 & 3 & 2 \\
  & 26.60 & $-6.71 \pm 0.69$ & 1 & 0 & 0 & 0 & 1 & 0 \\
  & 27.00 & $-6.71 \pm 0.69$ & 0 & 0 & 0 & 0 & 1 & 0 \\
1.6 \textless z \textless 2.0 & 23.88 & $-3.50 \pm 0.02$ & 29 & 17 & 51 & 27 & 589 & 31 \\
  & 24.12 & $-4.09 \pm 0.02$ & 27 & 19 & 34 & 17 & 338 & 44 \\
  & 24.38 & $-4.41 \pm 0.03$ & 17 & 5 & 18 & 9 & 137 & 52 \\
  & 24.62 & $-4.80 \pm 0.05$ & 3 & 3 & 9 & 6 & 53 & 40 \\
  & 24.88 & $-5.21 \pm 0.08$ & 1 & 1 & 1 & 1 & 20 & 29 \\
  & 25.12 & $-5.27 \pm 0.09$ & 0 & 1 & 0 & 5 & 12 & 18 \\
  & 25.38 & $-5.40 \pm 0.10$ & 0 & 2 & 0 & 1 & 6 & 5 \\
  & 25.62 & $-5.65 \pm 0.14$ & 0 & 1 & 0 & 1 & 7 & 3 \\
  & 25.88 & $-5.95 \pm 0.19$ & 0 & 0 & 0 & 0 & 3 & 2 \\
  & 26.20 & $-6.02 \pm 0.26$ & 0 & 0 & 0 & 0 & 2 & 1 \\
  & 26.60 & $-6.87 \pm 0.69$ & 0 & 0 & 0 & 0 & 1 & 3 \\
  & 27.00 & $-6.87 \pm 0.69$ & 0 & 0 & 0 & 0 & 1 & 0 \\
  & 27.40 & $-6.87 \pm 0.69$ & 0 & 0 & 0 & 1 & 0 & 0 \\
2.0 \textless z \textless 2.5 & 24.12 & $-3.96 \pm 0.02$ & 37 & 16 & 23 & 12 & 314 & 19 \\
  & 24.38 & $-4.45 \pm 0.03$ & 18 & 11 & 21 & 12 & 158 & 31 \\
  & 24.62 & $-4.77 \pm 0.04$ & 7 & 3 & 12 & 7 & 65 & 19 \\
  & 24.88 & $-5.10 \pm 0.06$ & 5 & 1 & 1 & 2 & 25 & 16 \\
  & 25.12 & $-5.27 \pm 0.08$ & 1 & 1 & 0 & 3 & 10 & 19 \\
  & 25.38 & $-5.46 \pm 0.09$ & 0 & 1 & 1 & 0 & 7 & 16 \\
  & 25.62 & $-5.51 \pm 0.10$ & 2 & 1 & 0 & 3 & 4 & 9 \\
  & 25.88 & $-5.72 \pm 0.13$ & 0 & 1 & 0 & 1 & 3 & 7 \\
  & 26.20 & $-5.86 \pm 0.19$ & 3 & 0 & 1 & 0 & 2 & 8 \\
  & 26.60 & $-6.69 \pm 0.49$ & 0 & 0 & 0 & 1 & 1 & 1 \\
  & 27.00 & $-6.69 \pm 0.49$ & 0 & 0 & 0 & 0 & 1 & 2 \\
  & 27.40 & $-6.69 \pm 0.49$ & 0 & 0 & 0 & 0 & 2 & 0 \\
2.5 \textless z \textless 3.3 & 24.38 & $-4.24 \pm 0.03$ & 20 & 11 & 33 & 9 & 251 & 13 \\
  & 24.62 & $-4.79 \pm 0.03$ & 18 & 5 & 16 & 5 & 113 & 14 \\
  & 24.88 & $-5.11 \pm 0.05$ & 2 & 4 & 7 & 1 & 47 & 15 \\
  & 25.12 & $-5.34 \pm 0.06$ & 5 & 1 & 1 & 1 & 17 & 22 \\
  & 25.38 & $-5.52 \pm 0.08$ & 2 & 1 & 1 & 0 & 8 & 11 \\
  & 25.62 & $-5.78 \pm 0.11$ & 1 & 1 & 2 & 0 & 8 & 11 \\
  & 25.88 & $-6.19 \pm 0.18$ & 0 & 1 & 0 & 0 & 3 & 19 \\
  & 26.20 & $-6.17 \pm 0.22$ & 1 & 0 & 0 & 0 & 2 & 18 \\
  & 26.60 & $-6.48 \pm 0.31$ & 0 & 0 & 0 & 0 & 2 & 6 \\
  & 27.00 & $-7.19 \pm 0.69$ & 0 & 0 & 0 & 0 & 1 & 1 \\
  & 27.40 & $-7.19 \pm 0.69$ & 0 & 0 & 0 & 0 & 1 & 0 \\
  & 27.80 & $-7.19 \pm 0.43$ & 0 & 0 & 0 & 0 & 0 & 1 \\
  & 28.20 & $-7.19 \pm 0.43$ & 0 & 0 & 0 & 0 & 0 & 1 \\
3.3 < z < 4.6 & 24.62 & $-4.90 \pm 0.43$ & 4 & 2 & 9 & 4 & 67 & 4 \\
  & 24.88 & $-5.62 \pm 0.43$ & 6 & 2 & 1 & 1 & 26 & 0 \\
  & 25.12 & $-5.93 \pm 0.43$ & 3 & 1 & 4 & 0 & 10 & 3 \\
  & 25.38 & $-6.17 \pm 0.43$ & 2 & 1 & 1 & 0 & 5 & 3 \\
  & 25.62 & $-6.69 \pm 0.43$ & 1 & 1 & 0 & 0 & 1 & 1 \\
  & 25.88 & $-6.01 \pm 0.43$ & 0 & 1 & 0 & 0 & 1 & 13 \\
  & 26.20 & $-5.77 \pm 0.43$ & 0 & 0 & 1 & 0 & 1 & 36 \\
  & 26.60 & $-5.90 \pm 0.43$ & 0 & 0 & 0 & 0 & 1 & 27 \\
  & 27.00 & $-6.35 \pm 0.43$ & 0 & 0 & 0 & 0 & 1 & 9 \\
  & 27.40 & $-6.88 \pm 0.43$ & 0 & 0 & 0 & 0 & 1 & 3 \\
  & 27.80 & $-7.36 \pm 0.43$ & 0 & 0 & 0 & 0 & 0 & 1 \\

\end{longtable}
\endgroup
\twocolumn


\section{Alternative Redshift Assignment Tests}
\label{appendix:redshift-tests}

This appendix examines how different redshift assignment method affects the RLFs and star-SFRDs) In addition to our primary PDF-based analysis with $z$>1, we test two alternative approaches, a uniform-$z$ assignment for non-XID sources and a single-best-$z$. These comparisons allow us to assess the robustness of our evolutionary fits and quantify systematic uncertainties arising from redshift incompleteness.

\begin{figure*}
    \centering
    \includegraphics[width=\textwidth]{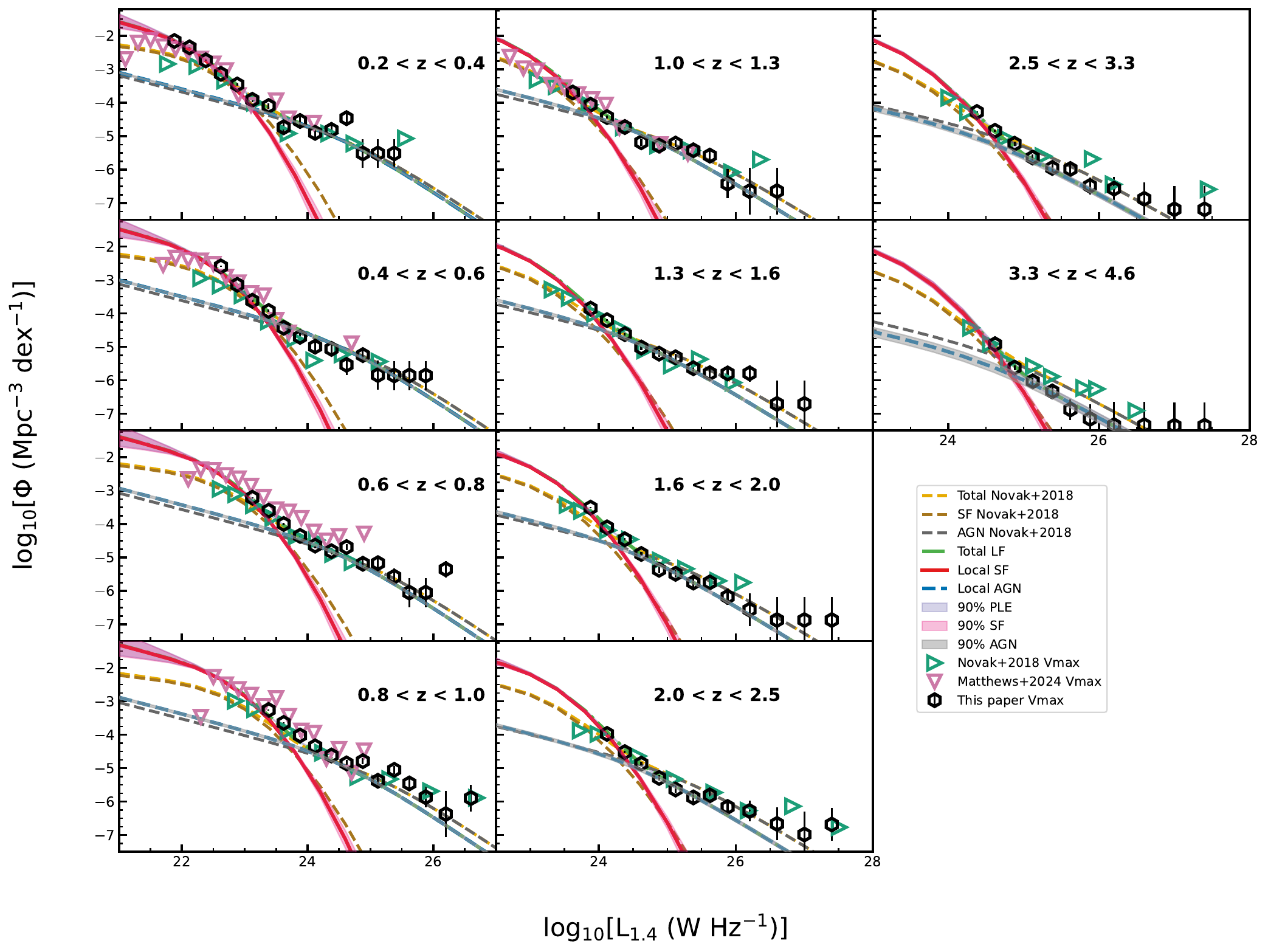}
    \caption{
        Total 1.4\,GHz radio luminosity function (LF) in ten redshift bins when redshifts of non-XID sources are assigned uniformly in $z$.
        Black hexagons: $1/V_{\max}$ measurements from this work, including statistical redshifts for radio sources without optical/NIR counterparts at $z>1$.
        Red solid and blue dashed curves: best-fit SF (PLE) and AGN (PDE) components; magenta and light-blue bands show their 90\% credible intervals.
        Yellow/brown/grey dashed curves: total, SF, and AGN LFs from \citet{Novak_2018} for comparison.
        For reference, $V_{\max}$ points from \citet{Novak_2018} and \citet{matthews2024confirmationsubstantialdiscrepancyradio} are also shown as green right-pointing and magenta down-pointing triangles, respectively.
    }
    \label{fig:rlf_uniform}
\end{figure*}

\begin{figure*}
    \centering
	\includegraphics[width=\textwidth]{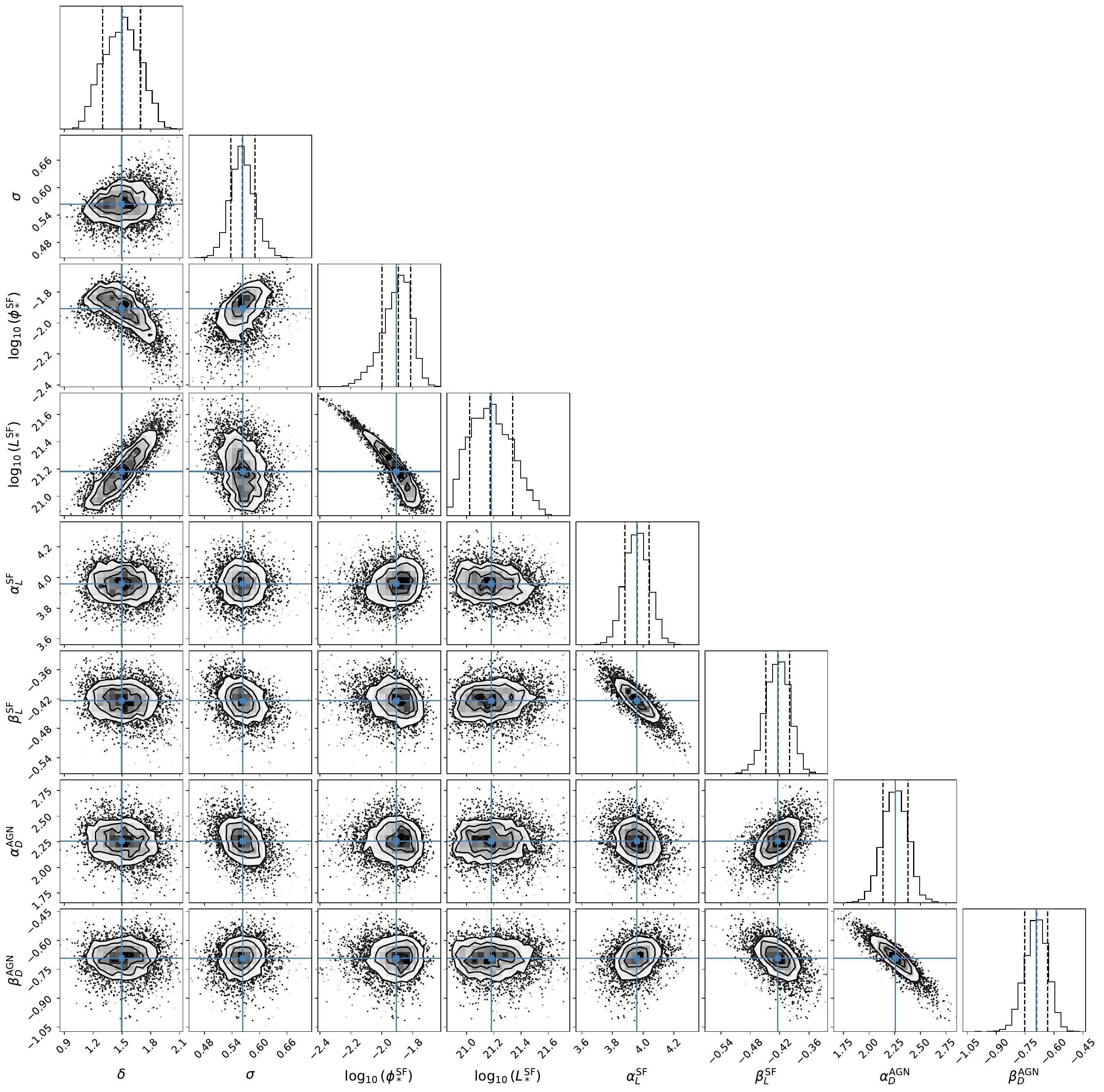}
    \caption{
        Corner plot displaying the posterior distributions of the eight best-fit parameters for the total radio LF evolution in the uniform-$z$ analysis.
    }
    \label{fig:corner_plot_1}
\end{figure*}

\begin{figure*}
    \centering
    \includegraphics[width=\textwidth]{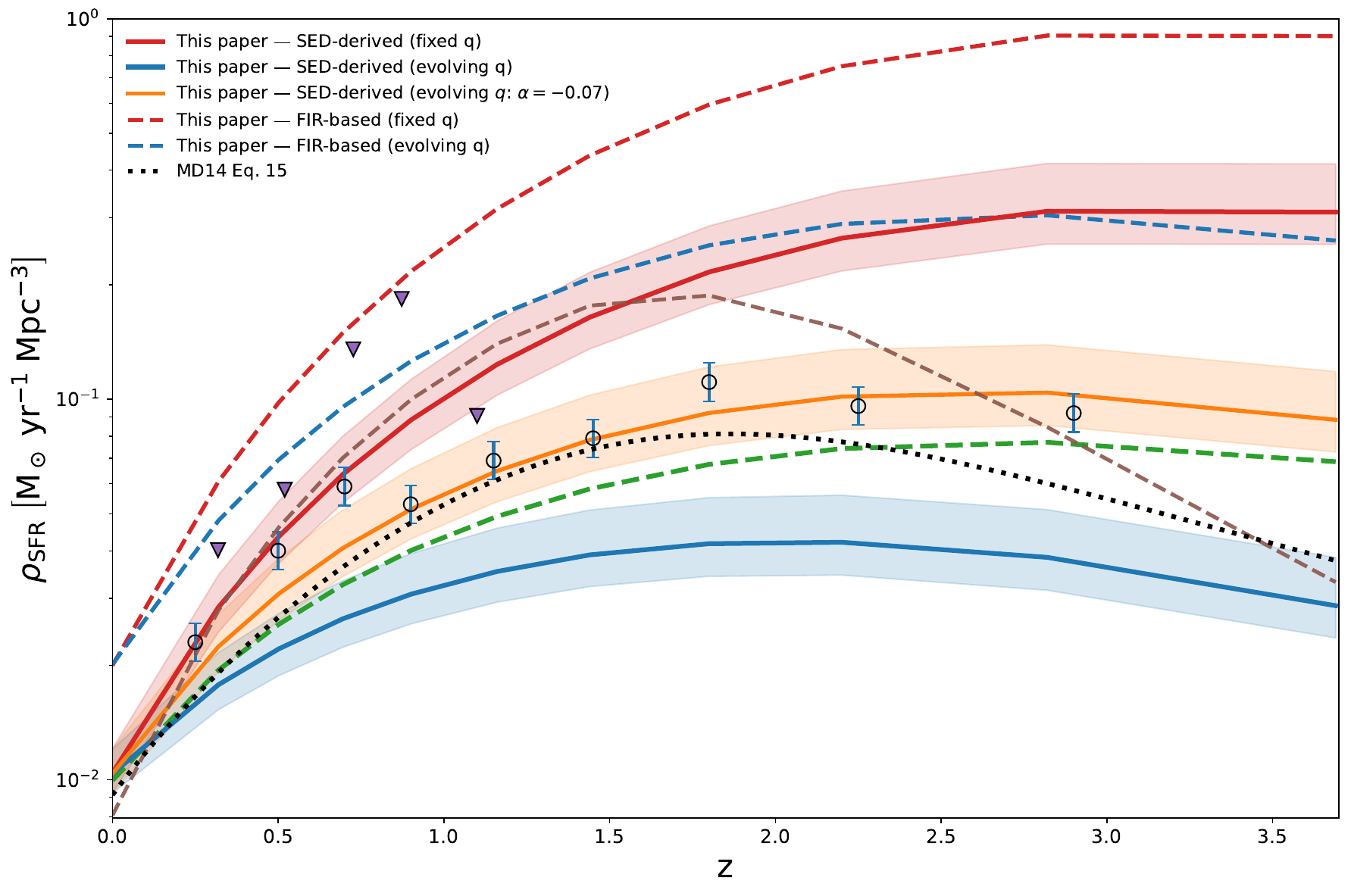}
    \caption{
    Star-formation rate density, $\rho_{\mathrm{SFR}}$, as a function of redshift where non-XID radio sources are assigned uniformly in $z$. Solid lines show the evolution assuming the SED-derived conversion to SFR from radio luminosity \citep{Cook2024}. The fixed-$q$ track is denoted by the red curve (with 68\% confidence band) and the evolving-$q$ track by the blue line (with 68\% confidence band).
    We also show an alternative line with $q_{\rm tot} \propto (1+z)^{-0.07}$ (orange). 
    For comparison, the dashed lines show the SFRD based on the FIR-based conversion from radio luminosity to SFR with fixed-$q$ (red) and evolving-$q$ (blue). The brown dashed curve shows the radio-only model from  \citet{Matthews_2021}. We also show the radio derived SFRD from \citet{Novak_2017} (green dashed line),  \citet{Cochrane2023} (blue error bars) and \citet{matthews2024confirmationsubstantialdiscrepancyradio} (purple inverted triangles). The black dotted line is the UV+IR compilation fit of \citet{Madau_2014}. We only show the confidence region for our baseline model for clarity.
    }
    \label{fig:sfrd_uniform}
\end{figure*}

\begin{figure*}
    \centering
    \includegraphics[width=\textwidth]{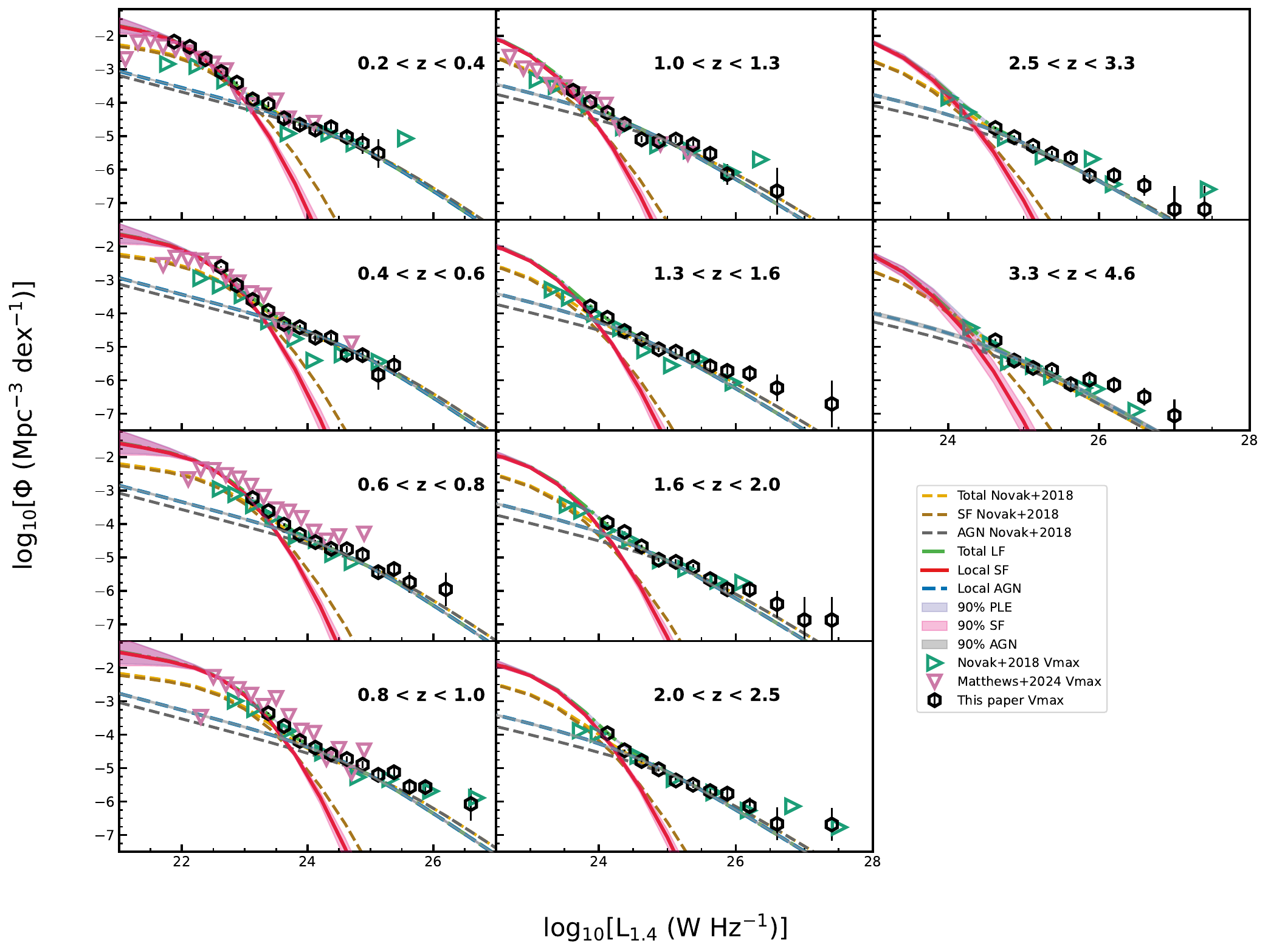}
    \caption{
        Total 1.4\,GHz radio luminosity function (LF) in ten redshift bins from the single-redshift analysis.
        Black hexagons: $1/V_{\max}$ measurements from this work, including statistical redshifts for radio sources without optical/NIR counterparts at $z>1$.
        Red solid and blue dashed curves: best-fit SF (PLE) and AGN (PDE) components; magenta and light-blue bands show their 90\% credible intervals.
        Yellow/brown/grey dashed curves: total, SF, and AGN LFs from \citet{Novak_2018} for comparison.
        For reference, $V_{\max}$ points from \citet{Novak_2018} and \citet{matthews2024confirmationsubstantialdiscrepancyradio} are also shown as green right-pointing and magenta down-pointing triangles, respectively.
    }
    \label{fig:rlf_single_appendix}
\end{figure*}

\begin{figure*}
    \centering
	\includegraphics[width=1.0\textwidth]{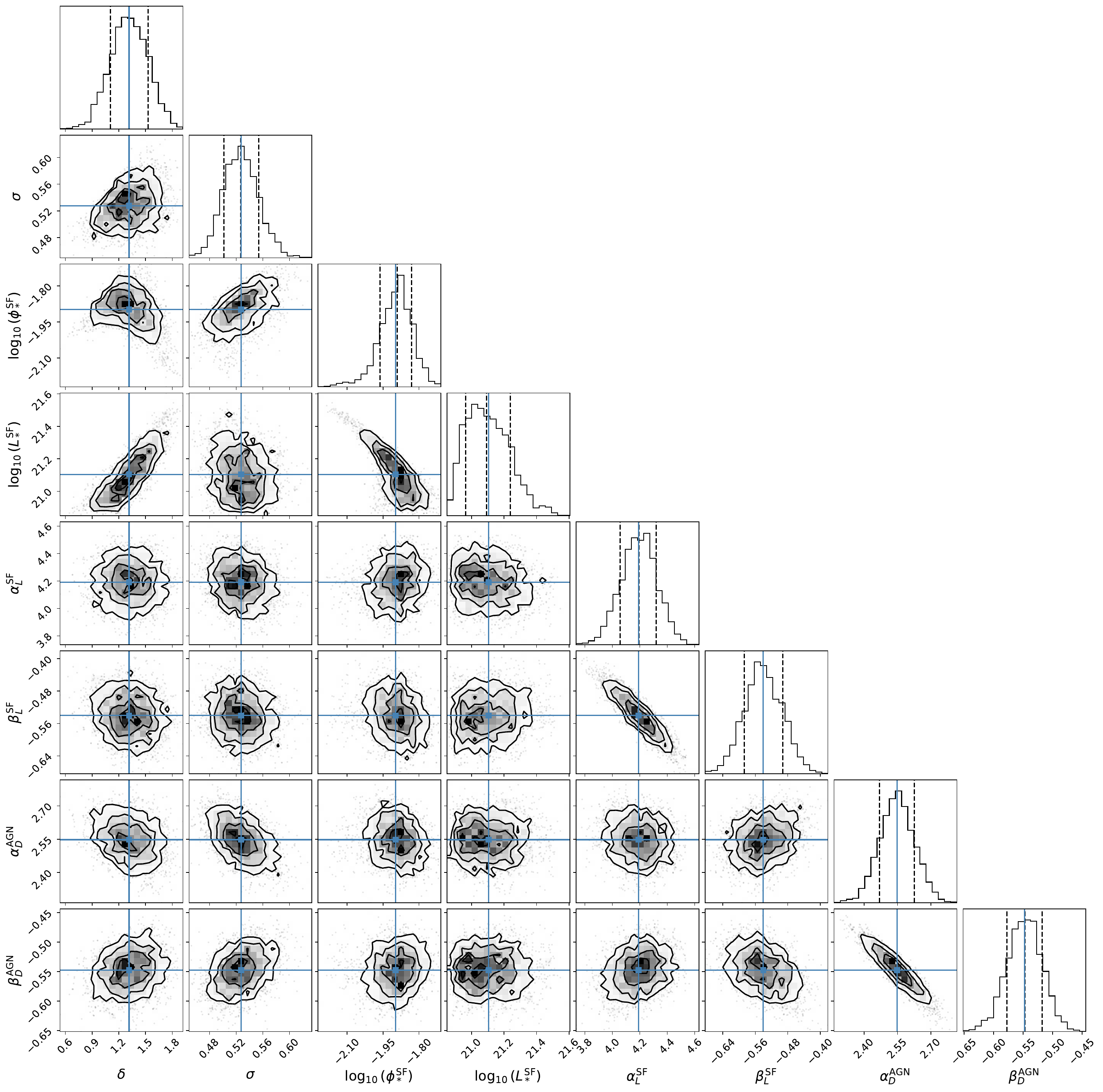}
    \caption{
        Corner plot displaying the posterior distributions of the eight best-fit parameters used to describe the total radio LF evolution in the single-redshift analysis.
    }
    \label{fig:single_corner_plot_appendix}
\end{figure*}

\begin{figure*}
    \centering
    \includegraphics[width=\textwidth]{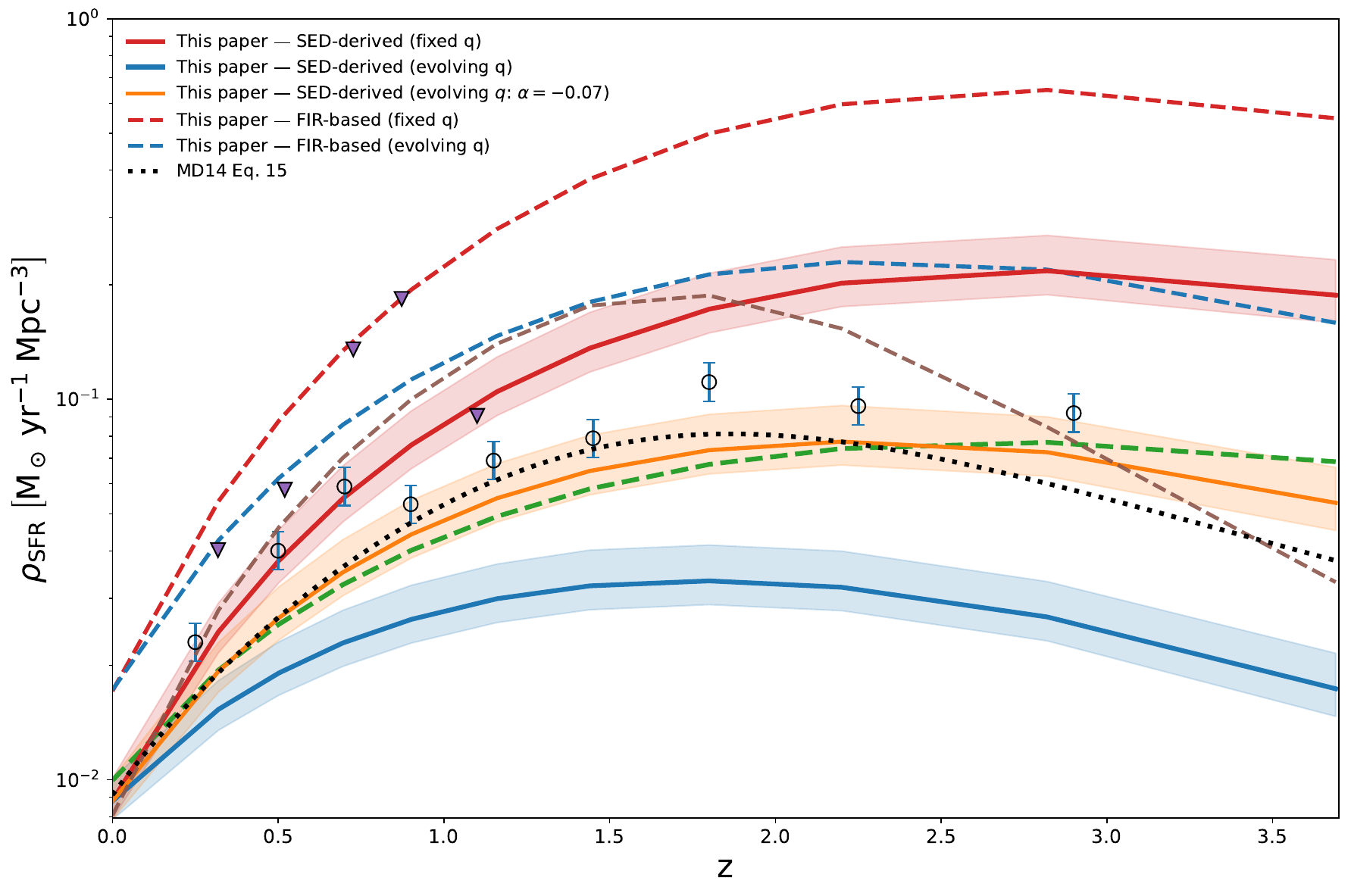}
    \caption{
    Star-formation rate density, $\rho_{\mathrm{SFR}}$, as a function of redshift from the single-redshift analysis. Solid lines show the evolution assuming the SED-derived conversion to SFR from radio luminosity \citep{Cook2024}. The fixed-$q$ track is denoted by the red curve (with 68\% confidence band) and the evolving-$q$ track by the blue line (with 68\% confidence band).
    We also show an alternative line with $q_{\rm tot} \propto (1+z)^{-0.07}$ (orange). 
    For comparison, the dashed lines show the SFRD based on the FIR-based conversion from radio luminosity to SFR with fixed-$q$ (red) and evolving-$q$ (blue). The brown dashed curve shows the radio-only model from  \citet{Matthews_2021}. We also show the radio derived SFRD from \citet{Novak_2017} (green dashed line),  \citet{Cochrane2023} (blue error bars) and \citet{matthews2024confirmationsubstantialdiscrepancyradio} (purple inverted triangles). The black dotted line is the UV+IR compilation fit of \citet{Madau_2014}. We only show the confidence region for our baseline model for clarity.
    }
    \label{fig:single_sfrd_appendix}
\end{figure*}

\clearpage
\section{RLFs and Evolution Parameters for COSMOS and XMM-LSS}
\label{appendix:cosmos-xmmlss}

This appendix presents RLFs derived separately for the COSMOS and XMM–LSS fields, together with their corresponding evolution parameters. These field-by-field measurements allow us to assess the consistency of the RLF shapes, evolution trends, and fitted parameters across the two MIGHTEE fields.

\begin{figure*}
    \centering
    \includegraphics[width=\textwidth]{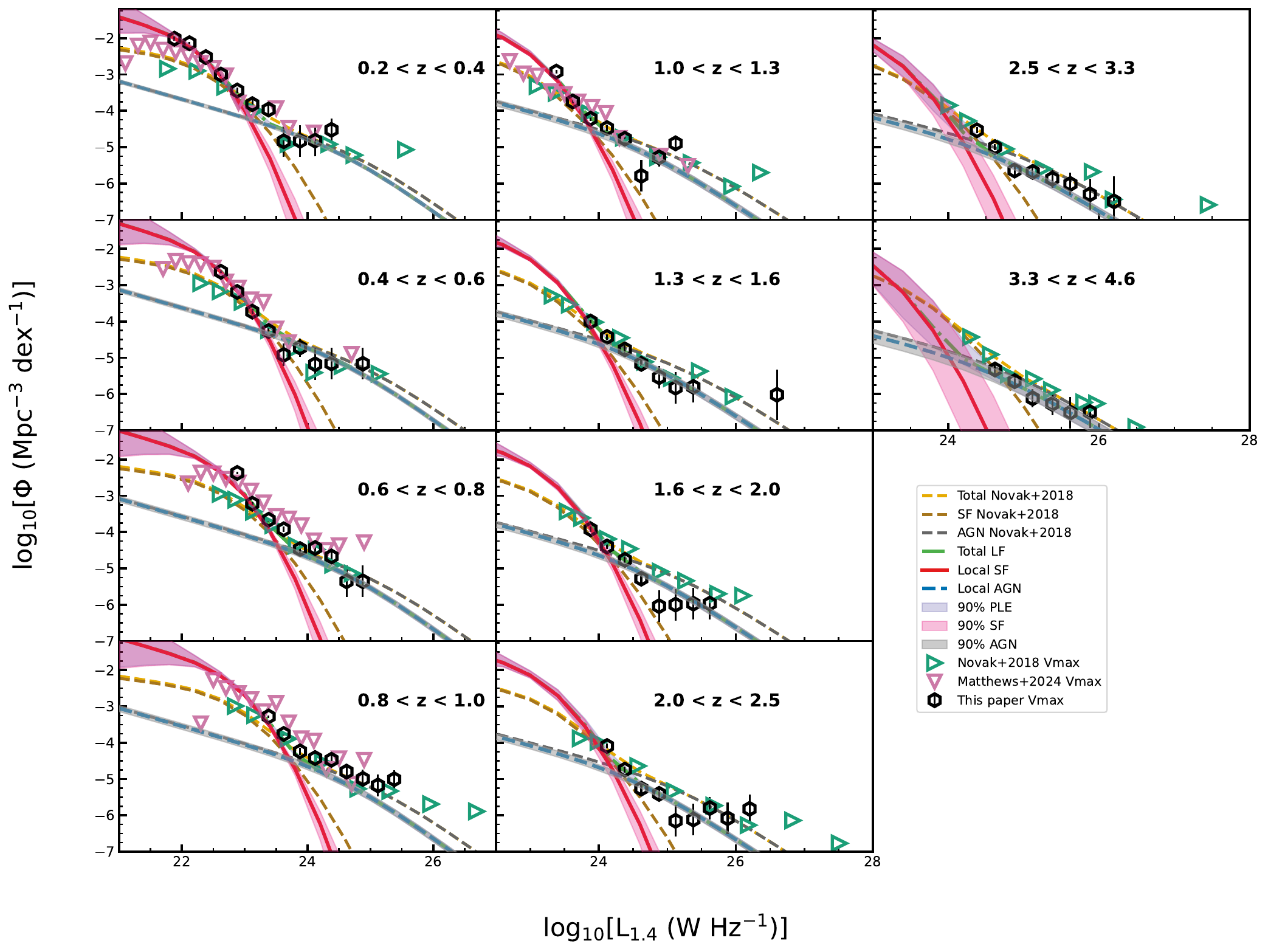}
    \caption{
        Total 1.4\,GHz radio luminosity function in ten redshift bins for the COSMOS field.
        Black hexagons: $1/V_{\max}$ measurements from this work, including statistical redshifts for radio sources without optical/NIR counterparts at $z>1$.
        Red solid and blue dashed curves: best-fit SF (PLE) and AGN (PDE) components for COSMOS; magenta and light-blue bands show their 90\% credible intervals.
    }
    \label{fig:rlf_single_cosmos}
\end{figure*}

\begin{figure*}
    \centering
    \includegraphics[width=\textwidth]{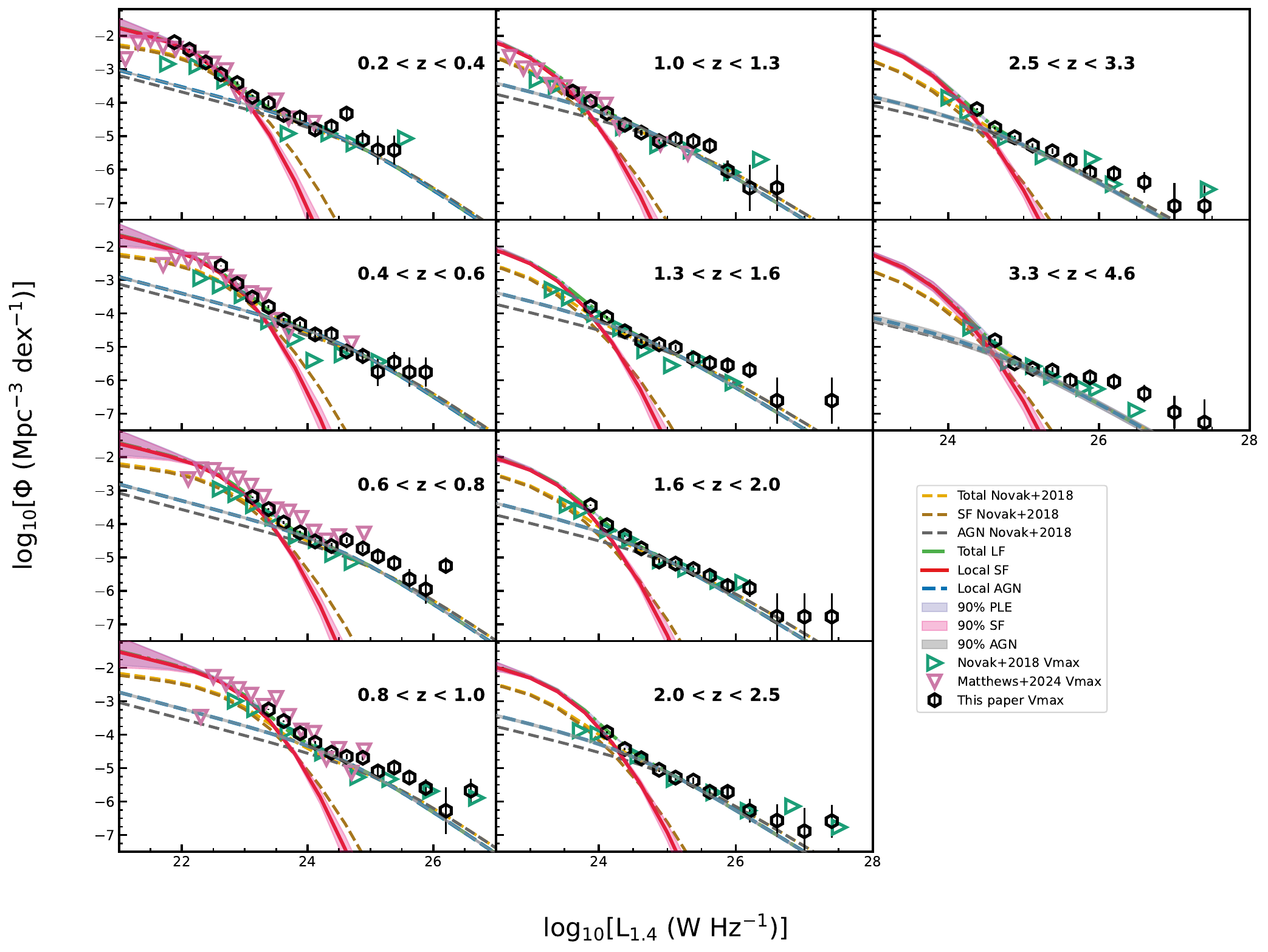}
    \caption{
        Total 1.4\,GHz radio luminosity function in ten redshift bins for the XMM-LSS field.
        Black hexagons: $1/V_{\max}$ measurements from this work, including statistical redshifts for radio sources without optical/NIR counterparts at $z>1$.
        Red solid and blue dashed curves: best-fit SF (PLE) and AGN (PDE) components for XMM-LSS; magenta and light-blue bands show their 90\% credible intervals.
    }
    \label{fig:rlf_single_xmmlss}
\end{figure*}

\
\vspace{6pt}


\bsp	
\label{lastpage}
\end{document}